\documentclass[aps,prf,reprint,superscriptaddress,onecolumn,nofootinbib]{revtex4-2}
\usepackage[usenames,dvipsnames]{xcolor}
\usepackage{lipsum}
\usepackage{graphicx}
\usepackage{array}
\usepackage{epstopdf}
\usepackage[justification=justified]{caption}
\usepackage{amsmath}
\usepackage[arrowdel]{physics}
\usepackage{multirow}
\usepackage{amsmath}
\usepackage{amsfonts}
\usepackage{amssymb}
\usepackage{mathrsfs}
\usepackage{mathtools}
\usepackage{tikz}
\usepackage{pgfplots}
\usepackage{csquotes}

\definecolor{color1}{RGB}{0,113,188}
\definecolor{color2}{RGB}{216,82,24}
\definecolor{color3}{RGB}{236,176,31}

\newsavebox\mycircle
\savebox\mycircle{                                                                   
  \begin{tikzpicture}                                                           
    \tikz\draw[line width=0.3mm, color=color1] (0.05,0.05)
circle (0.5ex); 
  \end{tikzpicture}                                                             
} 

\newsavebox\mycircleblack
\savebox\mycircleblack{                                                                   
  \begin{tikzpicture}                                                           
    \tikz\draw[line width=0.3mm, color=black] (0.05,0.05)
circle (0.5ex); 
  \end{tikzpicture}                                                             
} 

\newsavebox\myline
\savebox\myline{                                                                   
  \begin{tikzpicture}                                                           
    \tikz\draw[line width=0.3mm, color=color2] (0,-0.05)--(0.5,-0.05);
  \end{tikzpicture}                                                             
} 

\newsavebox\mylineblue
\savebox\mylineblue{                                                                   
  \begin{tikzpicture}                                                           
    \tikz\draw[line width=0.3mm, color=color1] (0,-0.05)--(0.5,-0.05);
  \end{tikzpicture}                                                             
} 

\newsavebox\mylineblack
\savebox\mylineblack{                                                                   
  \begin{tikzpicture}                                                           
    \tikz\draw[line width=0.3mm, color=black] (0,-0.05)--(0.5,-0.05);
  \end{tikzpicture}                                                             
}

\newsavebox\mylinedashed
\savebox\mylinedashed{                                                                   
  \begin{tikzpicture}                                                           
    \tikz\draw[dashed ,step=0.2, color=black] (0,-0.05)--(0.5,-0.05);
  \end{tikzpicture}                                                             
}

\newsavebox\mylinedashedblack
\savebox\mylinedashedblack{                                                                   
  \begin{tikzpicture}                                                           
    \tikz\draw[dotted ,step=0.2, color=black] (0,-0.05)--(0.5,-0.05);
  \end{tikzpicture}                                                             
}

\newsavebox\mylinedashedblue
\savebox\mylinedashedblue{                                                                   
  \begin{tikzpicture}                                                           
    \tikz\draw[dashed ,step=0.2, color=color1] (0,-0.05)--(0.5,-0.05);
  \end{tikzpicture}                                                             
}

\newsavebox\mylinedashedred
\savebox\mylinedashedred{                                                                   
  \begin{tikzpicture}                                                           
    \tikz\draw[dashed ,step=0.2, color=color2] (0,-0.05)--(0.5,-0.05);
  \end{tikzpicture}                                                             
}

\newsavebox\myboxblue                                                           
\savebox\myboxblue{                                                                   
  \begin{tikzpicture}                                                           
    \draw [line width=0.3mm, color=color1] (0.1,0.1) rectangle(0.3,0.3);                                       
  \end{tikzpicture}                                                             
}

\newsavebox\myboxred
\savebox\myboxred{                                                                   
  \begin{tikzpicture}                                                           
    \draw [line width=0.3mm, color=color2] (0.1,0.1) rectangle(0.3,0.3);                                       
  \end{tikzpicture}                                                             
}

\newsavebox\myboxblack                                                           
\savebox\myboxblack{                                                                   
  \begin{tikzpicture}                                                           
    \draw [fill=black] (0.1,0.1) rectangle(0.3,0.3);                                       
  \end{tikzpicture}                                                             
}  

\newsavebox\myboxgray                                                           
\savebox\myboxgray{                                                                   
  \begin{tikzpicture}                                                           
    \draw [fill=gray] (0.1,0.1) rectangle(0.3,0.3);                                       
  \end{tikzpicture}                                                             
}   

\newsavebox\myboxwhite
\savebox\myboxwhite{                                                                   
  \begin{tikzpicture}                                                           
    \draw [fill=white] (0.1,0.1) rectangle(0.3,0.3);                                       
  \end{tikzpicture}                                                             
}

\begin{document}

\author{Igor A. Maia}
\affiliation{Pprime Institute, CNRS, Universit\'e de Poitiers, ENSMA,Poitiers, France}
\author{Guillaume Br\`es}
\affiliation{Cascade Technologies Inc., Palo Alto, CA94303, USA}
\author{Lutz Lesshafft}
\affiliation{Laboratoire d'Hydrodynamique, CNRS - Ecole Polytechnique, Palaiseau, France}
\author{Peter Jordan}
\affiliation{Pprime Institute, CNRS, Universit\'e de Poitiers, ENSMA,Poitiers, France}

\title{The effect of a flight stream on subsonic turbulent jets}

\begin{abstract}


This study concerns a turbulent jet at Mach number $M_j=0.9$, subject to a uniform external flow stream at $M_f=0.15$. The analysis combines experimental and numerical databases, spectral proper orthogonal decomposition (SPOD) and linear  modelling. The experiments involve Time-Resolved, Stereo PIV measurements at different cross-sections of the jet. A companion large-eddy simulation was performed with the same operating conditions using the \enquote{CharLES} solver by Cascade Technologies in order to obtain a complete and highly resolved 3D database. We assess the mechanisms that underpin the reduction in fluctuation energy that is known to occur when a jet is surrounded by a flight stream. We show that this energy reduction is spread over a broad region of the frequency-wavenumber space and involves, apart from the known stabilization of the modal Kelvin-Helmholtz (KH) instability, the attenuation of flow structures associated with the non-modal Orr and lift-up mechanisms. Streaky structures, associated with helical azimuthal wavenumbers and very slow time scales, are the most strongly affected by the flight stream, in terms of energy attenuation and spatial distortion. The energy reductions are accompanied by a weakening of the low-rank behaviour of the jet dynamics. These trends are found to be consistent, to a great extent, with results of a local linear model based on the modified mean flow in the flight stream case.

\end{abstract}

\maketitle

\section{Introduction}
\label{sec:intro}

Recent research on turbulent shear flows has been driven by a need to obtain simplified descriptions of flow dynamics that would reduce the Navier-Stokes system to a form adapted to describe a subspace of essential mechanisms, with respect to an observable of interest (drag, momentum, noise, etc). Such mechanisms are often associated with large-scale (with respect to turbulence integral scales) organised motion in the form of coherent structures. It is their widespread presence in turbulent flows \citep{HussainCoherentStructures} that motivates the development of simplified models. 

Coherent structures are easily identifiable in unstable laminar flows, where the exponential growth of small disturbances underpins transition to turbulence. In that case, they can be modelled as instability waves using linear stability theory, where linearisation is performed about the laminar base state. In the turbulent regime, organised motion often persists and early observations in mixing layers \citep{Brown&Roshko1974} and jets \citep{MolloChristensen1967} revealed a remarkable resemblance with linear instability mechanisms found at lower Reynolds numbers. Eduction of coherent structures in high-Reynolds-number, turbulent jets, is challenging on account of their low fluctuation energy and their stochastic space-time organisation. That is why early attempts to study coherent structures in jets relied on external periodic forcing, which raises the coherent-structure energy above the background level and enhances their organisation \citep{CrowChampagne, Moore1977, HussainZaman1980_1, HussainZaman1980_2, HussainZamanpreferred, PetersenSamet}.

More recently, there has been considerable progress in the identification and modelling of coherent structures in unforced jets, thanks to progress in experimental techniques, the use of advanced signal processing approaches such as Proper Orthogonal Decomposition (POD) \citep{Lumley1967} and Spectral Proper Orthogonal Decomposition (SPOD), and linear mean-flow modelling. For instance, Kelvin-Helmholtz (KH) type wavepackets have been identified in the hydrodynamic pressure near-field \citep{SuzukiColonius, GudmundssonColonius, BreakeyPhysRevFluids} and in the velocity field \citep{Wavepacketsvelocity, CoherenceVincent} of unforced turbulent jets. The experimentally-educed wavepackets are found to be in good agreement with solutions to the Parabolised Stability Equations (PSE) in the initial jet region for azimuthal wavenumbers $m=0,1,2$ and frequencies in the range $0.3 \leqslant St \leqslant 1$, where $St=fU_j/D$ is the Strouhal number, (with $f$ being the frequency, $U_j$ the jet exit velocity and $D$ the nozzle diameter). Later, \citet{SasakiRapids} showed, using large-eddy simulation data, that agreement persists for Strouhal numbers as high as $St=4$ for the axisymmetric and first three helical azimuthal wavenumbers.

The above studies show compelling evidence for the existence of modal convective instability mechanisms in jets. More recently, attention has been turned to non-modal linear mechanisms such as the Orr \citep{Orr} and Lift-up \citep{Brandt2014} mechanisms, that give rise to different flow structures. These mechanisms, known to be important in the dynamics of wall-bounded flows (see the reviews by \cite{Jimenez2018} and \cite{Brandt2014}), were also observed in early experiments in jets \cite{BeckerMassaro, BrowandLaufer, Dimotakis1983, YuleJFM1978, Agui1988}. Recent studies have shown that they are dominant at very low Strouhal numbers, and their most salient features can be modelled through linear mean-flow analysis. \citep{Garnaud, Jeunetal, SemeraroAIAA2016, TissotJFM2017, SchmidtetalJFM2018, LesshafftPFR2019, NogueiraJFM2019, PickeringJFM2020}.

There is now a well-documented body of work concerning modal and non-modal mechanisms in jets, and linear mean-flow analysis provides a valuable framework for understanding the dynamics of coherent structures associated with these mechanisms, and eventually estimating and controlling them. One important caveat needs to be mentioned,
though: the validity of linear mean-flow analysis must be demonstrated \textit{a posteriori}, since the flow linearisation procedure does not result in an exact equation \cite{JordanColoniusReview}, and there is no guarantee that the same models will hold under different flow configurations. Turbulent jets in the presence of a flight stream, which are the focus of this work, are one example of a flow configuration for which the validity of linear mean-flow models needs to be explored through comparisons with data.

\subsection{Jets with flight stream}

Jets subject to a uniform external flight stream are a flow configuration of interest to the aeronautic industry, as it mimics the effect of forward flight in real aircraft. It is known that a flight stream modifies the mean flow development, producing a stretching of the potential core, reduction of the shear-layer thickness, turbulent kinetic energy and a reduction of the associated radiated sound levels \citep{TannaMorris1977}. Such mean-flow modifications have a stabilizing effect on the modal KH mechanism, as shown by \cite{MuchalkeHermann1982} with an inviscid locally parallel stability analysis. More recently, \cite{SoaresAIAA2020} extended the analysis by taking into account the jet mean-flow divergence in a PSE formulation. The results confirm a reduction in wavepacket growth rates with increasing flight stream velocity, followed by an increase in their convection velocity. The dynamics of axisymmetric wavepackets in the presence of a coflow have also been studied by \cite{Garnaud_AIAA_2013} using a global resolvent analysis. The flow response modes (both in the near-field and the far-field) were found to be in good qualitative agreement with (nonlinear) flow data, thus confirming that the linear-mean-flow framework can capture the effect of flight on axisymmetric coherent structures associated with modal growth mechanisms in jets.

\begin{figure}
\centering
\includegraphics[trim=0cm 0cm 0cm 0cm, clip=true,width=0.7\linewidth]{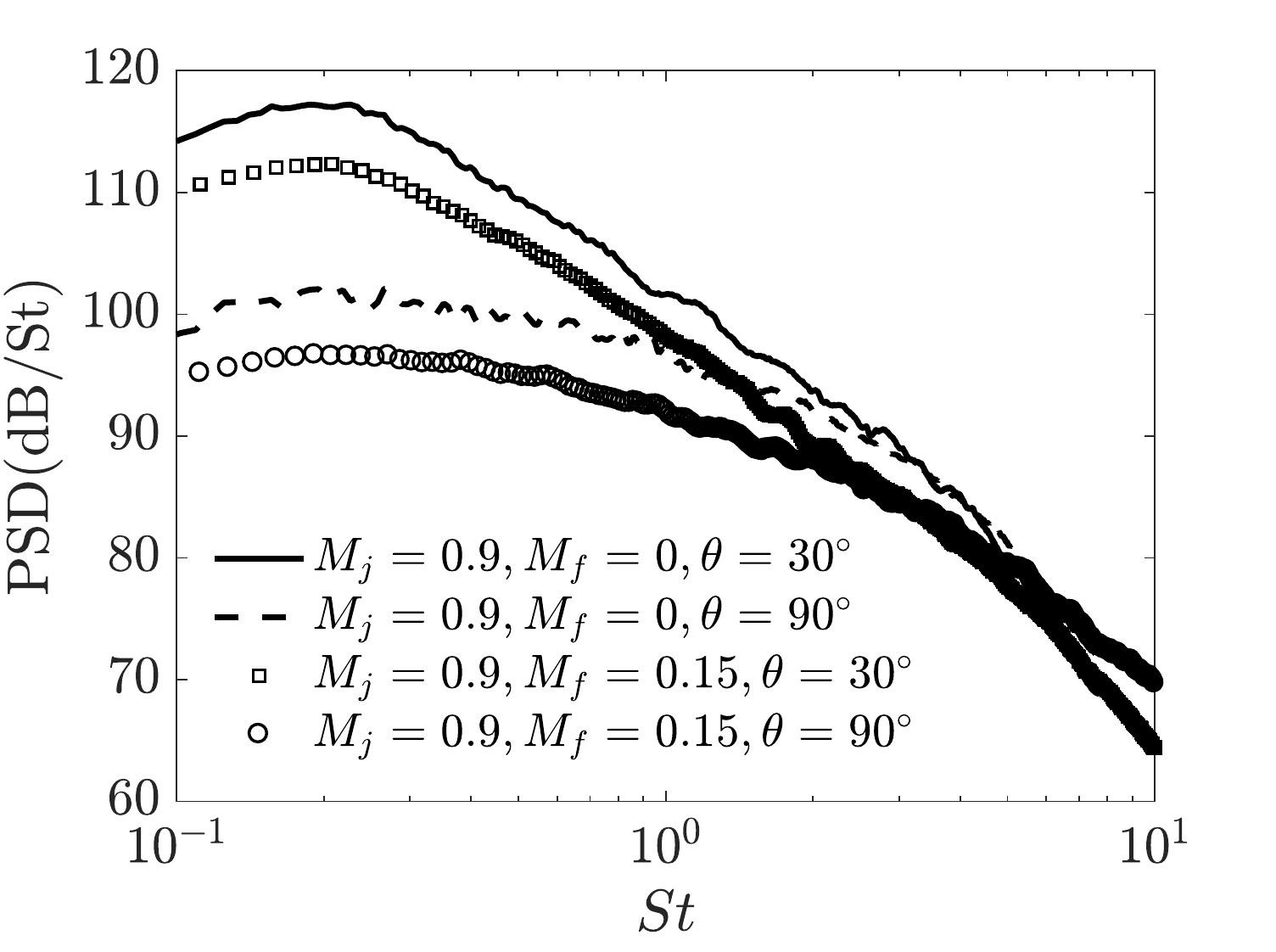}
\caption{Power spectral densities (PSD) of pressure measured in the acoustic field of jets with and without flight stream at polar angles of $\theta=30^{\circ}$ and $\theta=90^{\circ}$ from the jet axis. $St$ stands for Strouhal number.}
\label{fig1}
\end{figure}

Most studies on the effect of a flight stream on turbulent jets have focused on sound-radiation aspects \citep{VonGlahn, Cocking, Bushel, Packman, Plumbee, Bryce1984, MorfeyTester, VishwanathanFlight}. In this work we characterise the effect of a flight stream on the turbulent flow field of a subsonic jet. As discussed above, there is now an extensive characterisation in the literature of modal and non-modal linear mechanisms in jets in \enquote{static} conditions. And although the effect of the flight stream on the KH instability has been studied \citep{MuchalkeHermann1982, SoaresAIAA2020}, to the best of our knowledge no studies so far have characterised the changes in non-modal linear instability mechanisms (Orr and Lift-up) in flight conditions. Indeed, there has not yet been a demonstration that linear mean-flow-models can correctly capture the effect of flight on coherent structures associated with non-modal mechanisms. This is what motivates the present work. Figure \ref{fig1} shows spectra of acoustic pressure measured at polar angles of $\theta=30^{\circ}$ and $\theta=90^{\circ}$ from the jet axis, in the setup described in \S \ref{sec:expsetup}. The flight stream is seen to produce an effect at both angles for a broad range of Strouhal numbers, as a consequence of changes to the turbulent field. At low frequencies and small angles from the jet axis, represented here by $\theta=30^{\circ}$, the main sound-producing mechanism is associated with axisymmetric ($m=0$) jittering KH wavepackets \citep{CavalieriJitterJSV, AndreJFM2012, MaiaPRSA2019}. At sideline, higher azimuthal wavenumbers become dominant \citep{AndreJFM2012}. The observed reduction in Sound Pressure Levels (SPL) in the presence of the flight stream is consistent with Lighthill's eigth power law \citep{Lighthill} if we consider the centerline-to-freestream velocity difference, $dU = U_j-U_{\infty}$, as a characteristic flow velocity. $dU$ reduces from 0.9 to 0.75 with the flight stream. From Lighthill's law, the acoustic power sould be reduced by a factor of $10\mathrm{log}_{10}\left((0.75/0.9)^{8}\right) = -6.33$dB. For measurements taken at the same distance for both flow conditions, this translates into the same difference in SPL, and is close to what is reported in figure \ref{fig1}, for both measurement angles \footnote{We are grateful to an anonymous reviewer for pointing out the consistency with Lighthill's eighth power law.}. However, Lightill's law does not reveal the details of how the change in turbulent kinetic energy is organised in the frequency-wavenumber spectrum, or how it affects the dynamics of coherent structures which are likely associated with those changes. A thorough characterisation of frequency-wavenumber spectra with and without the flight stream is a key aspect of this study, and we will be interested in exploring to what degree the changes in the spectrum of turbulent velocity fluctuations can be associated with changes in the aforesaid categories of coherent structures.

In the case of a laminar jet, linear, fixed-point analysis of the effect of flight stream could be expected to reveal an attenutation of all shear-driven linear instability mechanisms, because of the reduction of shear that is produced by the flight stream. However, for fully turbulent jets, as mentioned above, it is not clear \textit{a priori} whether changes in the energy spectrum are associated with linear or nonlinear mechanisms. A direct connection between the flight stream and changes in coherent structures associated with linear mechanisms is therefore much less clear. A central objective of the present work is, therefore, in addition to an extensive characterisation of the changes produced by a flight stream, to explore the extent to which the observed changes can be explained by linear mean-flow mechanisms.

Our analysis combines experimental and numerical databases, modal decomposition and locally-parallel, linear mean-flow analysis. The experiments were performed at the Pprime Institute and consisted of Time-Resolved (TR), Stereo PIV measurements at different cross-sections of a jet at Mach number $M_{j}=0.9$ subject to a uniform external flight stream at $M_{f}=0.15$. Throughout the paper, the database is systematically compared to a case with no flight stream, $M_{f}=0$. Companion LES databases were generated for the same operating conditions using the \enquote{CharLES} solver by Cascade Technologies \citep{BresAIAA2017, BresJFM2018} in order to obtain a high-fidelity 3D database, allowing us to perform global SPOD.

The remainder of the paper is organised as follows: in \S \ref{sec:expsetup} we describe the experimental setup and provide details about flow conditions and PIV treatment. This is followed by a description of the numerical setup in \S \ref{sec:les_setup}. In \S \ref{sec:stats} we show the effect of the flight stream on first-order statistics and relevant mean-flow quantities. This part consists in a validation of the two databases with respect to previously reported results. In chapter \S \ref{sec:en_dist_m} and \S \ref{sec:local_spod} the effect of the flight stream is investigated in more depth through Fourier decompositions and local SPOD. Results of a local mean-flow model are presented in \S \ref{sec:resolvent} and interpreted in light of the PIV data. Global SPOD analysis using the LES data is discussed in \S \ref{sec:global_spod} and finally some concluding remarks are given in \S \ref{sec:conclusions}.

\section{Experimental setup}
\label{sec:expsetup}

The experiments were performed at the \textit{Bruit \& Vent} jet facility of the Pprime Institute in Poitiers, France. We performed measurements in jets at Mach number of $M_{j}=U_{j}/c_{\infty}=0.9$, where $U_{j}$ is the jet exit velocity and $c_{\infty}$ the ambient speed of sound. The corresponding Reynolds number was $Re = U_{j}D/\nu = 1 \times 10^{6}$, where $\nu$ is the kinematic viscosity and $D$ is the nozzle diameter, which was $50$mm. The jet was subject to an external uniform stream which could attain a maximum Mach number of $M_{f}=U_{f}/c_{\infty}=0.15$, where $U_{f}$ is its exit velocity. The flight stream comes from an outer convergent section that surrounds the main nozzle and finishes in a straight section of diameter $600$mm. The operating conditions in terms of nozzle-pressure ratio are $NPR=P_{t_{j}}/P_{\infty}=1.7$ for the main jet and $NPR=P_{t_{f}}/P_{\infty}=1.006$ for the flight stream, with $P_{t_{j,f}}$ the total pressure and $P_{\infty}$ the ambient pressure. The experiments were performed in isothermal conditions, the static temperatures of the main jet and the flight stream being controlled to ensure this. The internal and external boundary layers were tripped so as to produce turbulence, similar to what was done in previous studies in static conditions \citep{BresJFM2018}, at $3D$ upstream of the nozzle exit.

We performed a series of low-frequency-2D and Time-Resolved (TR), Stereo-PIV measurements. The former were used to characterise the effect of the flight stream in zero and first-order statistics, such as mean-flow distortion and reduction of turbulent kinetic energy, with increasing flight stream Mach number, varying from $M_{f}=0$ to $M_{f}=0.15$. In the 2D setup, the laser sheet is aligned with the jet axis, allowing a fine discretisation in the streamwise direction. The setup consisted of two Lavision Imager LX cameras and a Quantel Evergreen 532nm, 200mJ laser. The images had a resolution of 4920x3280 pixels, which allowed us to cover an $x-y$ plane in the range $x \in [0.15D,10D]$ $y \in [-1.9D,1.9D]$, resulting in one vector every 0.009D. A total of 10000 PIV images were acquired at a rate of $4$Hz, which was found to be sufficient to converge mean and rms fields. Both the main jet and the flight stream were seeded with glycerine smoke particles with diameters in the range 1-2$\mu$m, similar to what was done in previous PIV campaigns \citep{CoherenceVincent}. These experiments were used to characterise velocity statistics with four different flight stream levels corresponding to $M_{f}=0, 0.05, 0.1, 0.15$.

The characterisation of different instability mechanisms requires time-resolved flow data decomposed in azimuthal Fourier modes. This was achieved with the Stereo TR-PIV setup, which featured two Photron SAZ cameras and a 532nm, 2x60W continuum Mesa laser. The cameras were positioned in a forward scattering configuration, in order to assure maximal light intensities. The angle formed by the cameras and the laser sheet was 45$^{\circ}$ (with both cameras on the same side of the laser source), and Scheimpflug adaptors were used to ensure a correct focus on the entirety of the field of view. A resolution of 1024x1024 pixels was used to focus on a field of view in the range $y \in [-1D,1D]$, $z \in [-1D,1D]$, resulting in one vector every 0.026D. The image acquisition was performed in double frame mode at a frequency of $10$kHz, corresponding to a Strouhal number of $St=fU_j/D=1.6$, giving a total of 21000 images per plane measured. The laser sheet had a thickness of 3mm, and the time between laser pulses was set to 2.5$\mu$s, resulting in a maximum displacement of 4 pixels across the laser sheet. For both the 2D and TR-Stereo configurations, PIV  computations were carried out using a commercial software which performed a multi-pass iterative PIV algorithm \citep{Scarano2002}. The PIV interrogation area size was set to 64x64 pixels for the first pass, decreased to 16x16 pixels with an overlap of 50 between two neighbouring interrogation areas. Each instantaneous snapshot was interpolated into a polar grid, $r$-$\theta$ using a bi-cubic interpolation in order to perform a Fourier decomposition in azimuth \citep{CoherenceVincent}. The measurements in the TR-Stereo configuration were carried out for two flight stream conditions, $M_{f}=0$ and $M_{f}=0.15$, the same used in the numerical databases. Several PIV planes were measured in both conditions at different streamwise stations ranging from $x/x_{c}=0.04$ to $x/x_{c}=2$, where $x_{c}$ is the potential core length.

\begin{figure}
\centering
\includegraphics[trim=0cm 5cm 0cm 4cm, clip=true,width=\linewidth]{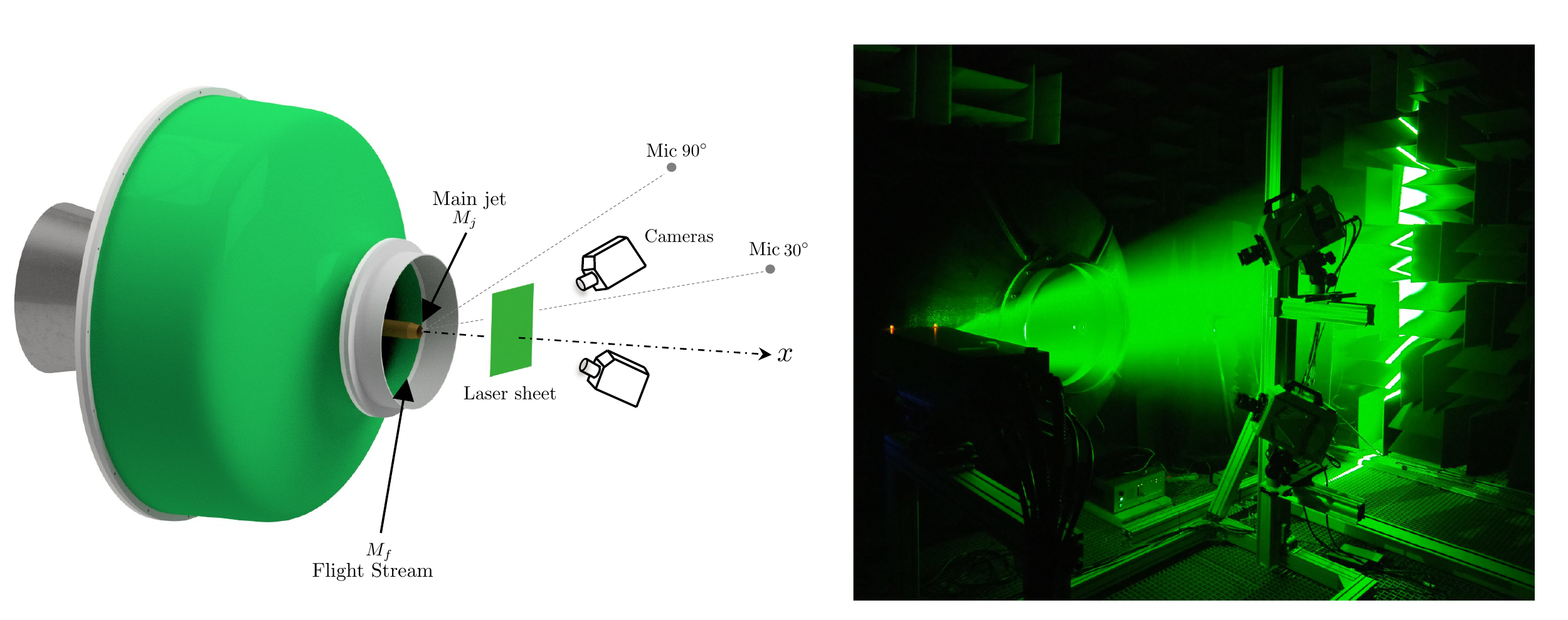}
\caption{Schematic of the jet facility showing the main elements of the Stereo PIV system, along with a picture of the experiment setup in the wind tunnel.}
\label{fig2}
\end{figure}

Figure \ref{fig2} shows a schematic of the jet facility and the TR-PIV setup along with a picture of the setup in the wind tunnel. A summary of the operating conditions of the experiments are shown in Table \ref{tb_exp}. Acoustic measurements were performed (in the absence of the PIV setup) at polar angles of $30^{\circ}$ and $90^{\circ}$ to the jet axis, at a distance of $50$D from the nozzle. Acoustic data was sampled at a rate of $St_{s}=32.7$ for 30s. PSDs reported in figure \ref{fig1} were computed with Welch's method, using 5860 blocks with 50\% overlap. Complementary Pitot tube and hot-wire measurements were performed for case 4 in order to characterise the boundary layers, and are described in more detail in \S\ref{sec:stats}.

\begin{table}
\centering
\begin{tabular}{c c c c c c c}
\multicolumn{7}{c}{}\\
\hline
 Case & $Re$ & $M_{j}$ & $M_{f}$  & $T_{j}/T_{f}$ & 2D-PIV & TR-Stereo PIV \\
1 & $10^6$ & $0.9$ & $0$ & 1 &\checkmark & \checkmark \\
2 & $10^6$ & $0.9$ & $0.05$ & 1 & \checkmark & - \\
3 & $10^6$ & $0.9$ & $0.10$ & 1 & \checkmark & - \\
4 & $10^6$ & $0.9$ & $0.15$ & 1 & \checkmark & \checkmark\\
\hline
\end{tabular}
 \caption{Operating conditions and experimental parameters of the experiments. The last two columns indicate which kind of PIV setup was used for each condition.}
 \label{tb_exp}
\end{table}

\section{Numerical setup}
\label{sec:les_setup}

To complement the experiments, the jet configurations with and without flight stream are investigated with high-fidelity large eddy simulations using the compressible flow solver "CharLES" developed at Cascade Technologies \citep{BresAIAA2017}. Results for the isothermal Mach 0.9 turbulent jet issued from the contoured convergent-straight nozzle of exit diameter $D=50$mm at $M_f = 0$ were initially reported by \cite{BresJFM2018}. The present work is an extension of the study for both $M_f = 0$ and $0.15$ with longer databases and higher sampling frequency. All the large eddy simulations feature localized adaptive mesh refinement, synthetic turbulence and wall modeling on the internal nozzle surface (and external nozzle surfaces at $M_f = 0.15$) to match the fully turbulent nozzle-exit boundary layers in the experiments. The LES methodologies, numerical setup and comparisons with measurements are described in more details in \cite{BresJFM2018} and are only briefly summarized here. 

The round nozzle geometry (with exit centered at $(0, 0, 0)$) is explicitly included in the axisymmetric computational domain, which extends from approximately $-10D$ to $50D$ in the streamwise (x) direction and flares in the radial direction from $20D$ to $40D$. The uniform external flight stream is imposed as upstream boundary condition outside the nozzle in the simulation. Note that a very slow coflow at Mach 0.009 is used in the LES at no flight stream condition to prevent any spurious recirculation and facilitate flow entrainment. Sponge layers and damping functions are applied to avoid spurious reflections at the boundary of the computational domain \citep{Freund-97, mani12}. The Vreman \citep{Vreman:2004p754} sub-grid model is used to account for the physical effects of the unresolved turbulence on the resolved flow.  

The nozzle pressure ratio and nozzle temperature ratio are $NPR = P_t/P_f = 1.7$ and $NTR = T_t/T_f = 1.15$, respectively, and match the experimental conditions. The jet is isothermal ($T_j/T_f=1.0$), and the jet Mach number is $M_j = U_j/c = 0.9$. For both experiment and simulation, the Reynolds number is $Re = \rho_j U_j D/\mu_j \approx 1\times10^6$.

In the experiment, transition is forced using a sandpaper strip located approximately 3$D$ upstream of the nozzle exit plane on the internal nozzle surface for all configurations, and on the external nozzle surface for $M_f = 0.15$. In the LES, synthetic turbulence boundary conditions are used to model these experimental boundary layer trips present on the internal and external nozzle surfaces. To properly capture the internal and external turbulent boundary layers, localized isotropic mesh refinement and wall modeling \citep{KawaiPoF2012, bodartwm2011} are applied on the interior and exterior surface from the boundary layer trip to the nozzle exit. All the other solid surfaces are treated as no-slip adiabatic walls. While several meshes were considered as part of a grid resolution study \citep{BresJFM2018}, the standard unstructured mesh containing approximately 16 million control volumes is used for the present work with $M_f =0$, and the case is simply referred to as \textit{BL16M\_M09} (i.e., extension of case \textit{BL16M\_WM\_Turb} from \cite{BresJFM2018}). For $M_f = 0.15$, the same isotropic near-wall mesh refinement used to capture the boundary layer inside the nozzle is applied outside of the nozzle. The mesh size is therefore increased to approximately 22 million control volumes, and the case is referred to as \textit{BL22M\_M09\_Mf015}.

Table~\ref{table:LEScase} lists the simulation parameters and settings for the LES runs with and without flight stream, including the time step $dt$, the total simulation time $t_{sim}$ for the collection of statistics and data (after the initial transient is removed), and the sampling period $\Delta t$ for the recording for the main LES databases. Note that the sampling period for the recording of the FW-H surface data for far-field noise predictions is $0.5\Delta t$.

To facilitate postprocessing and analysis, the LES data is interpolated from the original unstructured LES grid onto structured cylindrical grids in the jet plume and in the nozzle pipe. These structured cylindrical grids were originally designed for the grid with 16M control volumes, such that the resolution approximately corresponds to the underlying LES resolution. For both structured grids, the points are equally-spaced in the azimuthal direction to enable simple azimuthal decomposition in Fourier space.

\begin{table}
\begin{center}
\begin{tabular}{ l  c  c  c  c  c  c  c  c  }
\hline
       Case   name   &  Mesh size & $M_j$ & $M_f$ & $T_j/T_f$  & $Re$ & $dt c/ D$&  $ \Delta t  c /D$& $t_{sim} c /D$\\
    \textit {BL16M\_M09}           & $15.9\times10^6$   & 0.9  & 0  &     1.0&  $1 \times 10^6$ &      0.001 & 0.1 &3000   \\
    \textit{BL22M\_M09\_Mf015}        & $21.8\times10^6$  &  0.9 &0.15   &  1.0&  $1\times 10^6$ &0.001 & 0.1 &2000   \\
\hline
\end{tabular} 
\caption{Operating conditions and simulation parameters of the main LES, where $t_{sim}$ is the simulation time and $\Delta t$ is the sampling period of the database recording.}
\label{table:LEScase}
\end{center}
\end{table}

\section{Zero- and first-order statistics}
\label{sec:stats}

\begin{figure}
\centering
\includegraphics[trim=2cm 1.5cm 2cm 0cm, clip=true,width=\linewidth]{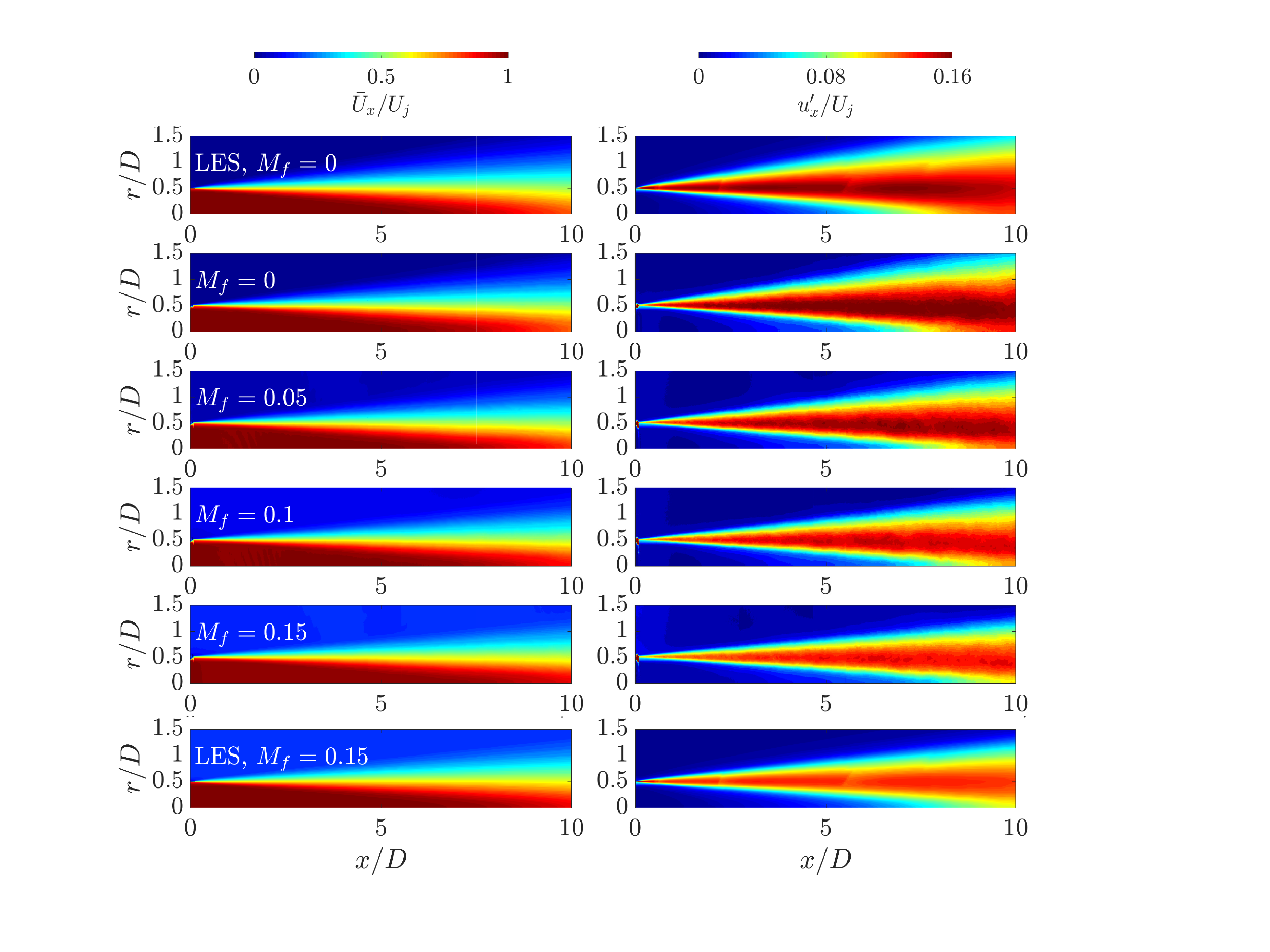}
\caption{Contours of mean, $\bar{U}_x$ (left), and rms, $u'_{x}$ (right), streamwise velocity for jets at $M_{j}=0.9$ and different flight stream Mach numbers. Velocities are normalised by $U_{j}$. The first and last contour maps correspond to LES data, and the intermediary ones to PIV results.}
\label{fig3}
\end{figure}

We start with an analysis of the effect of the flight stream on zero- and first-order velocity statistics, in order to provide a validation of the experimental and numerical databases. It is known that a flight stream affects the jet development by lengthening the potential core and reducing shear layer spreading and turbulence intensities \citep{TannaMorris1977}. These trends can be seen in figure \ref{fig3}, which presents contour plots of mean and rms streamwise velocity on a meridional plane, measured with the 2D-PIV setup, with increasing $M_{f}$. LES data for the case $M_{j}=0.9$, $M_{f}=0.15$ is also shown for comparison.  

\begin{figure}
\centering
\includegraphics[trim=0cm 3.5cm 0cm 3.2cm, clip=true,width=1\linewidth]{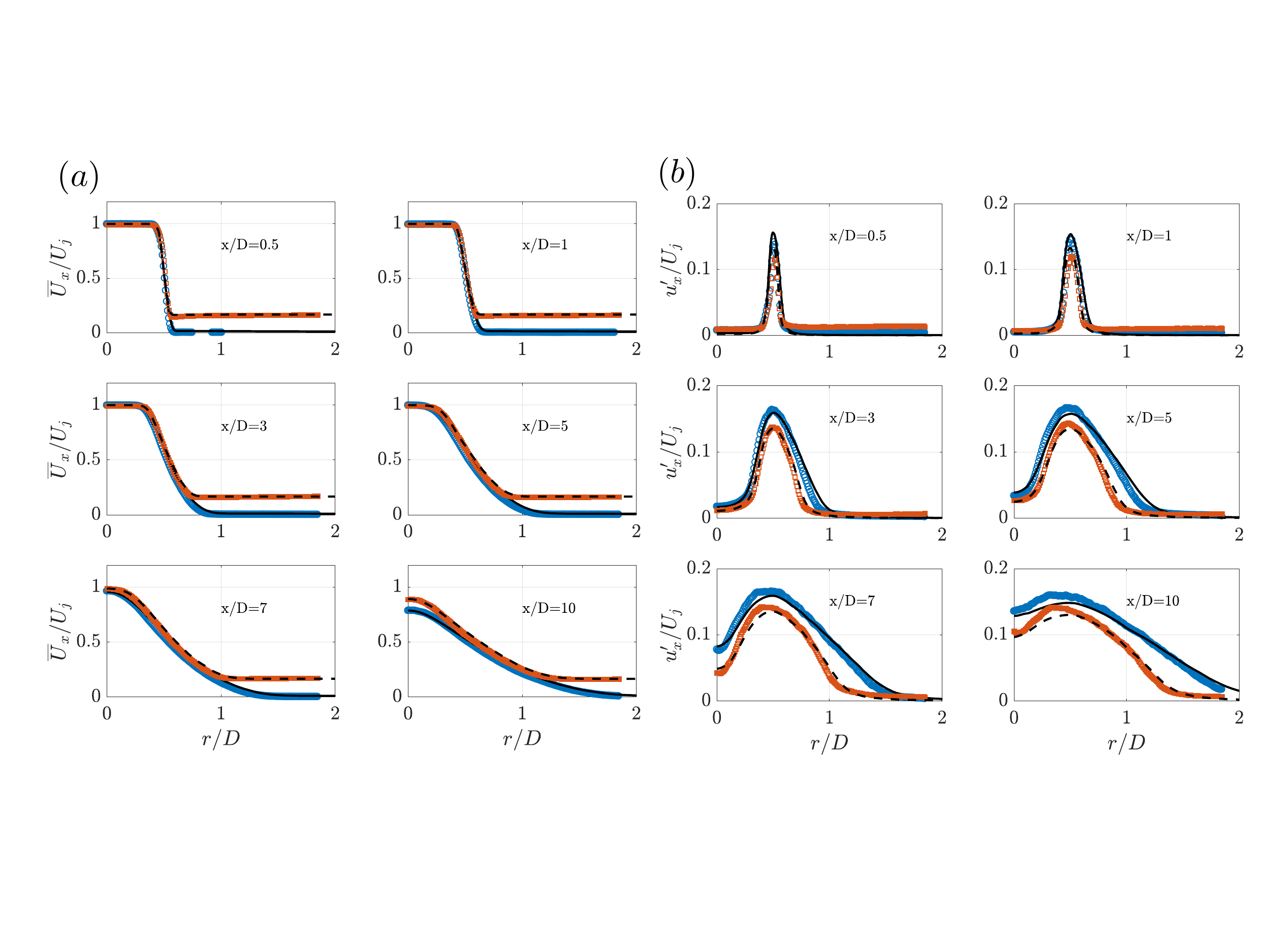}
\caption{Comparison between experimental and numerical mean(a) and rms(b) radial profiles of streamwise velocity component at different streamwise positions. \usebox{\mycircle}: exp. $M_{f}=0$; \usebox{\myboxred}: exp. $M_{f}=0.15$; \usebox{\mylineblack}:LES data for $M_{f}=0$, \textit {BL16M\_M09}; \usebox{\mylinedashed}: LES data for $M_{f}=0.15$, \textit{BL22M\_M09\_Mf015}.}
\label{fig4}
\end{figure}

The LES and experimental databases were found to be in excellent agreement, as can be seen in the radial profiles of mean and rms velocities shown in figure \ref{fig4} for different streamwise positions. LES data for the no-flight case, \textit {BL16M\_M09}, is also shown for comparison. Both jets exhibit a classic change from top-hat profiles in the near-nozzle region to bell-shaped profiles further downstream. However, the reduction in shear-layer thickness becomes apparent after a couple of jet diameters with increasing $M_f$. The rms profiles measured in the presence of the flight stream show amplitude reductions throughout the jet. These reductions are concentrated in radial positions around the peak in the initial jet region, but spread across the shear layer further downstream. 

The mean-flow shape changes with increasing flight stream levels. We assessed three important flow features at different conditions of flight stream Mach number, $M_{f}$: the potential core length, $x_{c}$, defined here as the streamwise position where $\bar{U}_{x}(x,0)=0.95U_{j}$, the centerline velocity decay and the shear-layer momentum thickness, $\delta_{\theta}$, defined as

\begin{equation}
\delta_{\theta}(x)=\int_{0}^{r_{0.05}} \frac{\tilde{U}_{x}(x,r)}{\tilde{U}_{x}(x,0)}\left(1-\frac{\tilde{U}_{x}(x,r)}{\tilde{U}_{x}(x,0)}\right)\mathrm{d}r,
\end{equation}
where $\tilde{U}_{x}$ is a normalised mean flow velocity,

\begin{equation}
\tilde{U}_{x}(x,r)=\frac{\bar{U}_{x}(x,r)-U_{f}}{\bar{U}_{x}(x,0)-U_{f}}.
\end{equation}
 
\begin{figure}
\centering
\includegraphics[trim=0cm 3cm 0cm 0cm, clip=true,width=\linewidth]{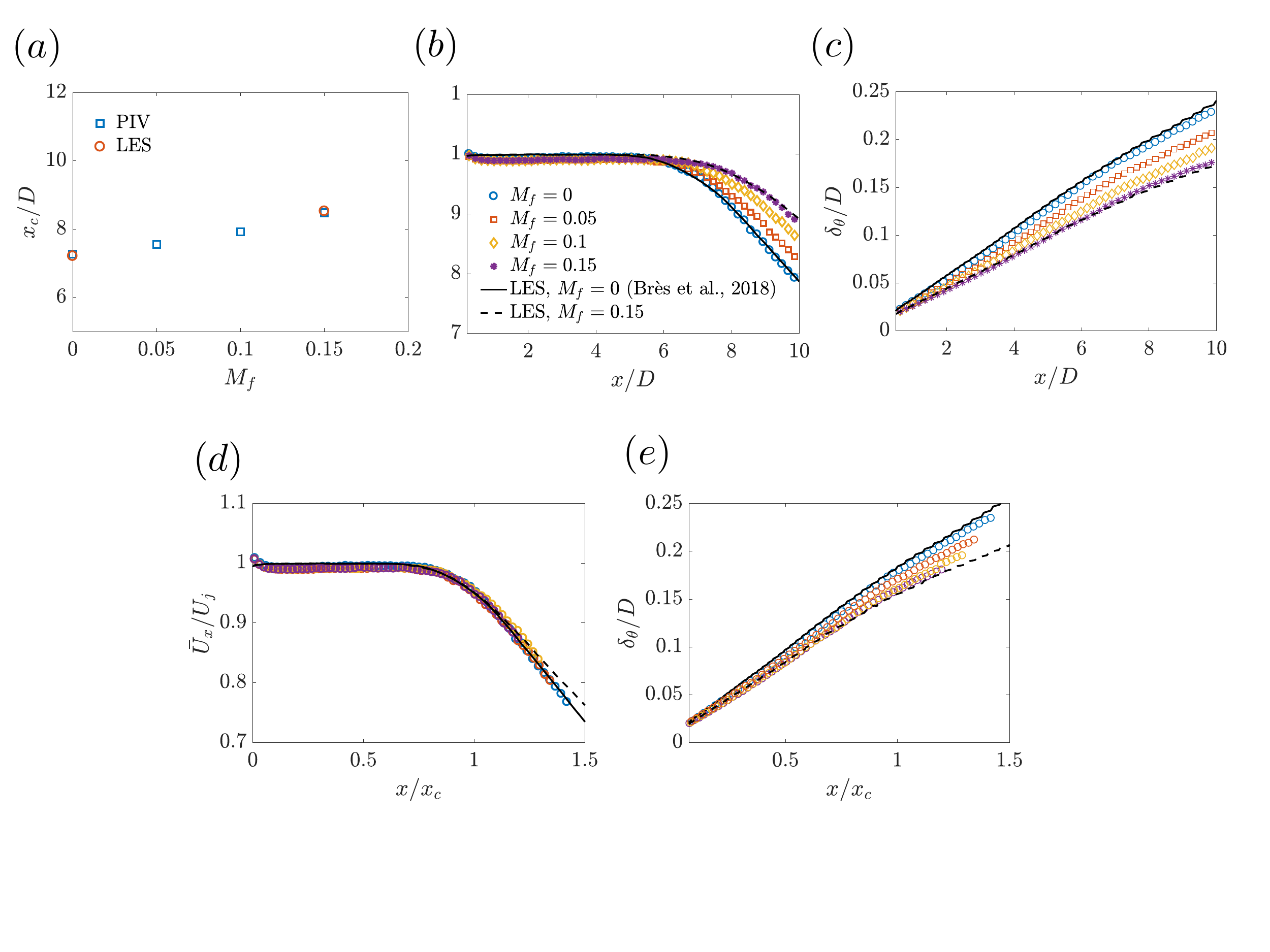}
\caption{Variation of mean-flow quantities with increasing flight stream velocity for a jet at $M_{j}=0.9$. (a) Potential core length; (b) Centerline velocity decay; (c) Streamwise evolution of the momentum thickness. (d) and (e) show centerline velocity and momentum thickness profiles when the streamwise coordinate is scaled by potential core length, $x_c$. Legends in (c), (d) and (e) are the same as in (b).}
\label{fig5}
\end{figure}

The evolution of these quantities with increasing $M_{f}$ is shown in figure \ref{fig5}. The potential core length grows approximately linearly with $M_{f}$ in the range of conditions tested, presenting an increase of 17\% between the no-flight-case and the case with $M_{f}=0.15$. The jet development is also affected downstream of the potential core, where the velocity decays at smaller rates in the presence of the flight stream. This delayed development is also manifest in momentum thicknesses, which are significantly reduced. Figures \ref{fig5}(d) and (e) show the evolution of centerline velocity and momentum thickness with the streamwise coordinate scaled by the case-dependent core length $x_c$. It can be seen that this scaling produces a collapse of the centerline velocity decay. The momentum thickness, on the other hand, does not collapse so well with this scaling, and the shear-layer remains thinner in the flight case, even when the flow stretching is taken into account.

\begin{figure}
\centering
\includegraphics[trim=0cm 5.5cm 0cm 4cm, clip=true,width=0.9\linewidth]{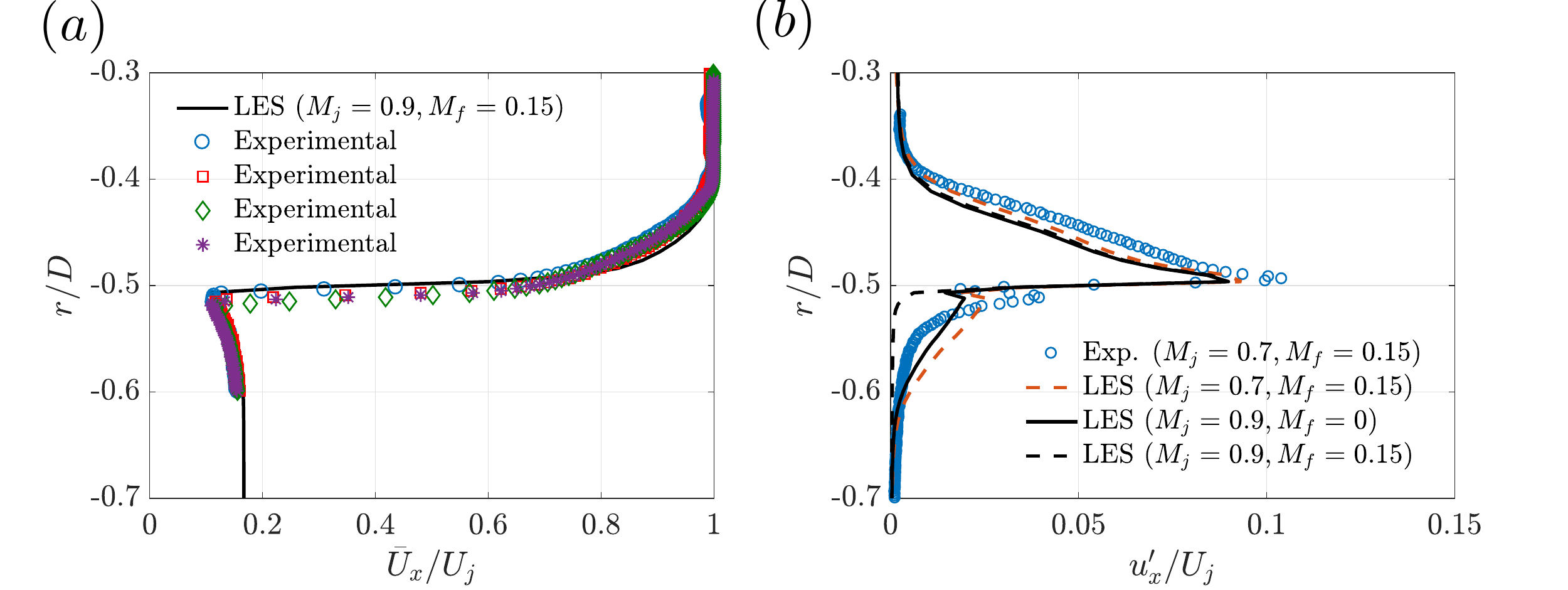}
\caption{Nozzle-exit mean (a) and rms (b) boundary layer profiles measured in the flight stream case, $M_f=0.15$. The different experimental profiles in (a) correspond to measurements performed at different azimuthal positions, in order to verify the homogeneity of the boundary layer. Rms profiles for the static case, $M_f=0$, are shown in (b) for comparison.}
\label{fig6}
\end{figure}

The boundary layer, which is now recognized as an important flow region underpinning jet dynamics and sound radiation \citep{BresJFM2018, KaplanJFM2021, LesshafftPFR2019} was also characterised through Pitot tube and hot-wire measurements. Figure \ref{fig6} shows mean and rms profiles for the flight stream case measured at $x/D=0.0024$, which was the closest position attainable without damaging the probes. The hot-wire had a length of 1.25mm and a diameter of 2.5$\mu$, and the measurements were performed with a 55M01 Dantec anemometer at a frequency of 30kHz ($St=4.8$). The homogeneity of the boundary layer was verified by measuring the profiles  at different azimuthal positions, which were found to be in be in good agreement with each other and the LES profile. Regarding the hot-wire measurements, at $M_{j}=0.9$ the \textit{in-situ} calibration of the probes was found to be very problematic, with large errors in the calibration coefficients due to compressibility effects and the large total temperature gradients in the thin initial shear layer. This issue was mitigated by performing the measurements at a lower Mach number, $M_{j}=0.7$, for which there was also another LES database available. As shown in figure \ref{fig6}(b), the differences between boundary layer profiles at $M_j=0.7$ and $M_j=0.9$ are slight. Rms profiles from the static and flight cases are found to be quite similar in the inner part of the shear layer. In the outer part, $|r/D|> 0.5$, rms values are higher in the presence of the flight stream, as expected. A second peak is seen in the $M_f=0.15$ profiles at $r/D \approx 0.51$, due to the external boundary layer. The LES profiles are found to be in good agreement with hot-wire data. The peak turbulence intensity and the shape of the curves are consistent with previously investigated turbulent jets.

Overall, the numerical and experimental databases for the jets at $M_{j}=0.9$ and $M_{f}=0$ are found to be in excellent agreement. Experimental data for the no-flight case, $M_{f}=0$, agrees with previously reported data \citep{BresJFM2018}. Furthermore, the trends in mean flow distortion and rms fields with the flight stream are consistent with data reported in the literature. We now proceed to a more detailed investigation of the effect of the flight stream on structures contained in the turbulent flow, which involves an analysis of energy distribution in the frequency-wavenumber space. 

\section{Energy distribution across azimuthal modes}
\label{sec:en_dist_m}

We first consider the energy distribution across azimuthal wavenumbers and assess how this is affected by the coflow. This is done by first interpolating the TR-PIV instantaneous fields $u(y,z,t)$ onto a polar grid $u(r,\theta,t)$ and then decomposing them into a Fourier series in $\theta$, obtaining a field $u(r,m,t)$, with $m$ the azimuthal wavenumber \citep{Wavepacketsvelocity,CoherenceVincent}. This field is then used to compute mean squared velocity fluctuations, $u_{x}^{'2}(r,m)$. 

We use the potential core length, $x_{c}$ as a mean-flow scaling parameter. Throughout the remainder of the study we make comparisons between the $M_{f}=0$ and $M_{f}=0.15$ cases at different streamwise positions on the basis of the normalised coordinate, $x/x_{c}$. In previous studies of jets subject to an external flow, attempts have been made to find \enquote{stretching factors} that, when used as a suitable scaling to the coordinate system, would make the flow field, the stability characteristics, and turbulent quantities and sound field independent of the external flow velocity. For instance, \cite{MICHALKE1979341} proposed a scaling factor proportional to the velocity difference which is used to model, with some degree of success, the reduction in SPL produced by the external flow. The factor is calibrated through a fitting procedure of acoustic data. In a similar spirit, \cite{MuchalkeHermann1982} have shown that a scaling can be found that accounts at the same time for the reduction of spatial growth rates of the Kelvin-Helmholtz instability and for the change in the range of unstable frequencies. The scaling factor, however, is wavenumber-dependent. Here we use the potential core length as a physical scaling factor, without delving much further in the search for a \enquote{universal} scaling of all aspects of jet dynamics and sound radiation. The suitability of this choice will be discussed in light of the results shown in the following.
 
\begin{figure}
\centering
\includegraphics[trim=0cm 0cm 0cm 0cm, clip=true,width=0.8\linewidth]{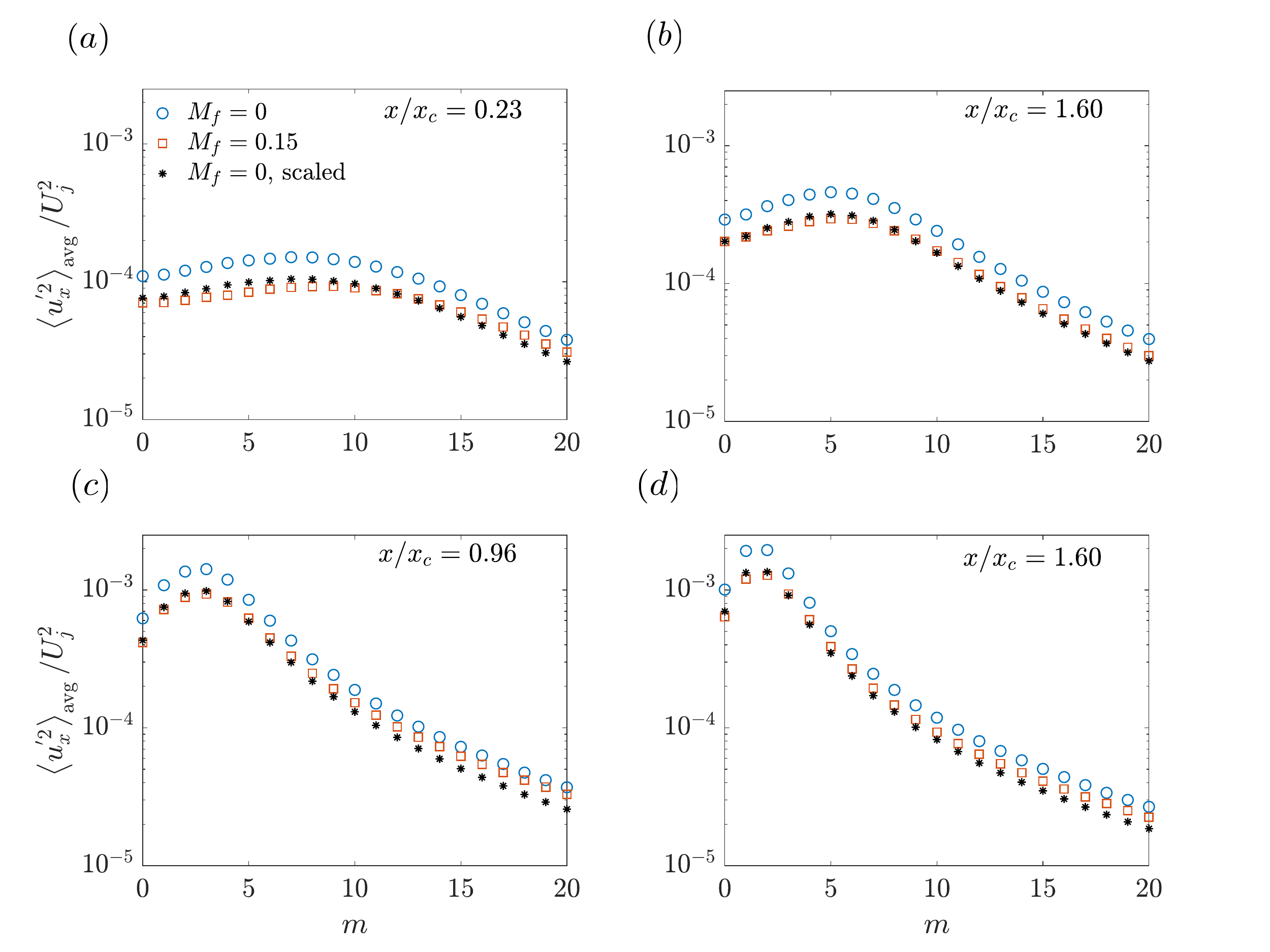}
\caption{Energy of streamwise velocity fluctuations integrated in $r$ as a function of azimuthal wavenumber, $m$, for different streamwise positions, $x/x_{c}$. The black asterisks represent the energy of the static case scaled by the factor, $(dU_f/dU_s)^2$, which represents the change in the centerline-to-freestream velocity in static and flight conditions.}
\label{fig7}
\end{figure}

Figure \ref{fig7} shows the distribution of energy for different $m$ at four different positions: $x/x_{c}=0.23, 0.45, 0.96, 1.6$. In order to provide a metric that represents that is representative of the energy at each $m$ and streamwise position, mean-squared velocities are averaged across the shear-layer, in the interval $0 \leqslant r/D \leqslant 2$. Comparisons of the averaged quantity, $\left<u_{x}^{'2}(m)\right>_{\mathrm{avg}}$, in the static and flight cases, illustrate the overall effect of the flight stream in the wavenumber energy distribution at each streamwise station, independent of radial position. At streamwise locations closer to the nozzle exit, the energy distribution is quite broadband, with at least 20 azimuthal modes having significant energy relative to the peak, which occurs around $m=7-8$ at $x/x_{c}=0.23$. As one moves downstream, the distribution becomes more narrowband and the energy peak is shifted towards lower $m$. This is a trend that has also been observed by past experimental \citep{CitrinitiGeorge2000,JungJFM2004, Wavepacketsvelocity} and numerical studies \citep{PickeringJFM2020}. Figure \ref{fig7} also reveals that the flight stream produces substantial reductions in energy levels for all azimuthal wavenumbers analysed. Also shown is a rescaling of the energy of the static case by the factor $(dU_f/dU_s)^2$, which represents the reduction in the centerline-to-freestream velocity in static and flight conditions. The scaled energy falls quite close to the values of the flight case, consistent with an expected global reduction of turbulence. Notice, however, that the rescaling works better around the peak energy, and some discrepancy is seen for high azimuthal wavenumbers. Notice also that at the position closest to the nozzle exit, there is a slight shift of the peak towards higher $m$ with the flight stream.

\begin{figure}
\centering
\includegraphics[trim=0cm 0cm 0cm 0cm, clip=true,width=0.5\linewidth]{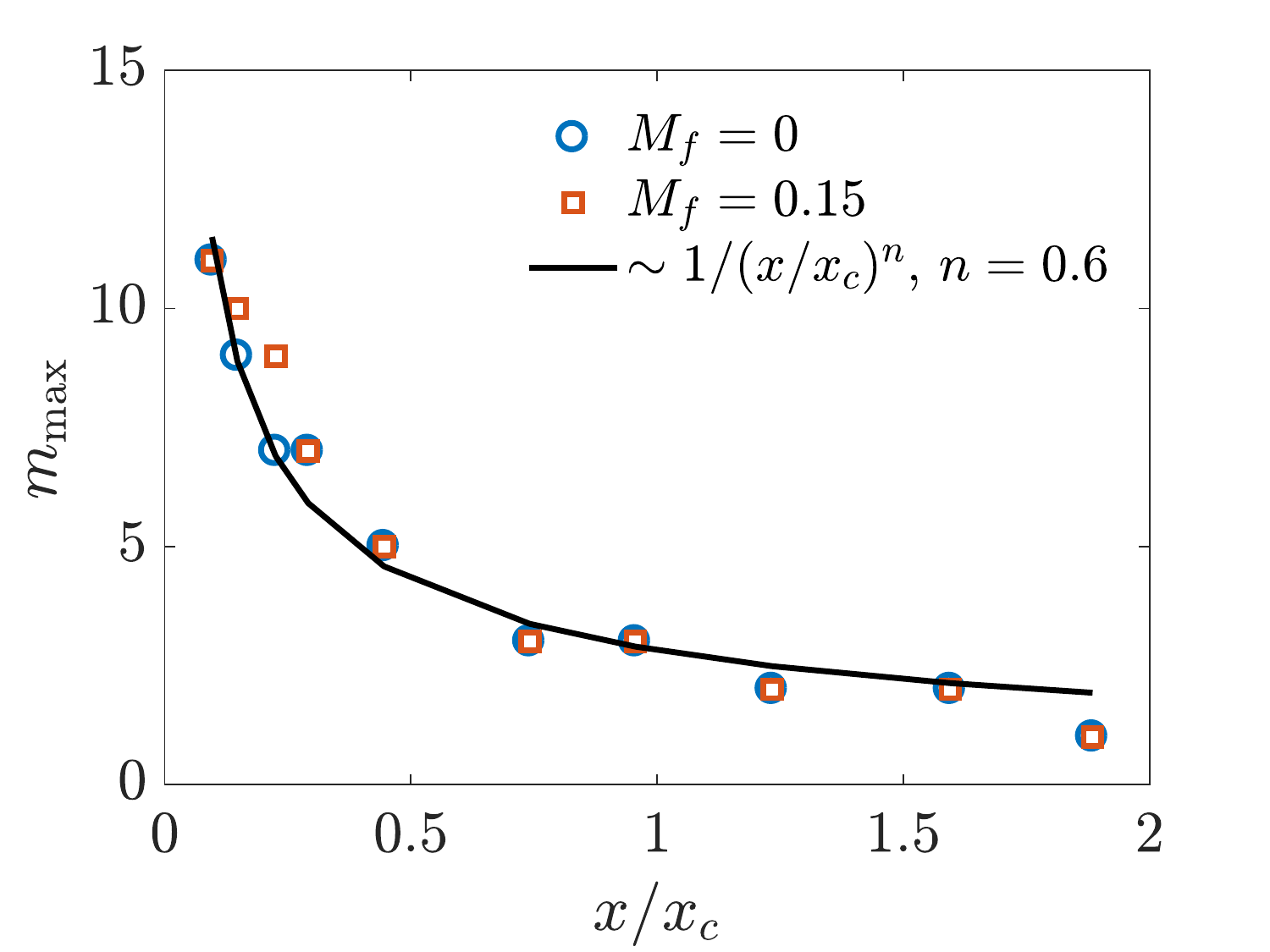}
\caption{Azimuthal wavenumber of the peak energy as a function of $x/x_{c}$. The black lines is a fit of the form $1/(x/x_c)^{n}$ for the $M_{f}=0$ and $M_{f}=0.15$, respectively. $n=0.6$ was found to provide a near-optimal fit for both cases.}
\label{fig8}
\end{figure}

The downstream evolution of the azimuthal wavenumber of the energy peak was also found to be similar, once normalised by $x_c$, with and without the flight stream, as seen in figure \ref{fig8}. In both cases, the peak evolves towards lower $m$ with downstream position, approximately scaling as $\sim1/(x/x_c)^{n}$. The exponent $n$ which provides the best fit with the data was found to be virtually the same for both jets. Far downstream, past the end of the potential core, mode $m=1$ becomes dominant in the experimental data. In the results of figures \ref{fig7} and \ref{fig8} one can note a subtle difference between near-nozzle and downstream regions. In the former, the peak wavenumber is different, whereas further downstream the trends are virtually identical for the two jets. 

\section{Frequency-wavenumber energy maps}
\label{sec:local_spod}

We now continue the analysis by further decomposing the velocity field into Fourier modes in time. The decomposed velocity field then becomes a function of frequency (expressed by the Strouhal number), azimuthal wavenumber and radial position, and we can assess the effect of the flight stream in the frequency-wavenumber plane. Next we perform a Spectral Proper Orthogonal Decomposition (SPOD) on $\hat{u}_x(St,m,r)$. SPOD decomposes the data into an orthogonal basis ranked in terms of an energy norm. This decomposition is instructive for turbulent flows because it acts like a filter for coherent structures: their dynamics are often well represented by the leading mode and its modal energy. Leading SPOD modes can frequently be associated with the Kelvin-Helmholtz, Orr and Lift-up mechanisms, once compared with linear mean-flow analysis, according to their region of dominance in the $St-m$ plane \citep{NogueiraJFM2019, PickeringJFM2020}. It is important to emphasise, though, that a clear demarcation between the linear mechanisms in the frequency-wavenumber plane does not exist. For example, in the low $St$ range, coherent structures are likely a mixture (for non zero $m$) of Orr structures, streaks and weak KH wavepackets. Likewise, for the $m=0$ mode, the change from KH to Orr-type structures with decreasing $St$ is gradual, with no clear cut transition. In that sense, different regions of the $St$-$m$ plane should be seen merely as indications of mechanism dominance, which is nonetheless useful insofar as away from the grey zones the distinction is relatively clear. With that in mind, in the following we analyse the modal energy maps of the leading SPOD mode, which in section \ref{sec:resolvent} will be associated with instability mechanisms studied through linear mean-flow analysis.

In the framework of SPOD, given the state vector, $\mathbf{q}=[\rho, u_{x},u_{r}, u_{\theta}, T]^{T}$, SPOD modes for a given azimuthal wavenumber and Strouhal number pair, $\mathbf{\Psi}_{m,\omega}$ are obtained through eigendecomposition of the cross-spectral-density (CSD) matrix, $\hat{\mathbf{S}}_{m,\omega}$,

\begin{equation}
\hat{\mathbf{S}}_{m,\omega}\mathbf{W}\mathbf{\Psi}_{m,\omega}=\mathbf{\Psi}_{m,\omega}\mathbf{\Lambda}_{m,\omega}.
\label{eq:spod}
\end{equation}
The cross-spectral density matrix is computed as $\hat{\mathbf{S}}_{m,\omega}=\hat{\mathbf{Q}}_{m,\omega}\hat{\mathbf{Q}}_{m,\omega}^{*}$, where $\hat{\mathbf{Q}}_{m,\omega}=[\hat{\mathbf{q}}_{m,\omega}^{(1)} \hat{\mathbf{q}}_{m,\omega}^{(2)} \cdots \hat{\mathbf{q}}_{m,\omega}^{(N_{blk})}]$ is the ensemble of $N_{blk}$ flow realisations at $(m,\omega)$, with $\hat{\mathbf{q}}_{m,\omega_{k}}^{(l)}$ denoting the $l$th realisation of the Fourier transforms in time and azimuthal direction at the frequency $\omega$ and wavenumber $m$. The eigenvalues, $[\lambda_{m,\omega}^{(1)}, \lambda_{m,\omega}^{(2)} \cdots \lambda_{m,\omega}^{nblk}]$ corresponding to the modal energy are organised in decreasing order in the diagonal matrix $\mathbf{\Lambda}_{m,\omega}$. The modes so obtained are orthogonal in a given inner product,

\begin{equation}
\left< \mathbf{q}_{1},\mathbf{q}_{2}\right>=\mathbf{q}_{1}^{*} \mathbf{W} \mathbf{q}_{2};
\label{eq:inner_product}
\end{equation}
where $\mathbf{W}$ is a weight matrix containing the numerical quadrature weights and choice of a given norm.

We first consider, using the experimental database, CSDs of different cross-sections of the jet. This \enquote{local} approach serves two main purposes: first, it allows us to analyse the evolution of the local organisation with increasing streamwise distance; and second, it allows us to study coherent structures in the initial jet region without their energy being masked by the most energetic structures that dominate the flow far downstream and tend to mask the upstream organisation when viewed using global SPOD. In order to explore the PIV dataset, we reduce the state vector to $\mathbf{q}=[u_{x}]^{T}$, and consider a matrix $\mathbf{W}$ that contains trapezoidal quadrature weights for the uniform PIV grid. The CSDs are computed using Welch's periodogram method. For the PIV data, we used blocks of 128 samples and 50\% overlap, resulting in a resolution of $\Delta St=0.0126$. For the LES data, larger blocks of 256 samples were used in order to achieve a resolution similar ($\Delta St = 0.017$) to that of the PIV. 

\begin{figure}
\centering
\includegraphics[trim=3cm 0cm 3cm 0cm, clip=true,width=1\linewidth]{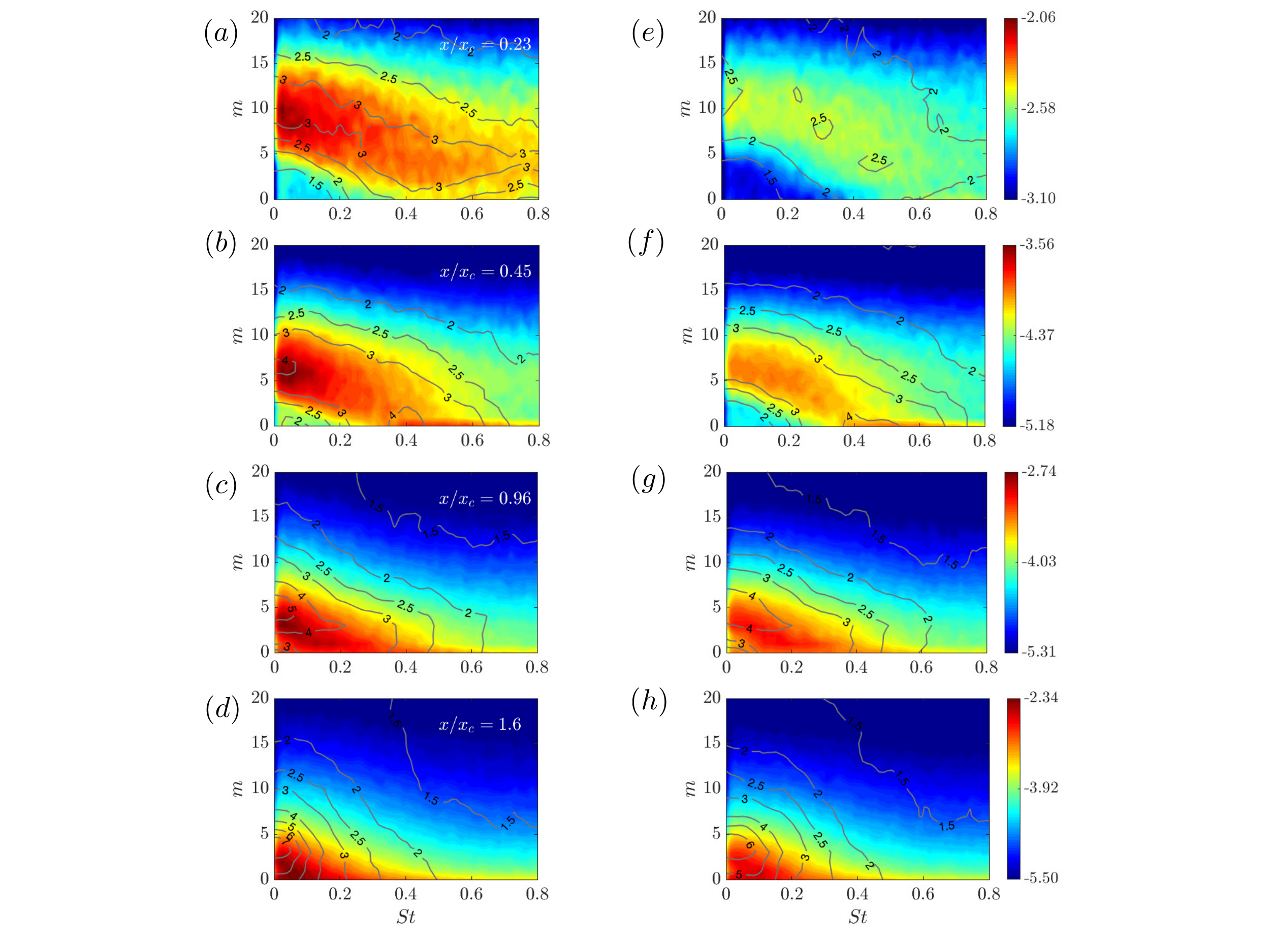}
\caption{Modal energy maps obtained through SPOD of PIV data. (a)-(d): maps for the $M_{f}=0$ case; (e)-(h): maps for the $M_{f}=0.15$ case. Colormaps correspond to the leading eigenvalue, $\mathrm{log}_{10}(\lambda_{1_{St,m}})$. Gray contours represent values of the ratio between the leading and first suboptimal eigenvalues, $\lambda_{1_{m,\omega}}/\lambda_{2_{m,\omega}}$}
\label{fig12}
\end{figure}

In the following, we show modal energy maps of the leading SPOD mode, which we associate with linear instability mechanisms, and infer changes in such mechanisms in the presence of the flight stream. This association is justified in regions of the spectrum where a large separation exists between its modal energy, $\lambda_1(m,St)$, and that of  the first suboptimal mode suboptimal mode, $\lambda_2(m,St)$ (even if a precise threshold for low-rank behaviour is unclear). However, we stress that such comparisons should be made with care whenever the leading and suboptimal modes have comparable amplitudes. We emphasise that, in the following, the distinction between mechanisms concerns the flow response, given by SPOD and resolvent response modes, because the optimal forcing (in the framework of resolvent analysis) associated with KH and Orr structures display similar structures. It has been shown that KH wavepackets are also optimally forced by Orr-like structures on the vicinity of the nozzle \citep{Garnaud, SchmidtetalJFM2018, LesshafftPFR2019}. Orr structures in the response, on the other hand, are characterised by lower phase speeds and extended spatial structure with respect to KH wavepackets. Figure \ref{fig12} shows modal energy maps of the leading SPOD modes, $\lambda_{1_{m,\omega}}$ at the four streamwise positions considered previously. Superposed on these maps are contours of the ratio between the leading and first suboptimal eigenvalues, $\lambda_{1_{m,\omega}}/\lambda_{2_{m,\omega}}$. In both flow conditions, the energy peak occurs in the $St \to 0$ limit for azimuthal wavenumbers $m>0$. Following the downstream development of the jet, the wavenumber associated with the energy peak decreases monotonically, as already revealed in figures \ref{fig7} and \ref{fig8}, but with the maximum energy always remaining in the $St \to 0$ zone. Downstream of the end of the potential core, mode $m=1$ eventually becomes dominant, as observed previously in static jet conditions \citep{CitrinitiGeorge2000,NogueiraJFM2019,PickeringJFM2020}. The flight stream is seen to produce a significant reduction of the energy levels, attenuating the first 15-20 azimuthal wavenumbers. Closer to the nozzle exit, this reduction is seen to take place for a broad range of Strouhal numbers and azimuthal wavenumbers. As the jet evolves downstream, the attenuation gradually becomes concentrated around the energy peak, which occurs at low $St$. This reveals that much of the reduction in turbulent kinetic energy observed in many previous studies, and described as a global attenuation of turbulence, is in fact underpinned by low-frequency, streak-like structures.

\begin{figure}
\centering
\includegraphics[trim=3cm 0cm 3cm 0cm, clip=true,width=1\linewidth]{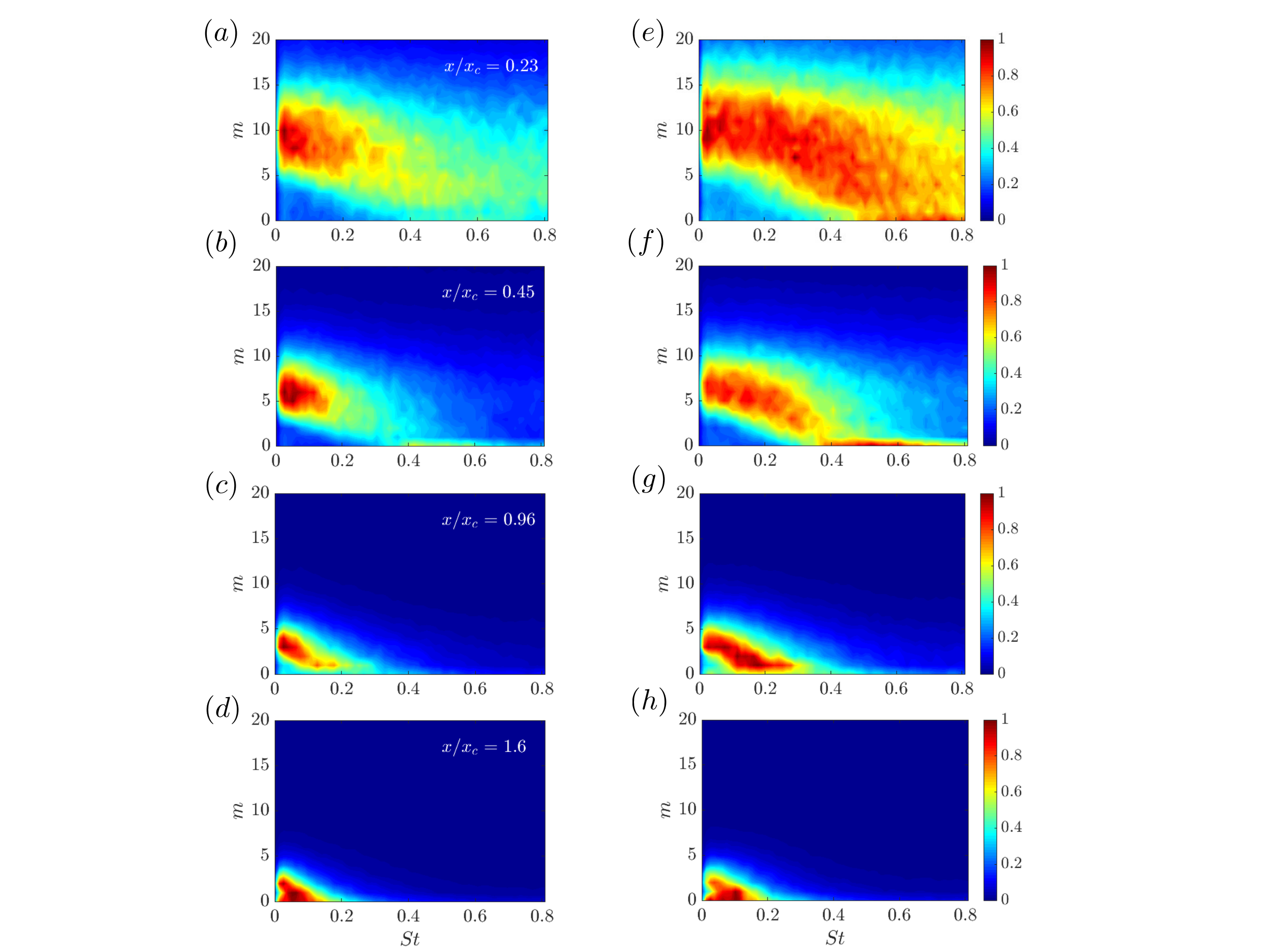}
\caption{ Normalised modal energy maps of the leading SPOD mode, $(\lambda_{1_{St,m}}/\mathrm{max}(\lambda_{1_{St,m}}))$ in the frequency-azimuthal wavenumber space for different streamwise positions based on PIV data. (a)-(d): maps for the $M_{f}=0$ case; (e)-(h): maps for the $M_{f}=0.15$ case.}
\label{fig13}
\end{figure}

Besides the energy attenuations, the flight stream also produces a significant effect on the eigenvalue separation. Close to the nozzle exit, $\lambda_{1_{m,\omega}}/\lambda_{2_{m,\omega}}$ is reduced for almost all the azimuthal wavenumbers in the range $St \gtrsim 0.2$. As the jet evolves downstream, this decrease in eigenvalue separation gets gradually concentrated at lower $St$ and lower $m$, following the shift in energy peak. This implies that the energy attenuation is accompanied by a weakening of the leading SPOD modes which describe the most energetic, coherent flow structures. These have been shown to be associated with modal and non-modal linear instability mechanisms \citep{NogueiraJFM2019, PickeringJFM2020}. In the initial jet region, modal growth mechanisms are strong, with KH wavepackets being convectively unstable for a broad range of frequencies starting from $St \gtrapprox 0.1$ and azimuthal wavenumbers $m<5$, as will be shown shortly by the linear mean-flow analysis. Non-modal mechanisms, which dominate the energy spectrum throughout the jet, give rise to Orr ($m=0$) and streaky, lift-up ($m>0$) structures, and have peak energy in the $St \to 0$ limit. The modal maps shown here suggest an attenuation of these three mechanisms, as revealed by both the energy reduction and the lower eigenvalue separation, which will be confirmed shortly following a comparison with a linear mean-flow model. Apart from the overall reduction in levels, it is also interesting to assess whether the flight stream changes the energy distribution in the $St-m$ plane. Figure \ref{fig13} shows modal energy maps normalised by their maxima. The normalisation reveals that, for a given $m$, the spectrum is much broader in the $St$ direction with the flight stream, especially upstream of the end of the potential core. Whereas in the $M_f=0$ case the peak in the spectrum is concentrated in the $St \to 0$ region, for the $M_f=0.15$ case it flattens and spreads to higher $St$. This shows that the effect of the flight stream is not limited to a simple rescaling of turbulence levels due to the reduced shear. In section \S \ref{sec:resolvent}, we interpret the broadening of the spectrum in light of stability characteristics of the flow. Interestingly, the azimuthal organisation of the energy is less affected by the flight stream. This can be illustrated by the different normalisation of modal energy shown in appendix \S \ref{appA}, and also by the scaling introduced in figures \ref{fig7} and \ref{fig8}.

The results presented in sections \S \ref{sec:stats}-\ref{sec:local_spod} provide a comprehensive view of the effect of the flight stream, starting from zero-th order statistics, down to a more detailed analysis, through successive Fourier and modal decompositions, of how these changes are arranged in the frequency-wavenumber space, and how they vary in the streamwise direction. Comparisons between the streamwise evolution of jets with and without the flight stream on the basis of the normalised coordinate, $x/x_c$, produce an interesting similarity in certain properties of the flow between the flight and static cases. The centerline velocity profiles for different flight stream velocities collapse when the normalisation is applied. The azimuthal organisation of energy for the two jets are found to be in good agreement for equal $x/x_c$. The shape of the spectra are quite similar, as shown in figures \ref{fig7} and \ref{fig14}. Furthermore, the most energetic azimuthal modes, $m_{\mathrm{max}} (x/x_c)$ are also very similar, decaying exponentially with increasing $x/x_c$ at a rate which is virtually the same in the flight and static cases. Discrepancies are, however, manifest in the near-nozzle region, where the azimuthal spectra for the flight case are flatter and the peak is shifted towards higher $m$. Interestingly, it is in the same region that the eigenvalue separation, $\lambda_{1_{m,\omega}}/\lambda_{2_{m,\omega}}$ is significantly reduced by the flight stream. This indicates a reorganisation of the flow in which fluctuation energy is underpinned by somewhat more \enquote{disorderly} motion at higher $m$ at the expense of the most coherent structures associated with the leading SPOD mode. 

Other flow properties scale less well with $x_c$. For instance, collapse of the momentum thickness as a function of $x/x_c$ is not observed. Normalised modal maps, $(\lambda_{1_{St,m}}/\mathrm{max}(\lambda_{1_{St,m}}))$ for the two jets are also found to be quite different, with the flight stream producing spectra which are broader in the $St$ direction at a given $x/x_c$. This behaviour is more pronounced in the initial jet region. As will be explained in the next section, these two trends are connected. The slower growth of the shear-layer leads to a broader range of unstable frequencies, which can be associated with the broader spectrum.

\section{Linear mean-flow analysis}
\label{sec:resolvent}

The results reported above raise the question as to what causes the observed energy attenuation. Is it simply a mean-flow modification effect (and in this case it is something that can be mimicked through a linear model), or is it rather due to a deeper (nonlinear) reorganisation of turbulence? We address this issue through a locally-parallel linear model, where linearisation is performed about the mean flow. As discussed in section \S \ref{sec:intro}, although in a laminar regime, with instability analysis performed about a fixed point, the reduction of shear produced by the flight stream is expected to attenuate modal and non-modal instabilities by linear mechanisms alone, this scenario cannot be accepted, \textit{a priori}, for the case of a turbulent jet. In the latter case non-linearities, other than those that underpin the mean flow, may play an important role in the changes observed in the energy spectrum. It is therefore important to assess to what degree linear mean-flow mechanisms are the cause of the different dynamics observed with the flight stream in this high-Reynolds-number, turbulent regime.

The analysis starts with the linearised Navier-Stokes equations written in an input-output form:

\begin{equation}
\frac{\partial \mathbf{q}'}{\partial t} + \mathcal{A}_{\bar{\mathbf{q}}}\mathbf{q'} = \mathbf{f},
\label{ns}
\end{equation}
where $\mathbf{q}'$ is a vector containing fluctuations (in a Reynolds-decomposition sense) of the state variables and $\mathcal{A}_{\bar{\mathbf{q}}}$ is the linearised Navier-Stokes operator. The subscript $_{\bar{\mathbf{q}}}$ denotes linearisation about the mean flow. $\mathbf{f}$ is a term representing the nonlinear Reynolds stresses, which are treated as an endogenous forcing term. In the locally-parallel framework, we assume flow perturbations of the form,

\begin{equation}
\mathbf{q}'(x,r,\theta,t)=\hat{\mathbf{q}}(r)\mathrm{exp}^{i(\alpha x - \omega t +m\theta)},
\label{ansatz}
\end{equation}
where the radial structure of the perturbations is given by $\hat{\mathbf{q}}(r)$, $\alpha$ and $m$ are streamwise and azimuthal wavenumbers, respectively, and $\omega$ is the frequency. Applying a Fourier transform in \ref{ns} and substituting the \textit{Ansatz} in \ref{ansatz}, yields

\begin{equation}
-i \omega \hat{\mathbf{q}} + (\mathcal{A}_{0} + \alpha\mathcal{A}_{1}+ \alpha^2 \mathcal{A}_{2})_{\bar{\mathbf{q}}} \hat{\mathbf{q}} = \hat{\mathbf{f}},
\label{ns_Fourier}
\end{equation}
where the linear operators $\mathcal{A}_{0}, \mathcal{A}_{1}$, and $\mathcal{A}_{2}$ contain terms issuing from zero-th, first and second order derivatives in $x$, respectively. The superscripts $\hat{}$ denote Fourier transformed quantities. Details about the linearisation procedure, the operators and the boundary conditions are given in Appendix \S \ref{appB}. 

We carried out the analysis in the initial jet region. The mean flow profiles that served as input to the model were based on experimental data, and fitted with the hyperbolic tangent profiles proposed by \cite{MuchalkeHermann1982}. At frequencies for which the flow experiences a strong modal convective instability, it is know that eigenanalysis based on a spatial stability formulation provides a suitable framework to describe coherent structures in jets \citep{JordanColoniusReview}. When the modal instability is weak other approaches should be used to study linear mechanisms. We address such cases with a model based on the response modes of the resolvent operator. Since the work of \citet{MckeonSharma}, resolvent analysis has been extensively used to identify optimal forcing and response mechanisms in laminar and turbulent flows and to model observed coherent structures.

\subsection{Eigenanalysis: spatial stability}

\begin{figure}
\centering
\includegraphics[trim=2cm 5cm 2cm 5cm, clip=true,width=1\linewidth]{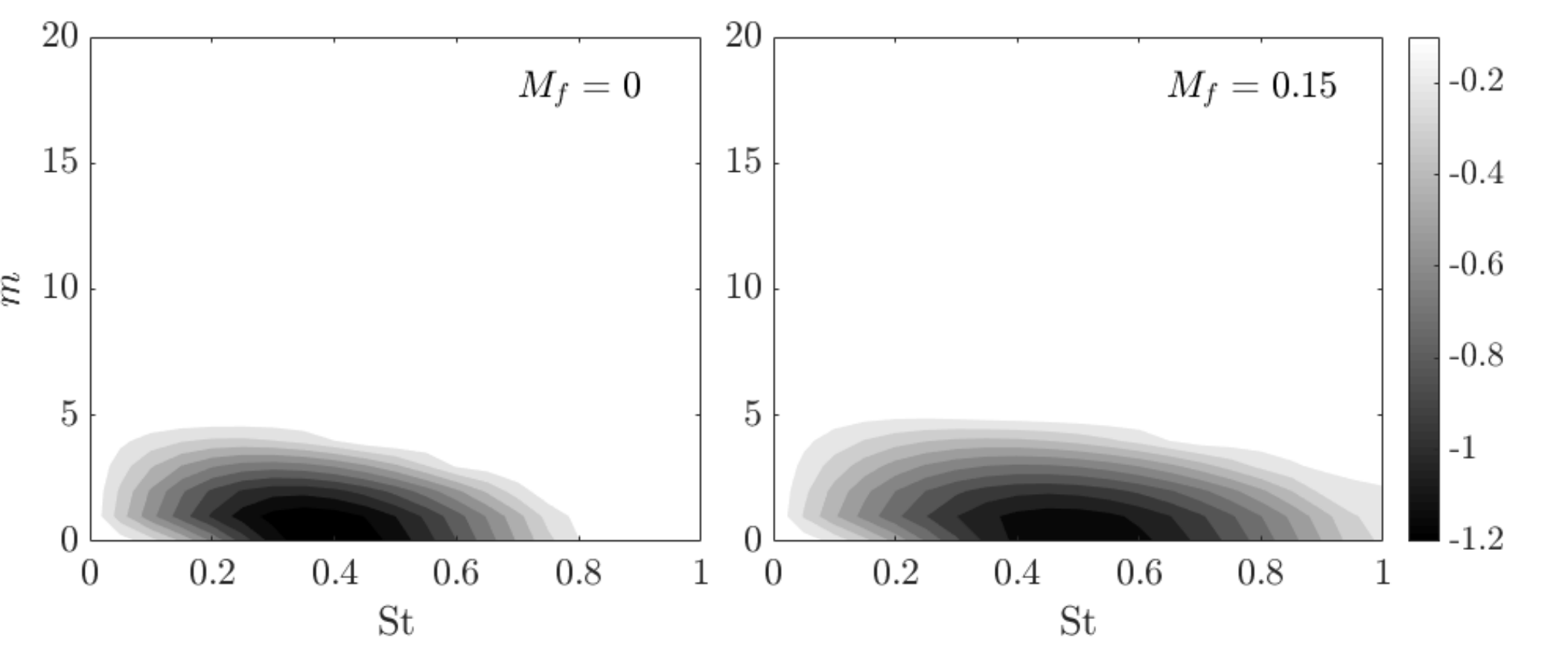}
\caption{Contours of the growth rates, $\alpha_i$, of the Kelvin-Helmholtz mode in the $St-m$ plane, computed at $x/x_c=0.23$.}
\label{fig15}
\end{figure}

In eigenanalysis, the nonlinear forcing terms are assumed negligible. The linearised Navier-Stokes system can then be recast in the form of an eigenvalue problem. We here consider the spatial stability problem, for which the eigenvalue problem is given by,

\begin{equation}
\mathbf{L}\hat{\mathbf{q}}=\alpha\mathbf{F}\hat{\mathbf{q}},
\label{eigen_spatial}
\end{equation}
where $L=-\omega\mathbf{I} +  \mathcal{A}_{0}$ and $F= -\mathcal{A}_{1}$. For high Reynolds numbers such as those considered here, \citet{RodriguezEuropean} have shown that $\alpha^2$ viscous terms can be neglected. The streamwise evolution of disturbances is governed by the sign of the imaginary part of the waveumber, $\alpha_i$. If $\alpha_i<0$, disturbances grow exponentially in the positive $x$ direction. In the following, we analyse the behaviour of the most unstable mode given by \ref{eigen_spatial}, which corresponds to the Kelvin-Helmholtz instability. Figure \ref{fig15} shows contours of $\alpha_i$ in the $St-m$ plane for instability analyses performed at $x/x_c=0.23$. A region of strong convective instability is seen for $m=0$-$4$ on a broad range of Strouhal numbers. Higher azimuthal wavenumbers were found to be stable for all Strouhal numbers analysed. As the jet evolves downstream, the region of convective instability gradually shifts to lower frequencies (not shown), following the thickening of the shear-layer. This is followed by a reduction of growth rates and eventual stabilization of the Kelvin-Helmholtz mode, as will be shown shortly; therefore, the $St \to 0$ region is always characterised by low growth rates. We note two important changes in the instability contours with the flight effect: a shift of the peak $\alpha_i$ to higher $St$ and a broader range of unstable frequencies. These changes are associated with the fact that the shear-layer thickness evolves at different rates for the two jets. As shown in figure \ref{fig5}, normalising the streamwise coordinate by $x_c$ corrects some of the discrepancy but does not eliminate it entirely, and the shear-layer in the $M_f=0$ still grows at a faster rate; therefore, for the same $x/x_c$, the jet with the flight stream has a smaller $\delta_{\theta}$. The thinner shear layer causes the most amplified mode to occur at a higher frequency, and the range of unstable frequencies to be broader. The broader range of amplified disturbances in turn offers more possibilities for nonlinear interactions between those disturbances to occur. This may explain why the flight stream produces a broader spectrum in the near-nozzle region, as seen in figure \ref{fig13}. 

\begin{figure}
\centering
\includegraphics[trim=0cm 5cm 0cm 3.5cm, clip=true,width=1\linewidth]{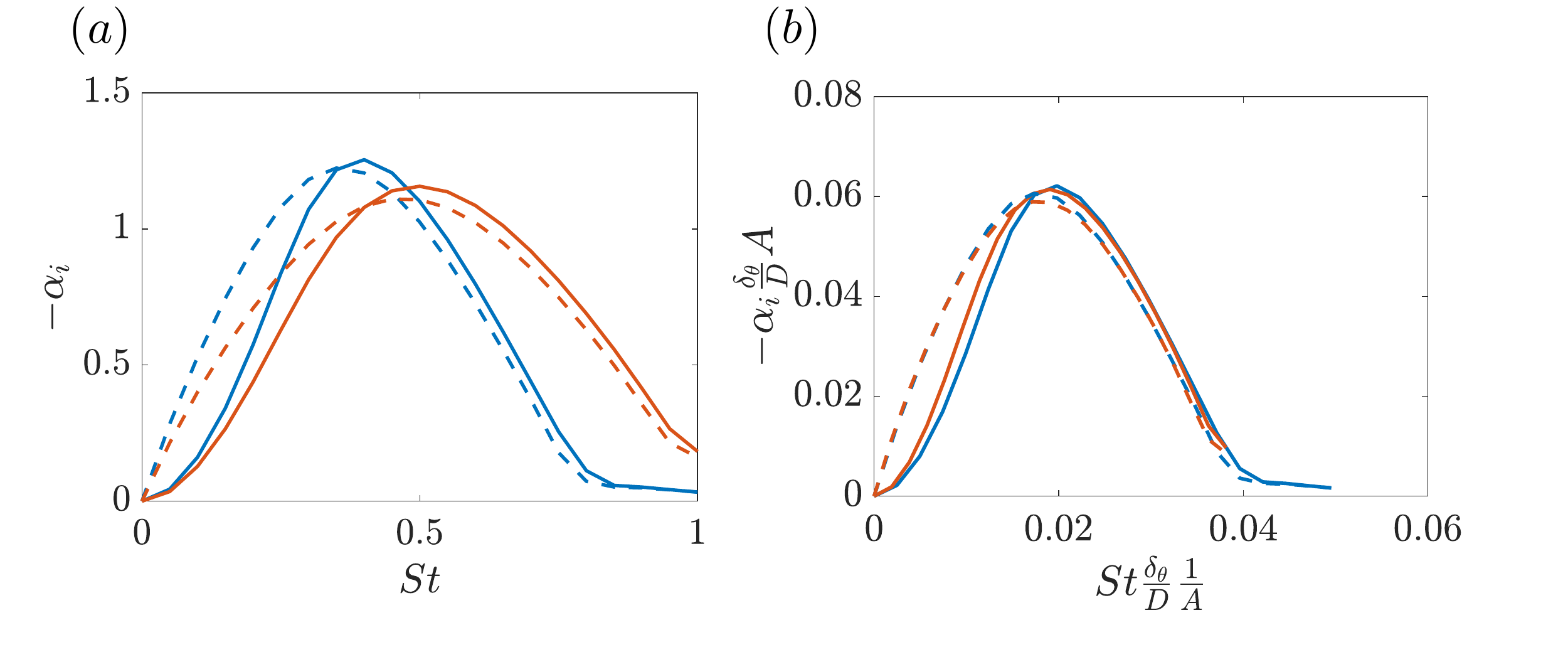}
\caption{Growth rates of the Kelvin-Helmholtz instability for azimuthal modes $m=0$ and $m=1$ at $x/x_c=0.23$. In (a), as a function of Strouhal number. In (b), the Strouhal number and growth rates are normalised using the local momentum thickness, $\delta_{\theta}/D$, and a stretching factor, $A$, which we define as the ratio between the potential core lengths in the flight and static cases. \usebox{\mylineblue}: $m=0$, $M_f=0$; \usebox{\mylinedashedblue}: $m=1, M_f=0$; \usebox{\myline}: $m=0$, $M_f=0.15$; \usebox{\mylinedashedred}: $m=1$, $M_f=0.15$.}
\label{fig16}
\end{figure}

\begin{figure}
\centering
\includegraphics[trim=0cm 5cm 0cm 4.5cm, clip=true,width=1\linewidth]{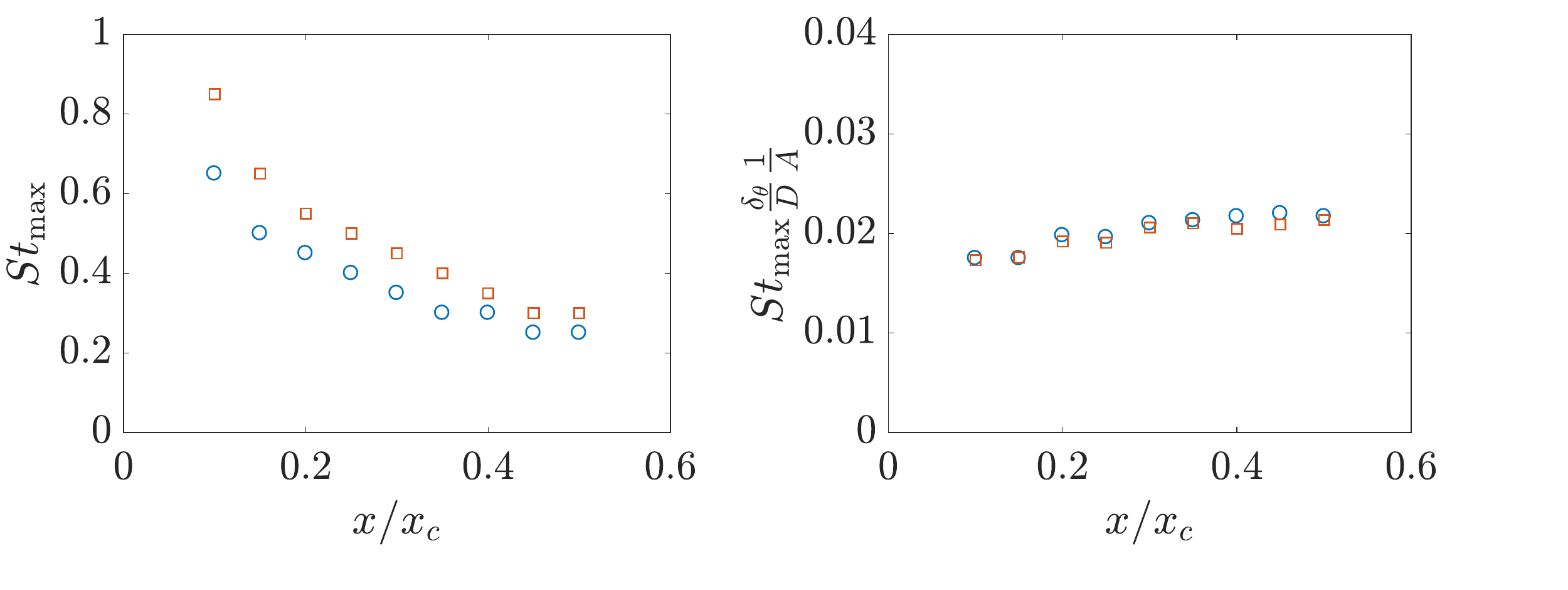}
\caption{(a) Strouhal number corresponding to the peak growth rate for the $m=0$ azimuthal wavenumber at different streamwise positions. In (b), the values are scaled by the local momentum thickness and the stretching factor, $A$, and a match is obtained for the flight and no-flight cases. Circles and squares correspond to the $M_f=0$ and $M_f=0.15$ cases, respectively.}
\label{fig17}
\end{figure}

These changes are followed by a decrease in the peak growth rates, as can  be more clearly seen in figure \ref{fig16} for the $m=0$ and $m=1$ azimuthal modes. In order to correct the discrepancy, \citet{MuchalkeHermann1982} proposed a scaling of the growth rates and frequencies by a \enquote{stretching factor}, derived from similarity considerations, $A = 1+ \Delta U /c_{n}U_{f}$, where $\Delta U = U_{j}-U_{f}$ and $c_n$ is a neutral phase velocity (we have adapted their variable nomenclature to avoid confusion with ours). This scaling was found to provide a reasonable match for the modified growth rates, $\alpha_i\delta_{\theta}A$ and frequencies, $\omega\delta_{\theta}/(\Delta U A)$ for jets in static and flight conditions. However, the stretching factor is not universal, but a function of $m$ and $\delta_{\theta}$. Furthermore, the match worsens with increasing shear-layer thickness, due to deviations from the hypotheses made in the derivation. Here we propose a scaling with a fixed stretching factor, given simply by the ratio between potential core lengths, $ A= (x_c)_f/(x_c)_s$, where the subscripts $f$ and $s$ refer to the static and flight cases, respectively. The modified growth rates and Strouhal numbers are thus given as $-\alpha_i \frac{\delta_{\theta}}{D} A$ and $St \frac{\delta_{\theta}}{D} \frac{1}{A}$. Figure \ref{fig16}(b) shows that the scaled curves are in excellent agreement, both in the frequency of the peak and its magnitude. Similar agreements were found for other $x/x_c$ in the initial jet region. Figure \ref{fig17} shows the streamwise evolution of the Strouhal number of the most amplified mode for the $m=0$ wavenumber. When corrected by the stretching parameters, the peak Strouhal numbers, $St_{max}\frac{\delta_{\theta}}{D} \frac{1}{A}$, are found to be virtually the same for two jets, and to be nearly independent of streamwise position. This scaling shows therefore that the frequency shift of the most unstable wavenumber is related to the shift in the potential core length. 

\begin{figure}
\centering
\includegraphics[trim=1cm 2cm 1cm 1cm, clip=true,width=0.8\linewidth]{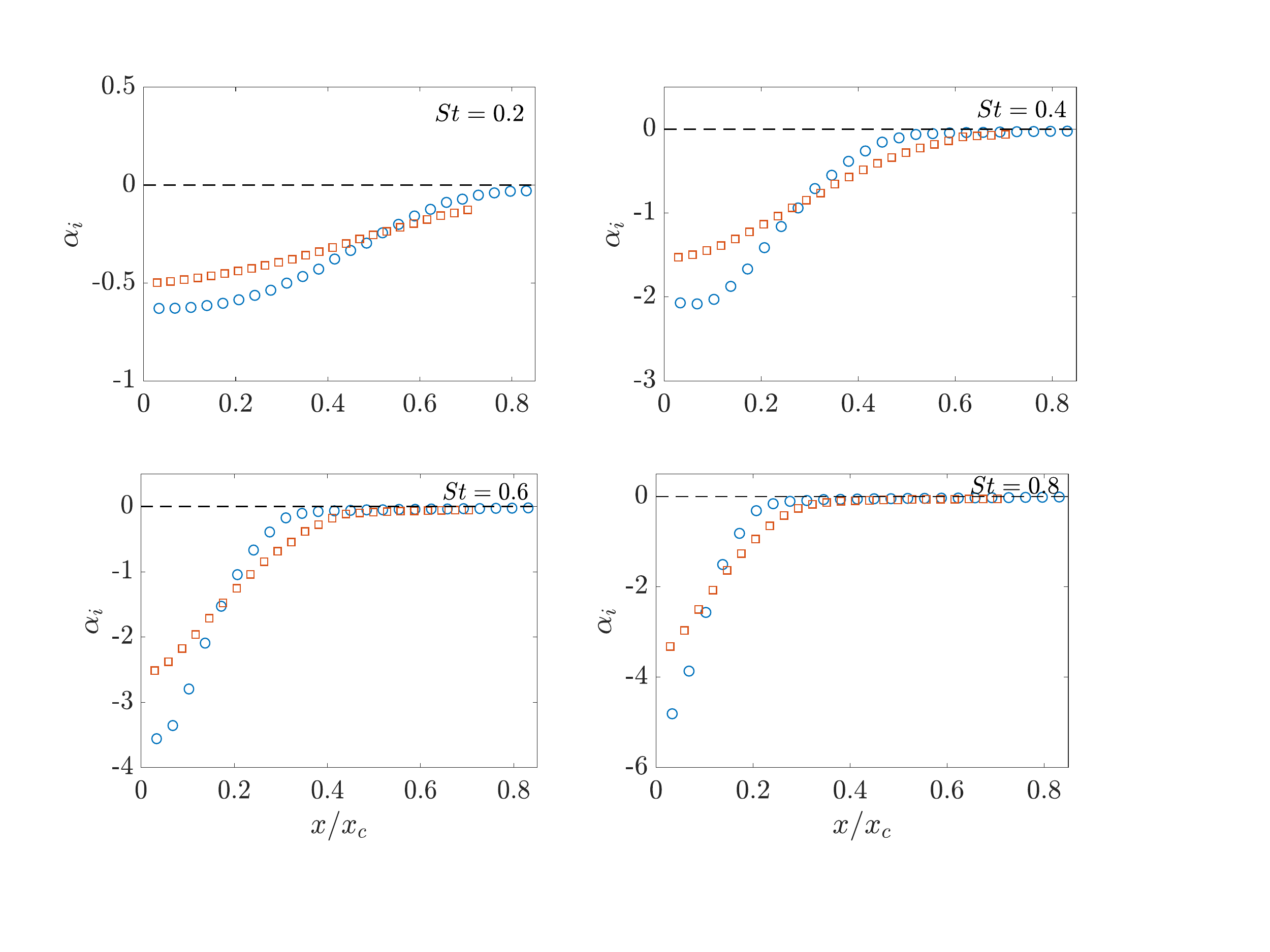}
\caption{Streamwise evolution of the growth rates of $m=0$ KH wavepackets for $St=0.2, 0.4, 0.6, 0.8$. Circles and squares correspond to the $M_f=0$ and $M_f=0.15$ cases, respectively. The black-dashed line corresponds to the $\alpha_i=0$, limit, when the KH mode becomes stable.}
\label{fig18}
\end{figure}

\begin{figure}
\centering
\includegraphics[trim=0cm 0cm 0cm 0cm, clip=true,width=0.5\linewidth]{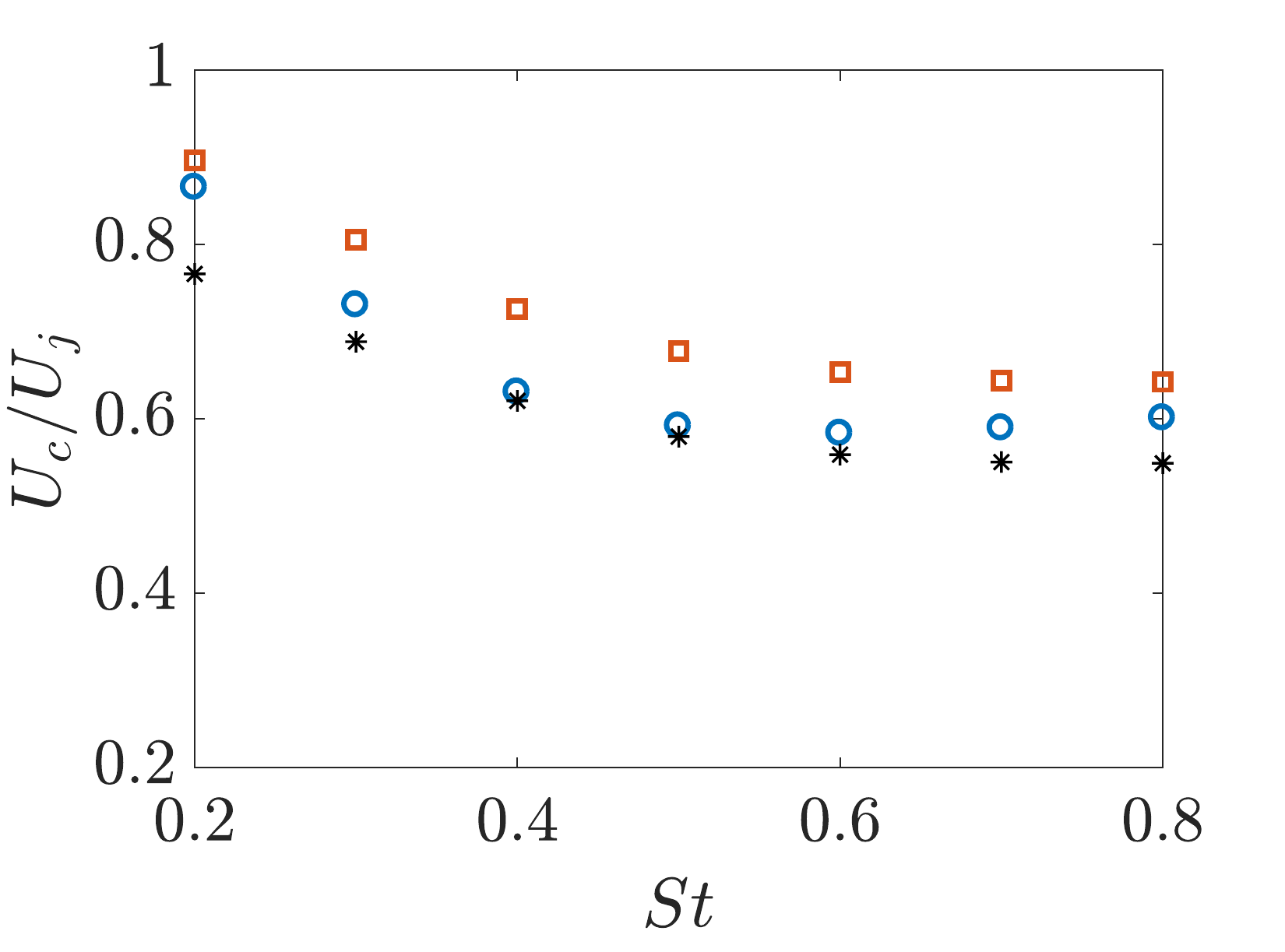}
\caption{Phase velocities of the $m=0$ KH instability as a function of $St$ computed at $x/x_c=0.23$. Circles: $M_f=0$; squares: $M_f=0.15$. The asterisks represent the phase velocities of the flight case scaled by the $A$ factor.}
\label{fig19}
\end{figure}

Figure \ref{fig18} shows the streamwise evolution of the axisymmetric KH growth rates for four Strouhal numbers within the unstable range. For a given $St$, growth rates in the flight case are lower close to the nozzle exit. However, without the scaling, they decay at a slower rate, due to the slower growth of the momentum thickness, and the stabilization predicted by the model occurs further downstream than in the static case. The changes in growth rates are followed by an increase in phase velocity, as shown in figure \ref{fig19} for the $m=0$ mode. Other azimuthal wavenumbers (not shown) display the same behaviour. Figure \ref{fig19} also shows that, in Strouhal numbers for which the KH instability is stronger ($St=0.4$-$0.6$ at $x_/x_c=0.23$), scaling the phase velocities of the flight stream case with $A$ makes them collapse with those of the static case.

The stabilization of the Kelvin-Helmholtz mechanism with the flight stream is consistent with the energy reduction at low $m$ seen in the modal energy maps depicted in figure  \ref{fig12} in the range $0.4 \leqslant St \leqslant 0.8$. It also explains the reduction in the $\lambda_{1_{m,\omega}}/\lambda_{2_{m,\omega}}$ ratio seen in the near-nozzle region. Previous studies showed the leading SPOD mode to be underpinned by the KH mechanism in this region \citep{Wavepacketsvelocity, SchmidtetalJFM2018, LesshafftPFR2019, PickeringJFM2020}. A less unstable KH mechanism then leads to a less dominant leading SPOD mode, and therefore a smaller separation with respect to suboptimal modes.

\subsection{$St \to 0$ limit: resolvent analysis}

As seen in figure \ref{fig15}, at low $St$ the growth rates are small, and therefore the KH modal instability mechanism is weak. Non-modal mechanisms, such as the Orr and lift-up mechanism then become dominant, but the spatial stability framework is not suitable to describe them. Therefore, in order to model the linear mechanisms in that zone we use a local model based on the resolvent operator. Equation \ref{ns_Fourier} can be rewritten as,

\begin{equation}
\hat{\mathbf{q}}_{\alpha,\omega,m} = \mathcal{C}\left(i\omega -\mathcal{A}_{0} - \alpha\mathcal{A}_{1} - \alpha^2 \mathcal{A}_{2} \right)_{\bar{\mathbf{q}}}^{-1}\mathcal{B}\hat{\mathbf{f}}_{\alpha,\omega,m},
\label{ns_fft}
\end{equation}
where matrices $\mathcal{B}$ and $\mathcal{C}$ can be used to restrict forcing and response to a desired subspace \citep{TissotJFM2017}. In a more compact form, we can write,

\begin{equation}
\hat{\mathbf{q}}_{\alpha,\omega,m}=\mathcal{R}_{\bar{\mathbf{q}},\alpha,\omega,m}\hat{\mathbf{f}}_{\alpha,\omega,m},
\end{equation}
where $\mathcal{C}\left(i\omega -\mathcal{A}_{0} - \alpha\mathcal{A}_{1} - \alpha^2 \mathcal{A}_{2} \right)_{\bar{\mathbf{q}}}^{-1}\mathcal{B}$ is the resolvent operator. For both the spatial stability problem and resolvent analysis discretisation in the radial direction is carried out using Chebyshev collocation points. The domain is extended to the far-field by mapping the original domain,  $r \in [-1, 1]$ to $r \in [0,\infty)$ using the function proposed by \cite{LesshaftHuerre2007} and concentrating most points in the potential core and shear layer. The goal of resolvent analysis is to seek an optimal forcing that maximizes the norm of the associated flow response,

\begin{equation}
\sigma^2 = \max_{\hat{\mathrm{f}}_{\alpha,\omega,m}}\frac{\left|\left| \mathcal{R}_{\bar{\mathbf{q}},\alpha,\omega,m}\hat{\mathbf{f}}_{\alpha,\omega,m} \right|\right|^{2}_{\mathbf{W}_{q}}}{\left|\left|\hat{\mathbf{f}}_{\alpha,\omega,m} \right|\right|^{2}_{\mathbf{W}_{f}}},
\end{equation}
where $\mathbf{W}_{q}$ and $\mathbf{W}_{f}$ are positive definite Hermitian matrices representing the weights of the  norms of response and forcing, respectively. The maximisation of the forcing can be achieved through a singular-value decomposition (SVD) of the resolvent operator,

\begin{equation}
\mathbf{W}_{q}^{-1/2} \mathcal{R}_{\bar{\mathbf{q}},\alpha,\omega,m} (\mathbf{W}_{f}^{-1/2})^{H} = \mathbf{U}\mathbf{\Sigma}\mathbf{V}^{H},
\end{equation}
where the superscript $^{H}$ denotes Hermitian transpose and $\mathbf{W}_{f,q}^{-1/2}$ are defined by a Cholesky decomposition. Forcing ($\mathbf{u}_{i}$) and response ($\mathbf{v}_{i}$) modes are defined as $\mathbf{v}_{i} = (\mathbf{W}^{-1/2})^{H}\mathbf{V}_{i}$ and $\mathbf{u}_{i} = (\mathbf{W}^{-1/2})^{H}\mathbf{U}_{i}$, with $i$ denoting the $i$-th column of $\mathbf{V}$ and $\mathbf{U}$. The singular values associated with each forcing-response pair, $\sigma_{i}$, are arranged in descending order in the diagonal matrix $\mathbf{\Sigma}$. 

Following \citet{Derghametal2013, NogueiraJFM2019, NogueiraetalTCFD2020}, we apply the following weighting function to the response,

\begin{equation}
\mathbf{W}_{q}=\mathrm{diag}\left[0.5(1-b)(1+\mathrm{tanh}(r_{p}/D)-r/D)+b\right]\mathbf{W}_{cheb},
\end{equation}
which is designed so as to localise responses inside the region of high shear and avoid the appearance of free-stream modes, mitigating the dependence of the results on domain size \citep{NogueiraetalTCFD2020}. $\mathbf{W}_{cheb}$ are the quadrature weights for the Chebyshev grid. The value of $r_{p}/D=1$ is chosen so that the weights are zero far from the sheared region. $b$ is a small positive parameter used to avoid zero weights at large $r/D$. No spatial restriction is applied to the forcing weight, so that $\mathbf{W}_{f}=\mathbf{W}_{cheb}$. With this choice of parameters, we seek an optimal forcing whose associated response is maximised inside the jet core and shear layer, $r/D \lesssim 1$. The locally-parallel resolvent analysis is carried out at a fixed streamwise position using the mean flow, $\alpha$, $\omega$ and $m$ as inputs.

For high-Reynolds-number flows such as the one considered here, recent studies have explored the use of eddy-viscosity models as a means to improve agreement between the model and coherent structures observed in flow data \citep{SchmidtetalJFM2018, Crouch_etal_JcP2007, Oberleithner_etal_JFM2014, Rukes_etal_2016, Tammisola_Juniper_2016, SchmidtetalJFM2018, HwangCossu2010, MorraetalArxiv2019, Pickeringetal_eddy_2021, Kuhn_etal_JFM2021}. The eddy-viscosity provides a
partial inclusion of non-linear effects in the linear operator, beyond the simple establishment of the mean-flow, and good agreement may be obtained in both cases with an effective Reynolds number that is substantially lower than the molecular one. Here we consider a radially-constant eddy-viscosity, following the approach of \citet{Kuhn_etal_JFM2021}. It is known that constant eddy-viscosity fields do not capture the intermittent boundary of the flow, where fluctuations change from rotational to potential. We do not expect the model to capture intermittency effects. However, it was shown by \citet{Kuhn_etal_JFM2021} that the use of radially-dependent eddy-viscosity only improves marginally the results of local mean-flow models. The magnitude of $\nu_T$ is determined from a least-square fit of the $\overline{u_{x}'u_{r}'}$ component of Reynolds tensor based on a Boussinesq model,

\begin{equation}
\tilde{\nu}_T = -\frac{\overline{u_{x}'u_{r}'}}{\mathrm{d}\overline{U}/\mathrm{d}r},
\end{equation}
with $\tilde{\nu}_T = \nu_T/(U_jD)$ being a non-dimensional kinematic viscosity. Considering mean velocity and Reynolds tensor profiles at $x/x_c=0.23$, data for the $M_f=0$ case yields $\tilde{\nu}_T = 0.0017$, which corresponds to a turbulent Reynolds number of $\Re_T = \tilde{\nu}_T^{-1} = 588$. For the sake of simplicity, we keep this value of $Re_T$ for the static and flight cases.

\begin{figure}
\centering
\includegraphics[trim=0cm 0cm 0cm 0cm, clip=true,width=0.45\linewidth]{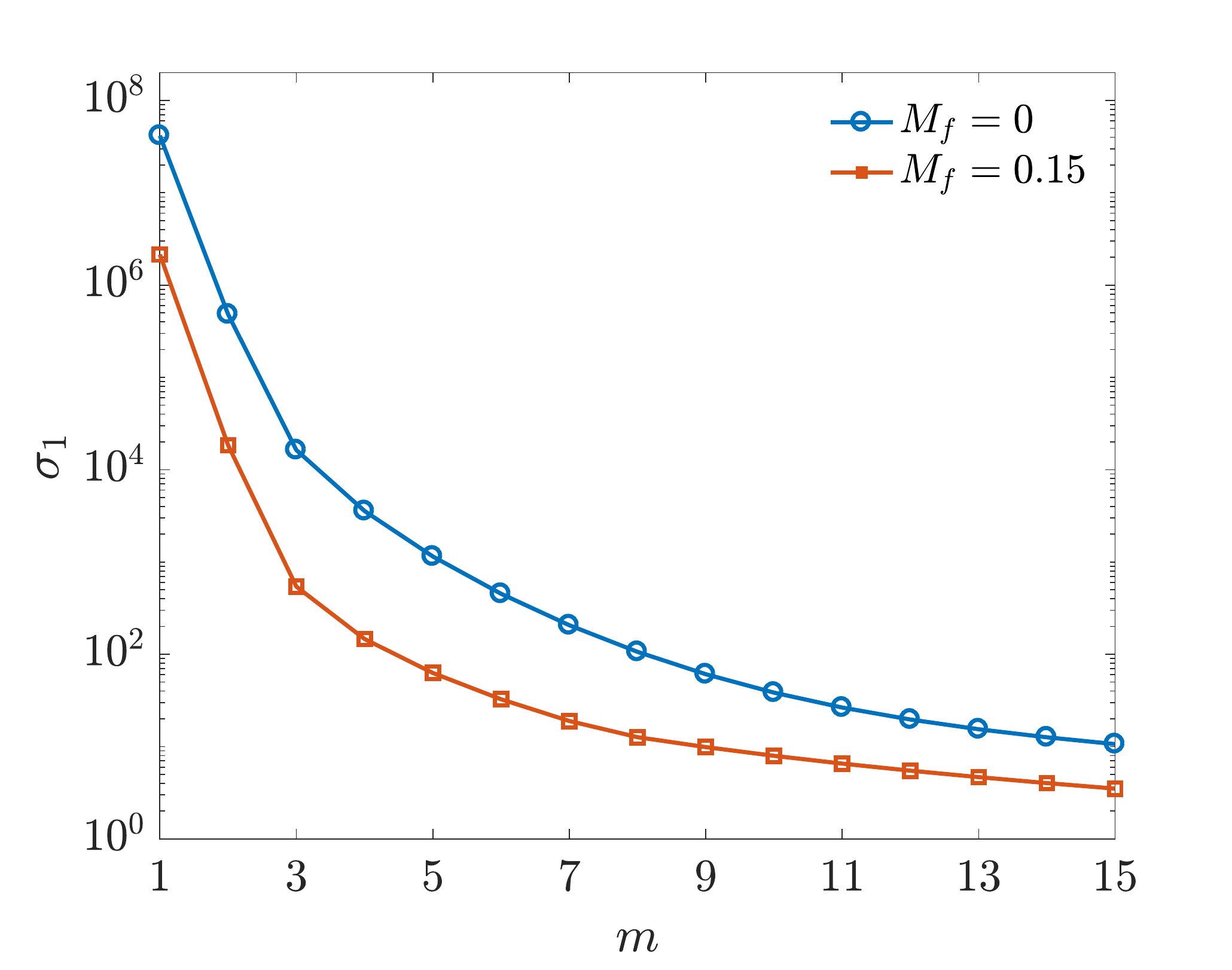}
\caption{Optimal resolvent gains, $\sigma_{1}$, as a function of azimuthal wavenumber, $m$, computed for $\omega$=0 and $\alpha=0$.}
\label{fig20}
\end{figure}

In the following we show results of resolvent analyses performed in the initial jet region. We consider a range of parameters at which the flow is expected to be dominated by streaky structures generated by the lift-up effect. These structures exist for non-zero azimuthal wavenumbers and are characterised by slow time scales and low streamwise wavenumbers. We therefore set $\omega=0$, $\alpha=0$ and $m>0$. Inspection of the energy maps in figure \ref{fig12} (b) and (f) show that, in the $St \to 0$ limit, the first ten to fifteen azimuthal wavenumbers observed in the experiment are weakened by the flight stream. This trend was found to be correctly captured by the resolvent model. Figure \ref{fig20} displays the leading resolvent gains as a function of $m$ computed with mean velocity profiles measured at $x/x_c=0.23$. The leading gain is reduced for all $m$ with the flight stream. 

The model does not predict correctly the azimuthal wavenumber of the peak energy. It is seen in figures  \ref{fig12} (b) and (f) that the peak energy occurs at $m=5$, whereas the model predicts gains that decay monotonically for $m\geqslant1$. This is a limitation of the local model, which predicts the gain and shape of optimally forced structures that, from a given streamwise position, will grow downstream as they are convected with the parallel mean flow. The highest gain computed for the model is for $m=1$. Indeed, further downstream the $m=1$ mode does become dominant; but because the local model does not take into account the upstream amplification of higher wavenumbers, it is not equipped to provide the correct shape of the energy spectrum at that position. For this reason, we limit our discussion of the resolvent-analysis to global differences in gain between the $M_{f}=0$ and $M_{f}=0.15$ cases. In that sense, the reduction in gains agrees with the trends of the energy maps based on the data, showing that the attenuation observed in the experiment may be associated with the attenuation of linear (mean-flow) growth mechanisms.

\begin{figure}
\centering
\includegraphics[trim=0cm 0cm 0cm 0cm, clip=true,width=1\linewidth]{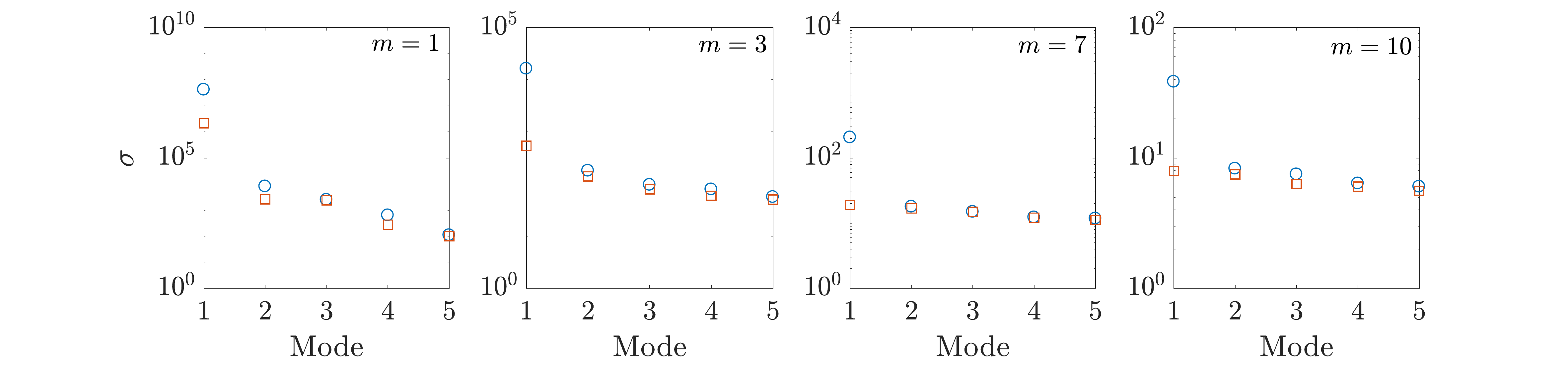}
\caption{Optimal and first 4 suboptimal resolvent gains, $\sigma_{1-5}$  computed for $\omega$=0 and $\alpha=0$. Circles: $M_f=0$; squares: $M_f=0.15$.}
\label{fig21}
\end{figure}

Figure \ref{fig21} shows a comparison between the leading gain and the first 4 suboptimal gains for selected $m$. It can be seen that, contrary to $\sigma_{1}$, the suboptimal gains are not strongly affected by the flight stream, and therefore the separation between $\sigma_{1}$ and $\sigma_{2}$ is significantly reduced. This suggests that the attenuation of the associated flow structures is produced mainly by a weakening of the optimal mechanism. This is consistent with results of the SPOD analysis, which showed a reduction of the modal energy separation, $\lambda_1/\lambda_2$, in the flight stream case, and suggests that the rank-reduction in the initial jet region is produced by a linear mechanism. This trend was also verified using other values of $Re_{T}$.

\begin{figure}
\centering
\includegraphics[trim=0cm 2cm 0cm 0cm, clip=true,width=1\linewidth]{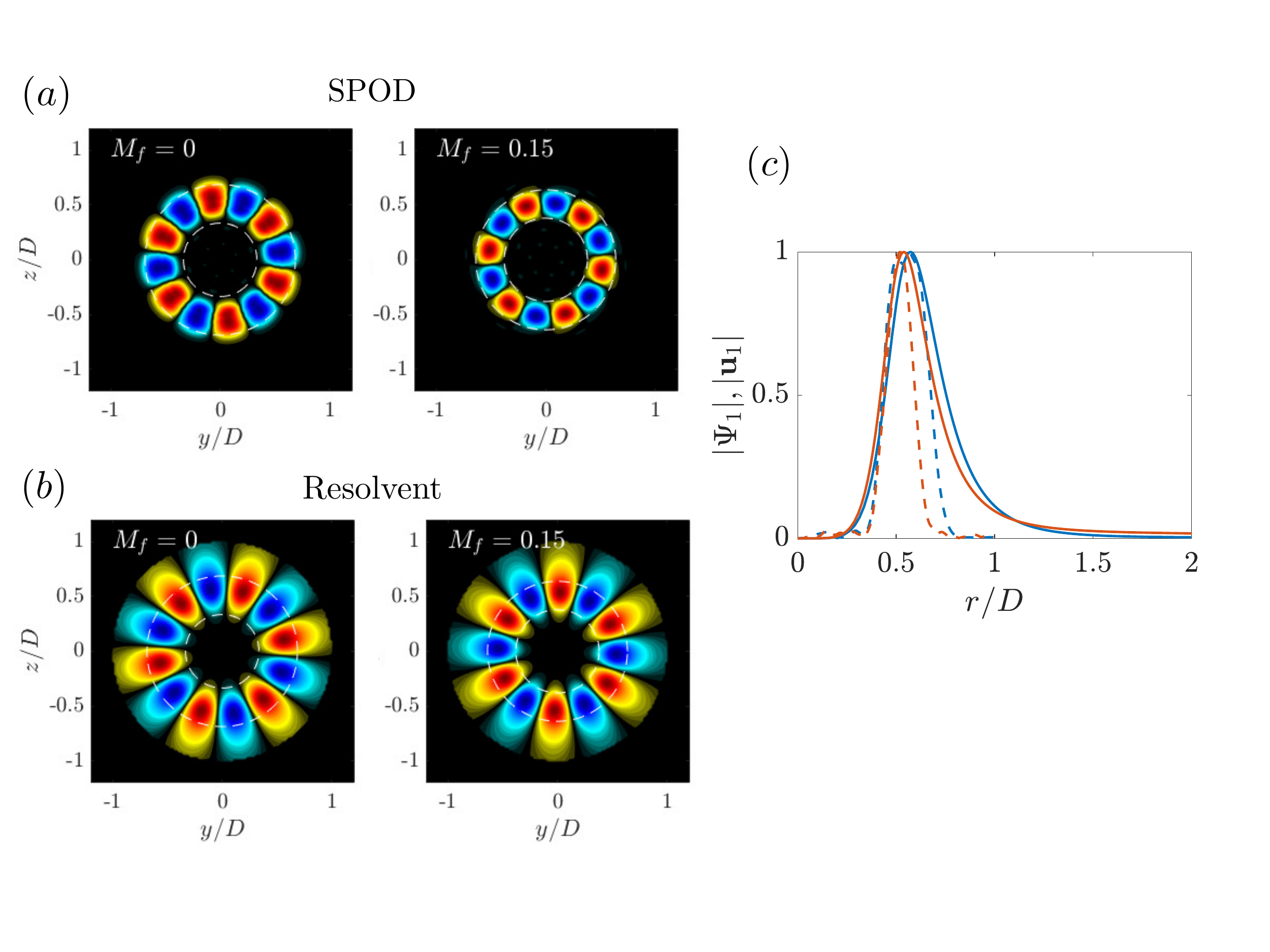}
\caption{Comparison between local streamwise velocity SPOD and resolvent response modes at $x/x_{c}=0.23$ associated with $m=6$ streaky structures. (a) and (b) show reconstructions of the real parts of local modes in the $y-z$ plane using the \textit{Ansatz} $\mathbf{\Psi}_{1},\mathbf{u}_{1}=\hat{u}_{x}(r)e^{im\theta}$. Contours are adjusted between 0 and 1. (c) show the absolute parts of the modes, $|\mathbf{\Psi}_{1}|,|\mathbf{u}_{1}|$; \usebox{\mylineblue}: leading resolvent mode for $M_{f}=0$; \usebox{\myline}: leading resolvent mode for $M_{f}=0.15$; \usebox{\mylinedashedblue}: leading SPOD mode for $M_{f}=0$; \usebox{\mylinedashedred}:  leading SPOD mode for $M_{f}=0.15$. The white dashed lines in (a) and (b) correspond to the positions where $\tilde{U}_{x}(x,r)=0.99$ and $\tilde{U}_{x}(x,r)=0.05$, roughly delimiting the shear-layer.}
\label{fig22}
\end{figure}

Figure \ref{fig22} shows SPOD and optimal resolvent response modes computed at $m=6$. The modes are reconstructed in the $y-z$ plane using the \textit{Ansatz} $\mathbf{\Psi}_{1},\mathbf{u}_{1}=\hat{u}_{x}(r)e^{im\theta}$, where $\hat{u}_{x}(r)$ is the radial structure issuing from the SPOD and resolvent calculations. The streaky structures possess no support in the jet core, and are confined to the shear-layer. Due to the thinner shear layers, in the flight stream case the streaks become more radially compact and  their peak shifts to a smaller $r/D$. These trends are well captured by the resolvent model. The resolvent modes display a slower decay in $r/D$ with respect to the SPOD, which is something that had also been observed by \cite{NogueiraJFM2019}. 

Overall, the linear mean flow models are able to mimic two important trends seen in the energy maps of experimental data: the stabilization of the KH and lift-up mechanisms and a weakening of the low-rank behaviour of the jet on in extended regions of the $St$-$m$ spectrum. 

\section{Global SPOD}
\label{sec:global_spod}

\begin{figure}
\centering
\includegraphics[trim=2cm 0cm 2cm 0cm, clip=true,width=1\linewidth]{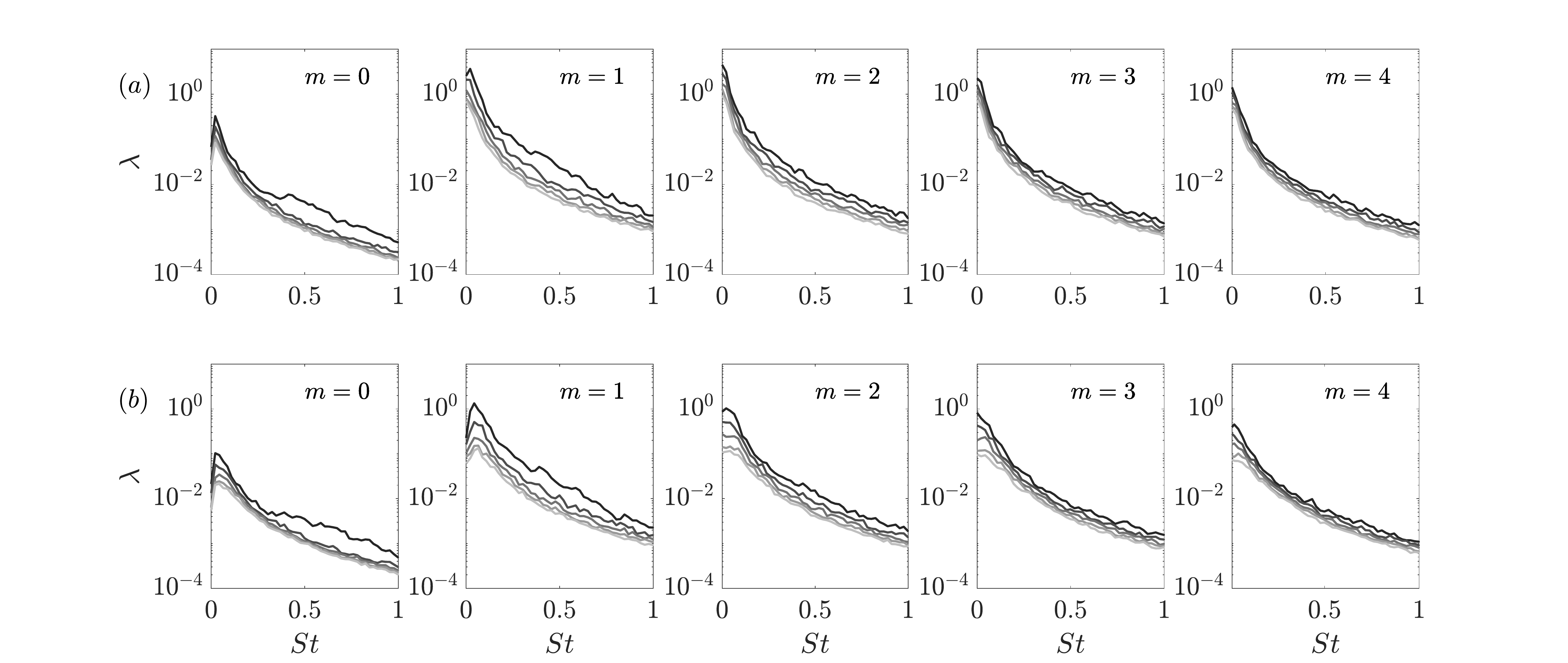}
\caption{Eigenvalue spectra of the global SPOD analysis modes performed with LES data. Color shading from black to white represent increasing mode numbers (\usebox{\myboxblack} \hspace{-3.5mm}\usebox{\myboxgray} \hspace{-3.5mm}\usebox{\myboxwhite}, $\lambda_{1}>\lambda_{2}>...\lambda_{5}$) (a) $M_f=0$; (b) $M_f=0.15$.}
\label{fig24}
\end{figure}

We now move from the local framework of PIV data to a global SPOD analysis performed with the LES database. Our goal is to compare modal energy maps and to quantify the distortion of global SPOD modes by the flight stream. We consider the complete state variable vector, $\mathbf{q}=[\rho, u_{x},u_{r}, u_{\theta}, T]^{T}$ in the entire available flow domain, $0 \leqslant x/D \leqslant 30$, $0\leqslant r/D\leqslant 6$. The computed modes are orthogonal with respect to Chu's energy norm \citep{Chu1965}.

\begin{equation}
\left< \mathbf{q}_{1},\mathbf{q}_{2}\right>_{E}=\iiint\mathbf{q}_{1}^{*}\mathrm{diag} \left(\frac{\bar{T}}{\gamma \bar{\rho}M_{j}^2},\bar{\rho}, \bar{\rho}, \bar{\rho}, \frac{\bar{\rho}}{\gamma(\gamma-1)\bar{T}M_{j}^2} \right)\mathbf{q}_{2} r\mathrm{d}x\mathrm{d}r \mathrm{d}\theta.
\label{eq:chu}
\end{equation}
Figure \ref{fig24} shows eigenvalue spectra for the first five azimuthal wavenumbers. Spectra for the axisymmetric mode display the known separation between the leading and first suboptimal mode for $St \gtrsim 0.3$ \citep{SchmidtetalJFM2018, LesshafftPFR2019}. The low-rank behaviour in that Strouhal number range becomes less and less marked with increasing wavenumber, a trend that was also observed by \cite{SchmidtetalJFM2018}, and which is consistent with the weakening of the KH instability at higher $m$, as can be seen in figure \ref{fig15}. Spectra for the static and flight cases follow the same general trends; however, some important difference are observed in the eigenvalue separation.

Figure \ref{fig24} shows contours of $\lambda_1/\lambda_2$ in the $St-m$ plane. It can be seen that peak values of $\lambda_1/\lambda_2$ are reduced by the flight stream in the KH-dominated zone, which was also observed in the local SPOD maps. Moreover, the low-rank zone is broader and shifted towards higher $St$. These trends are consistent with the results of the local stability analysis, which predicted a larger range of unstable frequencies in the flight case, and a shift of the most unstable mode towards higher $St$. Interestingly, in the low $St$ limit, the flight stream slightly increases the eigenvalue separation. This is contrary to what is observed in the PIV data up to $x/x_c=1.6$, and suggests that further downstream the flight stream changes the dynamics of streaks in a non trivial manner. The precise mechanism by which this happens is not yet clear, and is currently under investigation.

\begin{figure}
\centering
\includegraphics[trim=2cm 5cm 2cm 5cm, clip=true,width=0.8\linewidth]{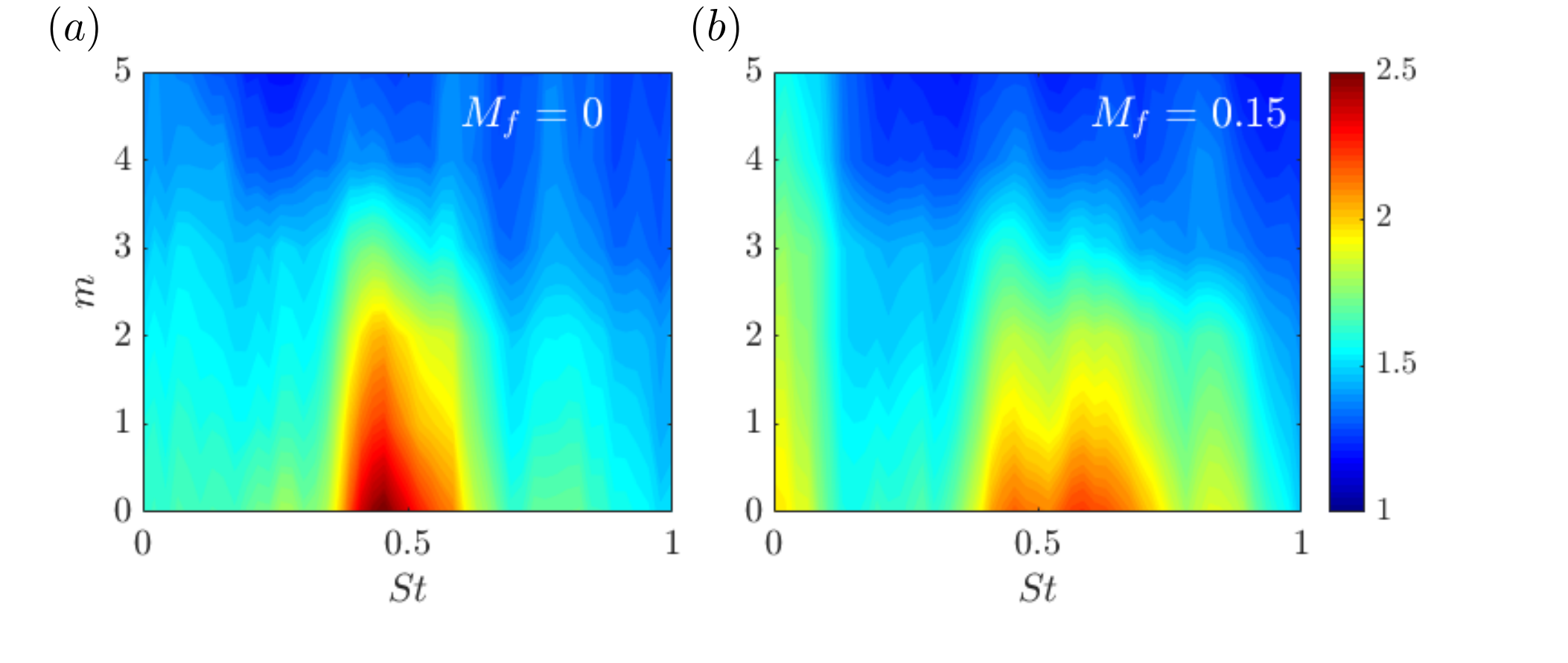}
\caption{Contours of the ratio between the leading and first suboptimal eigenvalues, $\lambda_{1_{m,\omega}}/\lambda_{2_{m,\omega}}$, from global SPOD for (a) the static case and (b) the flight case.}
\label{fig25}
\end{figure}

\begin{figure}
\centering
\includegraphics[trim=0cm 4cm 0cm 1cm, clip=true,width=1\linewidth]{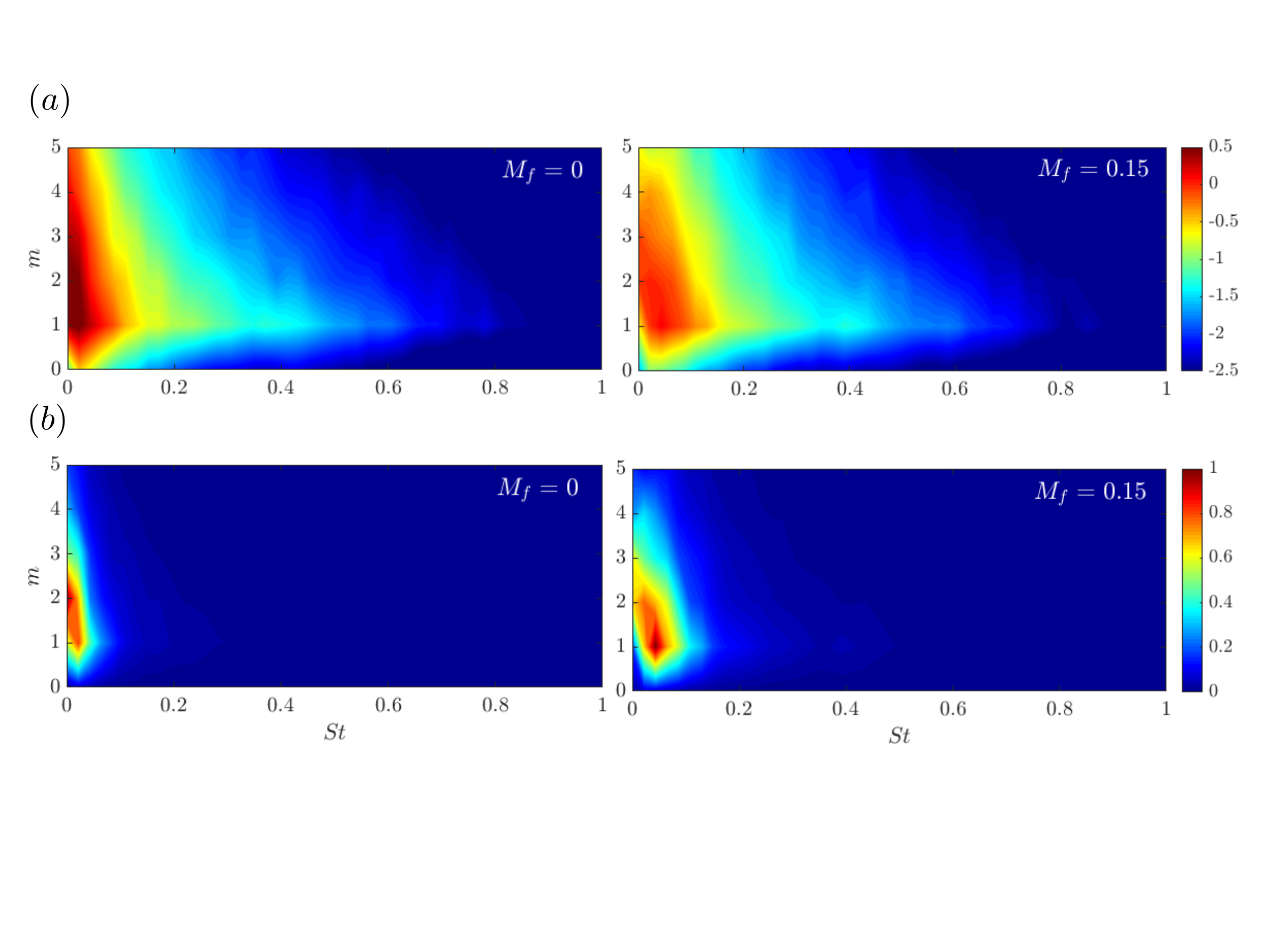}
\caption{Modal energy maps of the leading SPOD mode. In (a), contours of $\mathrm{log}_{10}(\lambda_{1_{St,m}})$ are shown. In (b), the maps are normalised by their maximum, $(\lambda_{1_{St,m}}/\mathrm{max}(\lambda_{1_{St,m}}))$.}
\label{fig26}
\end{figure}

The $St \to 0$ limit is also the region where the largest energy reductions occur, as clearly seen in the modal energy maps of the leading SPOD modes of figure \ref{fig26}. In the global framework, the modal energy spectrum is dominated, for a broad range of Strouhal numbers, by the $m=1$ mode (the dominance of the $m=1$ mode can also be observed in the results of figure \ref{fig12} at downstream positions). In the zero-frequency limit, $St \to 0$, $m=1$-$3$ modes are dominant. These trends are also found to hold in the presence of the flight stream. The attenuation of the $m=1$-$3$ streaks, which are globally the most energetic structures of the flow, is the most striking modification. In \ref{fig25} (b), the modal maps are normalised by their respective maximum values. The normalisation reveals a broader spectrum with the flight stream: while in the static case the peak is concentrated around $m=2$ and $St \to 0$, in the flight case it spreads over $1 \leqslant m \leqslant 3$ and up to $St=0.1$.

\begin{figure}
\centering
\includegraphics[trim=0cm 0cm 0cm 0cm, clip=true,width=1\linewidth]{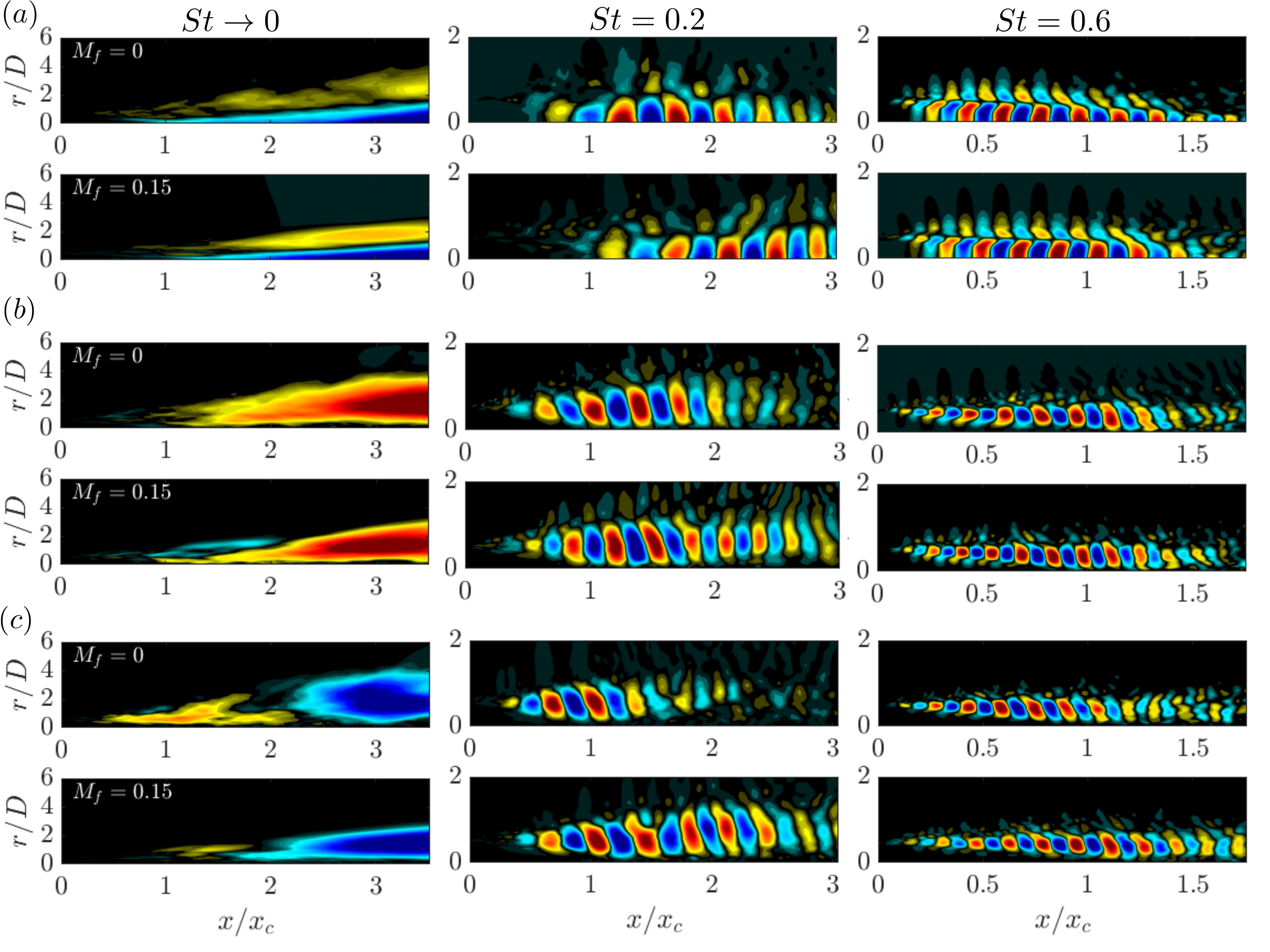}
\caption{Real part of global leading SPOD modes of streamwise velocity, $Re(\mathbf{\Psi}_{1}:\hat{u}_{x})$, at Strouhal numbers and wavenumbers representative of the Lift-up, Orr and Kelvin-Helmholtz instability mechanisms. (a), (b) and (c) are for azimuthal modes $m=0,2,3$, respectively.}
\label{fig27}
\end{figure}

Figure \ref{fig27} shows leading SPOD modes of streamwise velocity, $\mathbf{\Psi}_{1}:\hat{u}_{x}$, at different $St$ and $m$ representative of the Kelvin-Helmholtz, Orr and lift-up mechanisms, according to the linear mechanism maps of \cite{PickeringJFM2020}. The KH-associated structures, here represented by the modes at $St=0.6$, display the same characteristics described in previous studies \cite{SchmidtetalJFM2018, Garnaud, LesshafftPFR2019}: organised structures that grow in the first jet diameters, stabilise and decay towards the end of the potential core. As the mode order is increased, the structure of the Kelvin-Helmholtz wavepackets is seen to lose their support in the jet core and to become confined to the shear layer. That applies to structures found in both jets. The flight stream does not lead to any substantial change in the underlying spatial structure of the KH instability. The reduction in growth rates predicted by the linear model are not manifest in $\mathbf{\Psi}$, but rather in the modal energies. The linear model also predicts an increase in phase velocity with the flight stream, which in principle causes a change in the spatial wavelength of the modes; however, for Strouhal numbers around $St=0.4$-$0.6$, the change in phase velocity in the initial jet region was found to scale approximately with $x_c$, as shown in figure \ref{fig19}, so that when plotted as a function of $x/x_c$, the wavelengths of the leading SPOD modes look quite similar.

The structures associated with the Orr mechanism ($m=0$, $St=0,0.2$) are more spatially-extended and develop downstream of the end of the potential core \citep{SchmidtetalJFM2018, Garnaud, LesshafftPFR2019}. In the zero-frequency limit, the most energetic structures correspond to streaks, which are characterised by very large wavelengths, and a slow spatial development, attaining their maximum far downstream. Here the effect of the flight stream is seen to be more marked: since streaks have spatial support in the jet shear layer, the thinner shear-layer thicknesses in the $M_{f}=0.15$ case (see figures \ref{fig3} and \ref{fig5}) produce streaks which are less extended in the radial direction. For small, but non-zero Strouhal numbers up to $St \approx 0.2$ and small, but non-zero wavenumbers, the Orr mechanism is likely to be present, together with lift-up mechanism \cite{AKERVIK2008501, arratia_caulfield_chomaz_2013, hack_moin_2017}. This is reflected in the shapes of the SPOD modes at $St=0.2$, $m=3$, which bear some resemblance with the Orr structures found at $m=0$, although they have less support in the jet core, as do the streaks at $St \to 0$.

\begin{figure}
\centering
\includegraphics[trim=2cm 1cm 2cm 2cm, clip=true,width=0.6\linewidth]{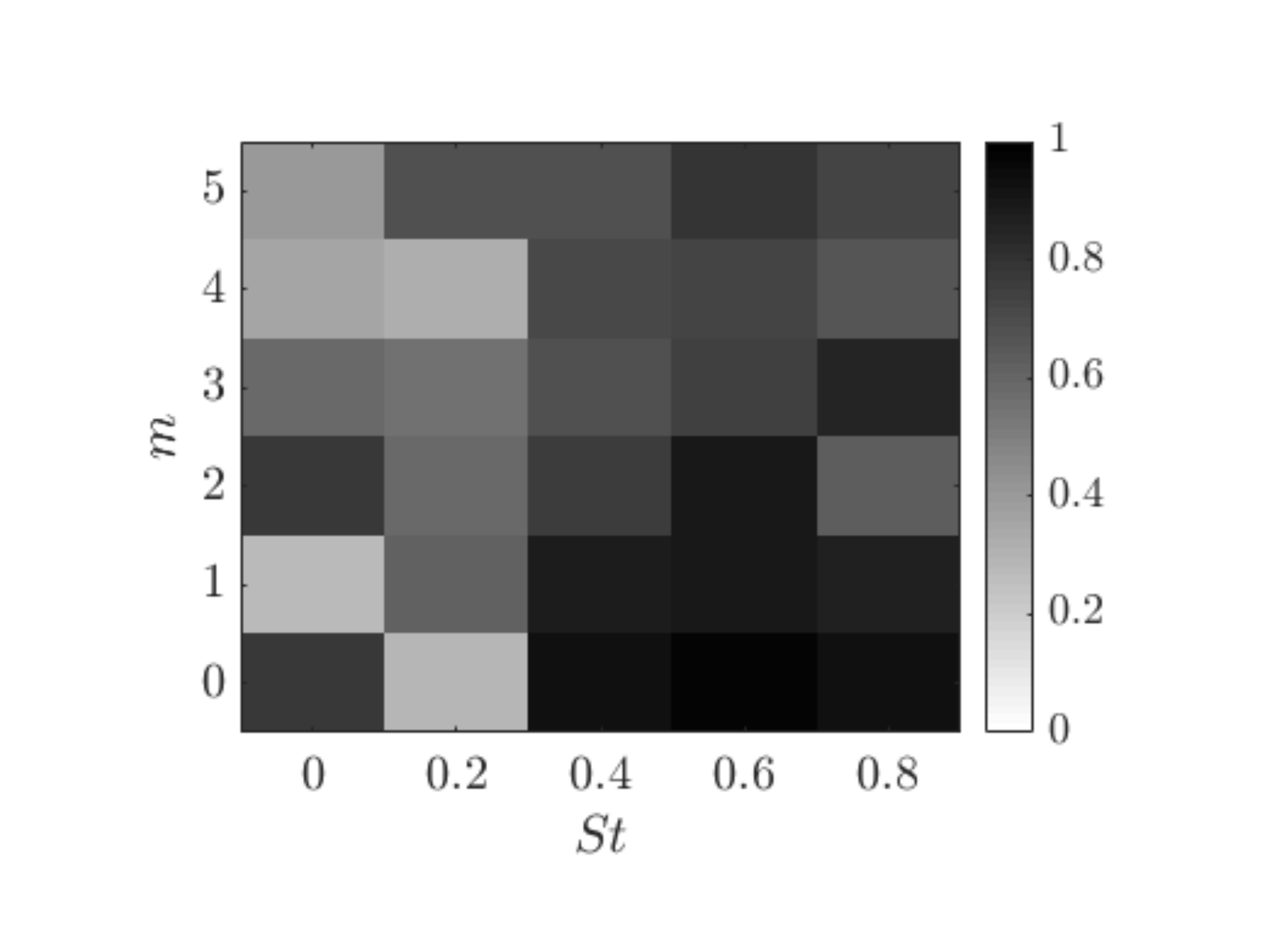}
\caption{$\beta$ metric for the alignment of global leading SPOD modes, $\mathbf{\Psi}_{1}$, with and without the flight stream.}
\label{fig28}
\end{figure}

The alignment between the SPOD modes computed in both flow conditions can be characterised in a quantitative manner through the following $\beta$ metric,


\begin{equation}
\beta = \left| \mathbf{\Psi}_{1} \mathbf{W} \mathbf{\Psi}_{1_{f}}\right|
\end{equation}
where the subscript $_{f}$ refers to the flight stream case. $\mathbf{\Psi}_{1}$ is the leading SPOD mode containing all flow variables. It represents a global alignment metric for the projection of the modes and varies from 0, if the modes are completely orthogonal, to 1, if they are perfectly aligned. The weight matrix $\mathbf{W}$ accounts for the quadrature weights and Chu's energy norm. In an attempt to see to what extent the potential core length collapses the overall flow organisation, prior to computing $\beta$, the modes and mean flows are scaled by $x_c$. Furthermore, when computing the alignment at $St=0.4, 0.6$ and $0.8$, the domain was restricted to $x/D=15$, because downstream of those positions KH wavepackets lose much of their spatial support and the modes become noisy, thus biasing the alignment. Using the same reasoning, alignment for $St=0.2$ was computed with a domain that extends up to $x/D=25$. Figure \ref{fig28} shows the values of $\beta$ as a function of $St$ and $m$. $\beta$ assumes very high values for Strouhal numbers and wavenumbers associated with KH wavepackets, confirming that their spatial shape is mostly controlled by the mean-flow stretching, and whose effect can be captured using the length of the potential core as a similarity. At Orr and lift-up dominated frequencies, the alignment between the modes is globally poorer than that of the KH mechanism, revealing that the associated flow structures are impacted in a more subtle way by the flight stream.

\section{Conclusions}
\label{sec:conclusions}

In this work, we present high-fidelity experimental and numerical databases of subsonic turbulent jets subject to flight streams. 2D-PIV experiments were first performed aiming at quantifying the evolution of first-order velocity statistics with increasing flight stream Mach numbers. The main effects of the flight stream on these were found to follow the trends reported in previous studies, namely: a stretching of the jet potential core, reduction of shear layer thicknesses and a reduction of turbulent intensities all over the jet domain. The main motivation of the present study is to perform an analysis of the energy distribution in a large region of frequency-wavenumber space. Such analysis requires time-resolved data in the $r$-$\theta$ plane; for that, extensive cross-plane, stereoscopic, time-resolved PIV experiments were performed in a streamwise region ranging from the near-nozzle region to twice the potential core length, $x_{c}$. Two flow conditions were chosen for the analysis, $(M_{j},M_{f})=(0.9,0)$ and $(M_{j},M_{f})=(0.9,0.15)$. Companion LES simulations are also performed, and found to be in excellent agreement with the experimental data. The experimental and numerical databases are used to perform local and global SPOD, respectively, and to compute modal energy maps. Apart from the well-known attenuation of the modal KH instability mechanism, \citep{MuchalkeHermann1982, SoaresAIAA2020}, here we also report, to the best of our knowledge for the first time, striking energy reductions in regions of the frequency-wavenumber space associated with the Orr and lift-up non modal mechanisms. The effect of the flight stream is found to be manifest in different aspects of coherent structure dynamics and energy organisation that go beyond the simple attenuation of turbulent energy due to the reduced shear.

The attenuation of coherent structures occurs from the very near-nozzle region and is accompanied by weakening of the low-rank dynamics of the jet, expressed by an eigenvalue separation between the leading and second SPOD modes. This shows that the most energetic, coherent structures that characterise the leading SPOD mode are less dominant with respect to suboptimal modes. Zero-frequency streaky structures associated with helical wavenumbers $m=1$-$4$ are globally the most affected by the flight stream, on account of both their energy attenuation and spatial distortion (see figures \ref{fig26} and \ref{fig28}). 

Throughout the study, a normalised coordinate, $x/x_{c}$, based on the potential core length is used to make careful comparisons between the static and flight cases at different streamwise positions. This normalisation, which attempts to account for the mean-flow stretching, reveals some similarities in important flow properties. For instance, the centerline velocity profiles collapse when plotted as a function of the scaled streamwise coordinate, and the azimuthal organisation of energy in the two cases also scales, to a great extent, with $x_c$. Furthermore, a stretching parameter, $A$, defined as the ratio of potential core lengths was shown to successfully correct the modifications in spatial growth rates of the Kelvin-Helmholtz instability in the static and flight cases. The scaled growth rates, $\alpha_i \delta_{\theta}A$, are shown to collapse when plotted as a function of scaled frequencies, $St \delta_{\theta}(1/A)$. The scaled frequency of the most amplified mode,$St_{max} \delta_{\theta}(1/A)$, is also seen to be the same in static and flight conditions. Global SPOD performed on the LES data showed that low-frequency streaky structures associated with helical wavenumbers are the most distorted by the flight stream. Such distortion is not simply an effect of the potential core stretching, as evidenced by the low values of the $\beta$ metric shown in figure \ref{fig28}. Such metric measures the alignment between SPOD modes of the static and flight cases, taking into account the mean flow scaling by $x_c$. At KH-dominated frequencies, on the other hand, $\beta$ is significantly higher, showing that the spatial organisation of KH wavapackets, unlike that or Orr and streaky structures, in largely established by the potential core length.

The most salient modifications on the energy spectrum are found to be well-captured by results of a locally-parallel model, based on the difference in the mean flows of the two jets. The attenuation of KH wavepackets and streaky structures are predicted by eigen- and resolvent analysis, respectively. Stabilization of the KH mechanism is consistent with the weakening of the leading SPOD mode in the range $0.4 \leqslant St \leqslant 0.8$ with respect to suboptimal modes, showing a deterioration of the low-rank behaviour of the jet in that zone. Moreover, resolvent analysis also predicts a reduction between the optimal gains, $\sigma_1$ and those of suboptimal modes at different $m$ in the near-nozzle region, consistent with the reduction in eigenvalue separation at $St \to 0$ (see figure \ref{fig12}). These results show that, despite the fully developed turbulence and the high Reynolds number, the changes in the flow dynamics produced by the flight stream are due, to a great extent, to a linear mean-flow effect rather than a more complex, nonlinear reorganisation of the flow.

Overall, the results of the present study show that the reduction in turbulent kinetic energy produced by the flight stream, and frequently evoked in the literature \citep{TannaMorris1977}, is largely underpinned by the weakening of different categories of coherent structures. Coherent structures are widely considered to underpin sound generation in jets, and therefore their attenuation is expected to be associated with the broadband changes in the acoustic spectra shown in figure \ref{fig1}. In this sense, the results reported here may be considered as a departure point for a deeper analysis of how associated sound-source mechanisms are impacted by the flight stream. However, educing and modelling sound-source mechanisms in a turbulent jet is an exceptionally delicate and complex task, given the acoustic inefficiency of the flow fluctuations that drive the sound field \citep{Karban_etal2022}, and is beyond the scope of this work. Future work will address implications of these results for sound radiation. Among other aspects, it is worth studying how the reorganisation of energy in the frequency-wavenumber space affects the sound direcitivity associated with different coherent structures, or how the dynamics of such structures is impacted by a flight stream if shear is held constant.

\section*{Acknowledgements}

The authors would like to thank Damien Eysseric for his invaluable work during the PIV measurements. The authors also acknowledge Drs. Matteo Mancinelli, Eduardo Martini, Petr\^onio Nogueira and Bruno Zebrowski for helpful discussions and insights concerning the mean-flow model. This work has received funding from the Clean Sky 2 Joint Undertaking (JU) under the European Union's Horizon 2020 research and innovation programme under grant agreement No 785303. Results reflect only the authors' view and the JU is not responsible for any use that may be made of the information it contains. The LES studies performed at Cascade were supported in part by NAVAIR SBIR project with computational resources provided by DoD HPCMP. I.A.M. also acknowledges support from the Science Without Borders program through the CNPq Grant No. 200676/2015-6.

\appendix

\section{Azimuthal normalisation of SPOD energy maps}\label{appA}

\begin{figure}
\centering
\includegraphics[trim=4cm 0cm 4cm 0cm, clip=true,width=0.9\linewidth]{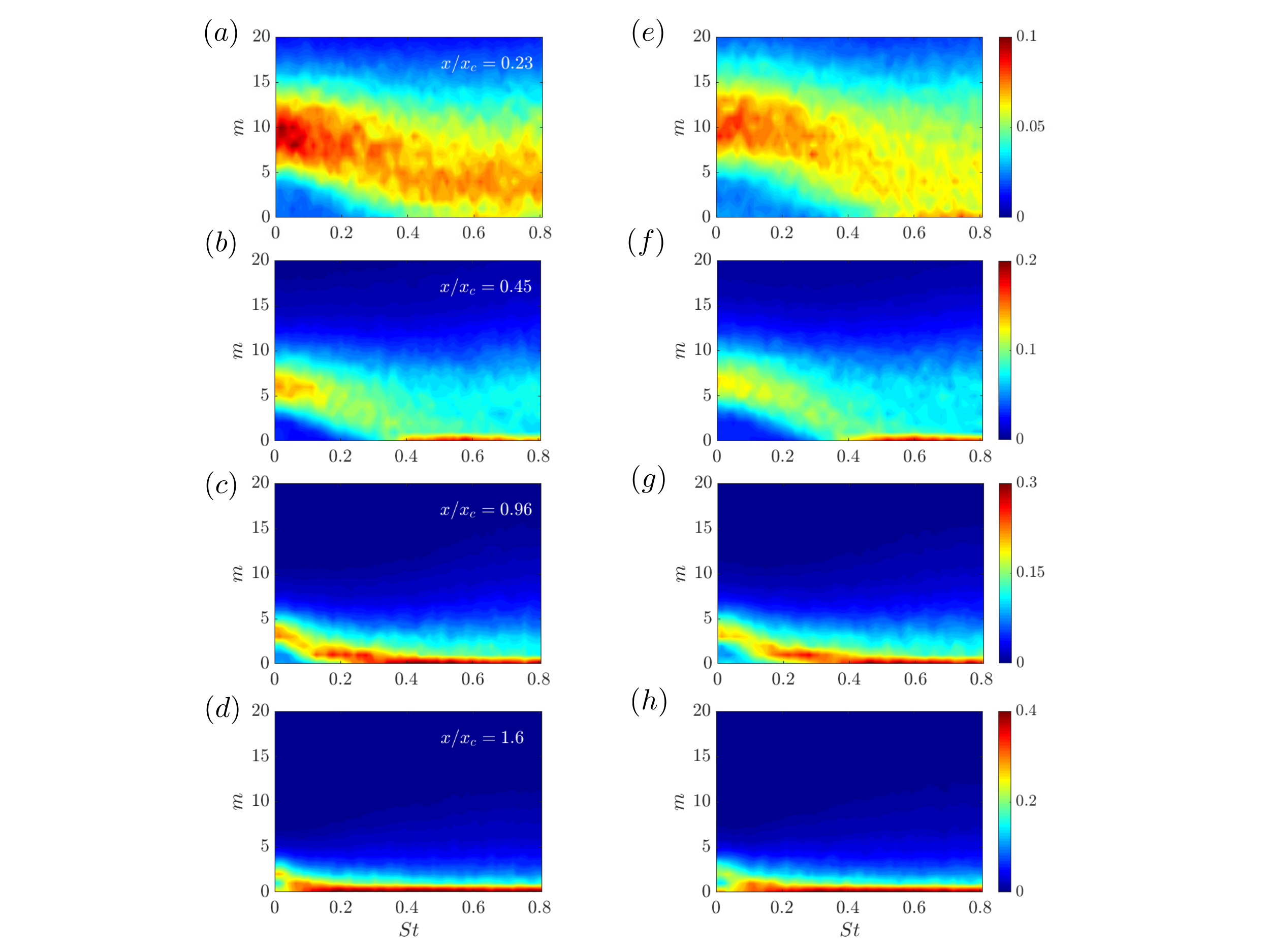}
\caption{Modal energy maps of the leading SPOD mode in the $St-m$ plane for different streamwise positions based on PIV data. Modal energy is normalised by the sum of energy across azimuthal modes $\sum_{m}\lambda_{1}(m,St)$ for each Strouhal number. (a)-(d): maps for the $M_{f}=0$ case; (e)-(h): maps for the $M_{f}=0.15$ case.}
\label{fig14}
\end{figure}

An alterative way of looking at the SPOD energy maps is obtained by normalising the modal energy at each $(St,m)$ pair by the by the sum of energy across all wavenumbers, $\sum_{m}\lambda_{1}(m,St)$,

\begin{equation}
\tilde{\lambda_{1}}=\lambda_{1}(m, St)/\sum_{m}\lambda_{1}(m,St),
\end{equation}
as proposed by \cite{PickeringJFM2020}. This metric, shown in figure \ref{fig14}, provides a more direct indication of the most energetic azimuthal wavenumber at each $St$. Close to the nozzle exit, helical modes (underpinned by the lift-up mechanism) are clearly dominant. As the jet evolves downstream, $m=0$ Kelvin-Helmholtz wavepackets grow exponentially and dominate the spectrum at $St\geqslant 0.4$. This is seen here more clearly than in figure \ref{fig12}. Further downstream, mode $m=1$ becomes important in a broad range of Strouhal numbers, and the Orr mechanism also starts to leave its signature at low Strouhal numbers. However, in the limit $St \to 0$, the streak mechanism ($m>0$) is always the dominant one. When normalised this way, the energy maps for the two jets were found to display a very similar organisation over an extensive streamwise range. At the very initial jet region, represented here by the plots at $x/x_c=0.23$, it can be seen, however, that the $M_f=0.15$ case has a broader spectrum in the $m$ direction. But the discrepancy with respect to the baseline case diminishes with increasing streamwise distance, and the maps for both jets are globally very much alike. 

\section{Linearised Navier-Stokes equations with a locally-parallel assumption}\label{appB}

The point of departure is the compressible Navier-Stokes equations:

\begin{equation}
\frac{\partial \rho}{\partial t} +\div (\rho \mathbf{u}),
\label{continuity}
\end{equation}

\begin{equation}
\rho\left[ \frac{\partial \mathbf{u}}{\partial t} + \mathbf{u}\cdot \grad \mathbf{u} \right] = -\grad p +\grad \left[ \left(\beta - \frac{2}{3}\mu \right) \div \mathbf{u} \right] +\div \left[ \mu \left( \grad \mathbf{u} +\left( \grad \mathbf{u} \right)^{T} \right) \right],
\label{momentum}
\end{equation}

\begin{equation}
\rho C_{v} \left[ \frac{\partial T}{\partial t} + \mathbf{u} \cdot \grad T \right] = -p \div \mathbf{u} + \Phi + k \laplacian T,
\label{energy}
\end{equation}
complemented by the state equation for an ideal gas. $\beta$ is the bulk viscosity, $C_{v}$ is the specific heat at constant volume and $k$ is the fluid thermal conductivity. For the sake of simplicity, we set $\beta=0$. Furthermore, since the jets under study here are isothermal, temperature gradients are negligible and the viscosity, $\mu$ was assumed to be constant throughout the domain. The energy dissipation term, $\Phi$, is given by,

\begin{equation}
\Phi= \left[ \left(\beta - \frac{2}{3}\mu \right) \div \mathbf{u} + \mu \left( \grad \mathbf{u} +\left( \grad \mathbf{u} \right)^{T} \right) \right] \left[ \mu \left( \grad \mathbf{u} +\left( \grad \mathbf{u} \right)^{T} \right) \right].
\end{equation}
The variables are then normalised using jet quantities: $\mathbf{u}^{+} = \mathbf{u}/U_{j}$, $p^{+} = p/(\rho_{j} U_{j}^2)$, $T^{+} = T/T_{j}$, $\rho^{+}=\rho/\rho_{j}$, $t^{+} = t U_{j}/D$, $\mathbf{x}^{+}=\mathbf{x}/D$. For the linearisation of equations \ref{continuity}-\ref{energy} flow variables are decomposed into a mean and a fluctuation component,

\begin{equation}
\mathbf{q}(x,r,\theta,t) = \bar{\mathbf{q}}(x,r,\theta) + \mathbf{q}'(x,r,\theta,t).
\end{equation}
Under the assumption of parallel flow, all streamwise derivatives of mean quantities are neglected, $\partial \bar{()}/\partial x = 0$. Using the axisymmetry of the jet, we can also set $\bar{\mathbf{u}} = \left[ \bar{U}_{x}(r), \bar{U}_{r}=0, \bar{U}_{\theta}=0 \right]$, $\bar{\rho}=\bar{\rho}(r)$ and $\bar{T}=\bar{T}(r)$. After all the simplifications, substitution of the \textit{Ansatz} in \ref{ansatz} into the linearised equations leads to the following relation,

\begin{equation}
-i \omega \left[\begin{array}{c}
\hat{\rho}\\
\hat{u}_{x}\\
\hat{u}_{r}\\
\hat{u}_{\theta}\\
\hat{T}\\
\end{array}\right] + (\mathcal{A}_{0} + \alpha\mathcal{A}_{1}+ \alpha^2 \mathcal{A}_{2})_{\bar{\mathbf{q}}} \left[\begin{array}{c}
\hat{\rho}\\
\hat{u}_{x}\\
\hat{u}_{r}\\
\hat{u}_{\theta}\\
\hat{T}\\
\end{array}\right] = \mathbf{f}.
\label{ns_3}
\end{equation}
The operators $\mathcal{A}_{0}$, $\mathcal{A}_{1}$ and $\mathcal{A}_{2}$, which compose the term $\mathcal{A}_{\bar{\mathbf{q}},\alpha,m}$ in equation \ref{ns_fft}, are:

\begin{equation}
    \resizebox{1\textwidth}{!}
     {%
     $\displaystyle
\mathcal{A}_{0}=\left[\begin{array}{ccccc}
0 & 0 & \partial_r\bar{\rho} +\bar{\rho}\left[ \mathbf{D}_{r_1} + \frac{1}{r} \right] & \bar{\rho}\frac{im}{r} & 0 \\
0 & -\frac{1}{Re}\left[ \mathbf{D}_{r_2} +\frac{1}{r}\mathbf{D}_{r_1} -\frac{m^2}{r^2} \right] & \bar{\rho}\partial_r\bar{U}_x & 0 & 0 \\
\frac{\bar{T}}{\gamma M_{j}^2}\mathbf{D}_{r_1} +\frac{1}{\gamma M_{j}^2}\partial_r\bar{T} & 0 & \frac{1}{Re}\left[ -\frac{4}{3} \left( \mathbf{D}_{r_2} +\frac{1}{r}\mathbf{D}_{r_1} - \frac{1}{r^2} \right) -\frac{m^2}{r^2} \right] & \frac{im}{3rRe} \left[ -\mathbf{D}_{r_1} + \frac{7}{r}\right] & \frac{\bar{T}}{\gamma M_{j}^2}\mathbf{D}_{r_1} + \frac{1}{\gamma M_{j}^2}\partial_r \bar{\rho} \\
\frac{im\bar{T}}{\gamma M_{j}^2 r} & 0 & -\frac{1}{3Re}\frac{im}{r}\left[\mathbf{D}_{r_1} + \frac{7}{r} \right] & \frac{1}{Re} \left[ \frac{4}{3}\frac{m^2}{r^2} -\mathbf{D}_{r_2} -\frac{1}{r}\mathbf{D}_{r_1} +\frac{1}{r^2} \right] & \frac{im \bar{\rho}}{r \gamma M_{j}^2} \\
0 & -\frac{2M_{j}\gamma(\gamma-1)}{Re}\partial_r \bar{U}_x \mathbf{D}_{r_1} & \bar{\rho} \partial_r \bar{T} +\bar{\rho}(\gamma -1)\bar{T} \left( \mathbf{D}_{r_1} + \frac{1}{r} \right) +\partial_r \bar{T} & \bar{\rho}(\gamma-1)\bar{T}\frac{im}{r} & -\frac{\gamma}{Re Pr}\left[ \mathbf{D}_{r_2} + \frac{1}{r}\mathbf{D}_{r_1} + \frac{m^2}{r^2} \right] \\
\end{array}\right]$%
     }
     \label{A0}
\end{equation}


\begin{equation}
\mathcal{A}_{1}=\left[\begin{array}{ccccc}
i\bar{U}_{x} & i\bar{\rho} & 0 & 0 & 0\\
i\frac{\bar{T}}{\gamma M_{j}^2} & i\bar{\rho}\bar{U}_{x} & -\frac{1}{3Re} \left[ i\mathbf{D}_{r_1} +\frac{1}{r} \right] & \frac{1}{Re} \left[ -\frac{2}{3} \frac{m^2}{r^2} + \frac{m}{r}  \right] & \frac{i\bar{\rho}}{M_{j}^2}\\
0 & -\frac{1}{3Re}i\mathbf{D}_{r_1} & i\bar{\rho}\bar{U}_{x} & 0 & 0\\
0 & \frac{1}{3Re} \frac{m}{r} & 0 & i\bar{\rho}\bar{U}_{x} & 0\\
0 & i\bar{\rho}(\gamma -1)\bar{T} & -\frac{2iM_{j}\gamma(\gamma-1)}{Re}\partial_r \bar{U}_x & 0 & i\bar{\rho}\bar{U}_{x}\\
\end{array}\right],
\label{A1}
\end{equation}

\begin{equation}
\mathcal{A}_{2}=\left[\begin{array}{ccccc}
0 & 0 & 0 & 0 & 0\\
\frac{4}{3Re} & 0 & 0 & 0 & 0\\
0 & 0 & \frac{1}{Re} & 0 & 0\\
0 & 0 & 0 & \frac{1}{Re} & 0\\
0 & 0 & 0 & 0 & \frac{\gamma}{Re Pr}\\
\end{array}\right],
\label{A2}
\end{equation}
where $Pr$ is the Prandtl number, which is taken as 0.7, and $\gamma=1.4$ is the specific heat ration. The superscripts $^{+}$ have been dropped for simplicity. $\mathbf{D}_{r_1}$ $\mathbf{D}_{r_2}$ are the first- and second-order Chebyshev derivation matrices. $\partial_{r} \bar{()}$ denotes the radial derivative of the mean field. As mentioned in \S \ref{sec:resolvent}, for the resolvent analysis, the molecular Reynolds number, $Re$ is later replaced by a turbulent Reynolds number, $Re_{T}$. Each element of $A_{0}$, $A_{1}$ $A_{2}$ is a $N \times N$ matrix, where $N$ is the number of Chebyshev points used. We found that 600 points were sufficient to attain converged results. The observation and forcing restrictions matrices, $\mathcal{C}$ and $\mathcal{B}$, respectively, in Equation \ref{ns_fft}, are equal to the identity matrix of dimension $5N \times 5N$. Dirichlet boundary conditions were applied in the far-field, $r/D \to \infty$. At the jet centerline, $r/D=0$, following \citet{Khorrami_etal_JCP1989} and \citet{LesshaftHuerre2007} the following symmetry boundary conditions were imposed:

\begin{equation}
  \begin{drcases}
    \frac{\partial \hat{\rho}}{\partial r} = 0 & \\
    \frac{\partial \hat{u}_{x}}{\partial r} = 0 & \\
    \hat{u}_{r} = 0 & \\
    \hat{u}_{\theta} = 0 & \\
    \frac{\partial \hat{T}}{\partial r} = 0 & \\
  \end{drcases}
  \text{for $m$=0},
\end{equation}

\begin{equation}
  \begin{drcases}
    \hat{\rho}= 0 & \\
    \hat{u}_{x} = 0 & \\
    \frac{\partial \hat{u}_{r}}{\partial r} = 0 & \\
    \hat{u}_r + i\hat{u}_{\theta} = 0 & \\
    \hat{T} = 0 & \\
  \end{drcases}
  \text{for $m$=1},
\end{equation}

\begin{equation}
  \begin{drcases}
    \hat{\rho}= 0 & \\
    \hat{u}_{x} = 0 & \\
    \hat{u}_{r} = 0 & \\
    \hat{u}_{\theta} = 0 & \\
    \hat{T} = 0 & \\
  \end{drcases}
  \text{for $m  \geqslant 2$},
\end{equation}
where the conditions for $\hat{\rho}$ and $\hat{T}$ for $m=0$ were derived assuming $\hat{u}_r$ to be an odd function of $r$.

\bibliographystyle{unsrtnat}
\bibliography{bibfile}

\begin{thebibliography}{80}
\providecommand{\natexlab}[1]{#1}
\providecommand{\url}[1]{\texttt{#1}}
\expandafter\ifx\csname urlstyle\endcsname\relax
  \providecommand{\doi}[1]{doi: #1}\else
  \providecommand{\doi}{doi: \begingroup \urlstyle{rm}\Url}\fi

\bibitem[Hussain(1986)]{HussainCoherentStructures}
A.~K. M.~F. Hussain.
\newblock Coherent structures and turbulence.
\newblock \emph{Journal of Fluid Mechanics}, 173:\penalty0 303--356, 1986.

\bibitem[Brown and Roshko(1974)]{Brown&Roshko1974}
G.~L. Brown and A.~Roshko.
\newblock On density effects and large structure in turbulent mixing layers.
\newblock \emph{Journal of Fluid Mechanics}, 64:\penalty0 775--816, 1974.

\bibitem[Mollo-Christensen(1967)]{MolloChristensen1967}
E.~Mollo-Christensen.
\newblock Jet noise and shear flow instability seen from an experimenter's
  viewpoint.
\newblock \emph{Journal of Applied Mechanics}, 34:\penalty0 1--7, 1967.

\bibitem[Crow and Champagne(1971)]{CrowChampagne}
S.~Crow and F.~Champagne.
\newblock Orderly structure in jet turbulence.
\newblock \emph{Journal of Fluid Mechanics}, 48:\penalty0 547--591, 1971.

\bibitem[Moore(1977)]{Moore1977}
C.~J. Moore.
\newblock The role of shear-layer instability waves in jet exhaust noise.
\newblock \emph{Journal of Fluid Mechanics}, 80:\penalty0 321--367, 1977.

\bibitem[Zaman and Hussain(1980)]{HussainZaman1980_1}
K.~B. M.~Q. Zaman and A.~K. M.~F. Hussain.
\newblock Vortex pairing in a circular jet under controlled excitation. part 1.
  general jet response.
\newblock \emph{Journal of Fluid Mechanics}, 101\penalty0 (3):\penalty0
  449–491, 1980.
\newblock \doi{10.1017/S0022112080001760}.

\bibitem[Hussain and Zaman(1980)]{HussainZaman1980_2}
A.~K. M.~F. Hussain and K.~B. M.~Q. Zaman.
\newblock Vortex pairing in a circular jet under controlled excitation. part 2.
  coherent structure dynamics.
\newblock \emph{Journal of Fluid Mechanics}, 101\penalty0 (3):\penalty0
  493–544, 1980.
\newblock \doi{10.1017/S0022112080001772}.

\bibitem[Hussain and Zaman(1981)]{HussainZamanpreferred}
A.~K. M.~F. Hussain and K.~B. M.~Q. Zaman.
\newblock The 'preferred' mode of the axisymmetric jet.
\newblock \emph{Journal of Fluid Mechanics}, 110:\penalty0 39--71, 1981.

\bibitem[Petersen and Samet(1988)]{PetersenSamet}
R.~A. Petersen and M.~M. Samet.
\newblock On the preferred mode of jet instability.
\newblock \emph{Journal of Fluid Mechanics}, 194:\penalty0 153--173, 1988.

\bibitem[Lumley(1967)]{Lumley1967}
J.~L. Lumley.
\newblock The structure of inhomogeneous turbulent flows.
\newblock In \emph{Atmospheric Turbulence and Radio Wave Propagation (ed. A. M.
  Yaglom and V. I. Tartarsky)}, pages 166--177, Nauka, Moscow, 1967.

\bibitem[Suzuki and Colonius(2006)]{SuzukiColonius}
T.~Suzuki and T.~Colonius.
\newblock Instability waves in a subsonic round jet detected using a near-field
  phased microphone array.
\newblock \emph{Journal of Fluid Mechanics}, 565:\penalty0 197--226, 2006.

\bibitem[Gudmundsson and Colonius(2011)]{GudmundssonColonius}
K.~Gudmundsson and T.~Colonius.
\newblock Instability wave models for the near-field fluctuations of turbulent
  jets.
\newblock \emph{Journal of Fluid Mechanics}, 689:\penalty0 97--128, 2011.

\bibitem[Breakey et~al.(2017)Breakey, Jordan, Cavalieri, Nogueira, L\'eon,
  Colonius, and Rodr\'{\i}guez]{BreakeyPhysRevFluids}
D.~Breakey, P.~Jordan, A.~V.~G. Cavalieri, P.~Nogueira, O.~L\'eon, T.~Colonius,
  and D.~Rodr\'{\i}guez.
\newblock Experimental study of turbulent-jet wave packets and their acoustic
  efficiency.
\newblock \emph{Phys. Rev. Fluids}, 2, 2017.

\bibitem[Cavalieri et~al.(2013)Cavalieri, Rodr\'iguez, Jordan, Colonius, and
  Gervais]{Wavepacketsvelocity}
A.~V.~G. Cavalieri, D.~Rodr\'iguez, P.~Jordan, T.~Colonius, and Y.~Gervais.
\newblock Wavepackets in the velocity field of turbulent jets.
\newblock \emph{Journal of fluid mechanics}, 730:\penalty0 559--592, 2013.

\bibitem[Jaunet et~al.(2017)Jaunet, Jordan, and Cavalieri]{CoherenceVincent}
V.~Jaunet, P.~Jordan, and A.~V.~G. Cavalieri.
\newblock Two-point coherence of wavepackets in turbulent jets.
\newblock \emph{Physical Review Fluids}, 2\penalty0 (024604), 2017.

\bibitem[Sasaki et~al.(2017)Sasaki, Cavalieri, Jordan, Schmidt, Colonius, and
  Br\`es]{SasakiRapids}
K.~Sasaki, A.~V.~G. Cavalieri, P.~Jordan, O.~T. Schmidt, T.~Colonius, and
  G.~Br\`es.
\newblock High-frequency wavepackets in turbulent jets.
\newblock \emph{Journal of Fluid Mechanics}, 830, 2017.

\bibitem[Orr(1907)]{Orr}
W.~Orr.
\newblock The stability or instability of steady motions of a perfect liquid
  and of a viscous liquid. part i: a perfect liquid.
\newblock In \emph{Proc. Royal Irish Acad. Sec. A: Math. Phys. Sci.}, pages
  9--68, 1907.

\bibitem[Brandt(2014)]{Brandt2014}
L.~Brandt.
\newblock The lift-up effect: the linear mechanism behind transition and
  turbulence in shear flows.
\newblock \emph{European Journal of Mechanics (B/Fluids)}, 47:\penalty0 80--96,
  2014.

\bibitem[Jim\'enez(2018)]{Jimenez2018}
J.~Jim\'enez.
\newblock Coherent structures in wall-bounded turbulence.
\newblock \emph{Journal of Fluid Mechanics}, 842, 2018.

\bibitem[Becker and Massaro(1968)]{BeckerMassaro}
H.~A. Becker and T.~A. Massaro.
\newblock Vortex evolution in a round jet.
\newblock \emph{Journal of Fluid Mechanics}, 31\penalty0 (3):\penalty0
  435--448, 1968.

\bibitem[Browand and Laufer(1975)]{BrowandLaufer}
F.K. Browand and J.~Laufer.
\newblock The roles of large scale structures in the initial development of
  circular jets.
\newblock In \emph{Symposia on Turbulence in Liquids}, University of
  Missouri-Rolla, 1975.

\bibitem[Dimotakis et~al.(1983)Dimotakis, Miake-Lye, and
  Papantoniou]{Dimotakis1983}
P.E. Dimotakis, R.~C. Miake-Lye, and D.~A. Papantoniou.
\newblock Structure and dynamics of round turbulent jets.
\newblock \emph{Phys. Fluids}, 26\penalty0 (11):\penalty0 3185--3192, 1983.

\bibitem[Yule(1978)]{YuleJFM1978}
A.~J. Yule.
\newblock Large-scale structure in the mixing layer of a round jet.
\newblock \emph{Journal of Fluid Mechanics}, 89\penalty0 (3):\penalty0
  413--432, 1978.

\bibitem[Ag\"u\'i and Hesselink(1988)]{Agui1988}
J.~C. Ag\"u\'i and L.~Hesselink.
\newblock Flow visualization and numerical analysis of a coflowing jet: a
  three-dimensional approach.
\newblock \emph{Journal of Fluid Mechanics}, 191:\penalty0 19--45, 1988.

\bibitem[Garnaud et~al.(2013{\natexlab{a}})Garnaud, Lesshafft, Schmid, and
  Huerre]{Garnaud}
X.~Garnaud, L.~Lesshafft, P.~Schmid, and P.~Huerre.
\newblock The preferred mode of incompressible jets: linear frequency response
  analysis.
\newblock \emph{Journal of Fluid Mechanics}, 716:\penalty0 189--202,
  2013{\natexlab{a}}.

\bibitem[Jeun et~al.(2016)Jeun, Nichols, and Jovanovic]{Jeunetal}
J.~Jeun, J.~W. Nichols, and M.~R. Jovanovic.
\newblock Input-output analysis of high-speed axisymmetric isothermal jet
  noise.
\newblock \emph{Physics of Fluids (1994-present)}, 28(4)\penalty0 (047101),
  2016.

\bibitem[Semeraro et~al.(2016)Semeraro, Jaunet, Jordan, Cavalieri, and
  Lesshafft]{SemeraroAIAA2016}
O.~Semeraro, V.~Jaunet, P.~Jordan, A.~V.~G. Cavalieri, and L.~Lesshafft.
\newblock Stochastic and harmonic optimal forcing in subsonic jets.
\newblock In \emph{Proceedings of the 21st AIAA/CEAS Aeroacoustics Conference
  and Exhibit}, Lyon, France, 2016. AIAA.

\bibitem[Tissot et~al.(2017)Tissot, Zhang, Jr., Cavalieri, and
  Jordan]{TissotJFM2017}
G.~Tissot, M.~Zhang, F.~C.~Laj\'us Jr., A.~V.~G. Cavalieri, and P.~Jordan.
\newblock Sensitivity of wavepackets in jets to nonlinear effects: the role of
  the critical layer.
\newblock \emph{Journal of Fluid Mechanics}, 811:\penalty0 95--137, 2017.

\bibitem[Schmidt et~al.(2018)Schmidt, Towne, Rigas, Colonius, and
  Brès]{SchmidtetalJFM2018}
O.~T. Schmidt, A.~Towne, G.~Rigas, T.~Colonius, and G.~A. Brès.
\newblock Spectral analysis of jet turbulence.
\newblock \emph{Journal of Fluid Mechanics}, 855:\penalty0 953--982, 2018.

\bibitem[Lesshafft et~al.(2019)Lesshafft, Semeraro, Jaunet, Cavalieri, and
  Jordan]{LesshafftPFR2019}
L.~Lesshafft, O.~Semeraro, V.~Jaunet, A.~V.~G. Cavalieri, and P.~Jordan.
\newblock Resovlent-based modelling of coherent structures wave packets in a
  turbulent jet.
\newblock \emph{Physical Review Fluids}, 4:\penalty0 063901, 2019.

\bibitem[Nogueira et~al.(2019)Nogueira, Cavalieri, Jordan, and
  Jaunet]{NogueiraJFM2019}
P.~Nogueira, A.~V.~G. Cavalieri, P.~Jordan, and V.~Jaunet.
\newblock Large-scale streaky structures in turbulent jets.
\newblock \emph{Journal of Fluid Mechanics}, 873:\penalty0 211--237, 2019.

\bibitem[Pickering et~al.(2020)Pickering, Rigas, Nogueira, Cavalieri, Schmidt,
  and Colonius]{PickeringJFM2020}
E.~Pickering, G.~Rigas, P.~Nogueira, A.~V.~G. Cavalieri, O.~Schmidt, and
  T.~Colonius.
\newblock Lift-up, kelvin-helmholtz and orr mechanisms in turbulent jets.
\newblock \emph{Journal of Fluid Mechanics}, 896:\penalty0 A2, 2020.

\bibitem[Jordan and Colonius(2013)]{JordanColoniusReview}
P.~Jordan and T.~Colonius.
\newblock Wave packets and turbulent jet noise.
\newblock \emph{Annual Review of Fluid Mechanics}, 45:\penalty0 173--195, 2013.

\bibitem[Tanna and Morris(1977)]{TannaMorris1977}
H.~Tanna and P.~Morris.
\newblock In-flight simulation experiments on turbulent jet mixing noise.
\newblock \emph{Journal of Sound and Vibration}, 53\penalty0 (3):\penalty0
  343--359, 1977.

\bibitem[Michalke and Hermann(1982)]{MuchalkeHermann1982}
A.~Michalke and G.~Hermann.
\newblock On the inviscid instability of a circular jet with external flow.
\newblock \emph{Journal of Fluid Mechanics}, 114:\penalty0 343--359, 1982.

\bibitem[Soares et~al.(2020)Soares, Cavalieri, Kopiev, and
  Faranosov]{SoaresAIAA2020}
L.~F. Soares, A.~V.~G. Cavalieri, V.~Kopiev, and G.~Faranosov.
\newblock Flight effects on turbulent-jet wave packets.
\newblock \emph{AIAA Journal}, 58\penalty0 (9), 2020.

\bibitem[Garnaud et~al.(2013{\natexlab{b}})Garnaud, Sandberg, and
  Lesshafft]{Garnaud_AIAA_2013}
X.~Garnaud, R.~D. Sandberg, and L.~Lesshafft.
\newblock Global response to forcing in a subsonic jet: instability wavepackets
  and acoustic radiation.
\newblock In \emph{19th AIAA/CEAS Aeroacoustics Conference},
  2013{\natexlab{b}}.

\bibitem[Glahn et~al.(1973)Glahn, Groesbeck, and Goodykoontz]{VonGlahn}
U.~Von Glahn, D.~Groesbeck, and J.~Goodykoontz.
\newblock Velocity decay and acoustic characteristics of various nozzle
  geometries with forward velocity.
\newblock In \emph{6th Fluid and PlasmaDynamics Conference}. 1973.
\newblock \doi{10.2514/6.1973-629}.

\bibitem[Cocking and Bryce(1975)]{Cocking}
B.~Cocking and W.~Bryce.
\newblock Subsonic jet noise in flight based on some recent wind-tunnel tests.
\newblock In \emph{2nd Aeroacoustics Conference}. 1975.
\newblock \doi{10.2514/6.1975-462}.

\bibitem[Bushell(1975)]{Bushel}
K.~Bushell.
\newblock Measurement and prediction of jet noise in flight.
\newblock In \emph{2nd Aeroacoustics Conference}. 1975.
\newblock \doi{10.2514/6.1975-461}.

\bibitem[Packman et~al.(1976)Packman, Ng, and Paterson]{Packman}
A.~B. Packman, K.~W. Ng, and R.~W. Paterson.
\newblock Effect of simulated forward flight on subsonic jet exhaust noise.
\newblock \emph{Journal of Aircraft}, 13\penalty0 (12):\penalty0 1007--1013,
  1976.
\newblock \doi{10.2514/3.58741}.

\bibitem[Plumblee(1976)]{Plumbee}
H.~E. Plumblee.
\newblock Effects of forward flight on turbulent jet mixing noise.
\newblock Technical Report CR-2702, NASA, 1976.

\bibitem[Bryce(1984)]{Bryce1984}
W.~Bryce.
\newblock The prediction of static-to-flight changes in jet noise.
\newblock In \emph{9th Aeroacoustics Conference}. 1984.
\newblock \doi{10.2514/6.1984-2358}.

\bibitem[Morfey and Tester(1977)]{MorfeyTester}
C.~L. Morfey and B.~J. Tester.
\newblock Noise measurements in a free jet flight simulation facility: shear
  layer refraction and facility-to-flight corrections.
\newblock \emph{Journal of Sound and Vibration}, 54\penalty0 (1):\penalty0
  83--106, 1977.

\bibitem[Vishwanathan and Czech(2011)]{VishwanathanFlight}
K.~Vishwanathan and M.~Czech.
\newblock Measurement and modeling of effect of forward flight on jet noise.
\newblock \emph{AIAA journal}, 49\penalty0 (1), 2011.

\bibitem[Cavalieri et~al.(2011)Cavalieri, Jordan, Agarwal, and
  Gervais]{CavalieriJitterJSV}
A.~V.~G. Cavalieri, P.~Jordan, A.~Agarwal, and Y.~Gervais.
\newblock Jittering wave-packet models for subsonic jet noise.
\newblock \emph{Journal of Sound and Vibration}, 330:\penalty0 4474--4492,
  2011.

\bibitem[Cavalieri et~al.(2012)Cavalieri, Jordan, Colonius, and
  Gervais]{AndreJFM2012}
A.~V.~G. Cavalieri, P.~Jordan, T.~Colonius, and Y.~Gervais.
\newblock Axisymmetric superdirectivity in subsonic jets.
\newblock \emph{Journal of Fluid Mechanics}, 704:\penalty0 338--420, 2012.

\bibitem[Maia et~al.(2019)Maia, Jordan, Cavalieri, and Jaunet]{MaiaPRSA2019}
I.~A. Maia, P.~Jordan, A.~V.~G. Cavalieri, and V.~Jaunet.
\newblock Two-point wavepacket modelling of jet noise.
\newblock \emph{Proceedings of the Royal Society A}, 475\penalty0 (20190199),
  2019.

\bibitem[Lighthill(1952)]{Lighthill}
M.~J. Lighthill.
\newblock On sound generated aerodynamically i. general theory.
\newblock \emph{Proceedings of the Royal Society of London A: Mathematical,
  Physical and Engineering Sciences}, 211:\penalty0 564--787, 1952.

\bibitem[Br\`es et~al.(2017)Br\`es, Ham, Nichols, and Lele]{BresAIAA2017}
G.~A. Br\`es, F.~E. Ham, J.~W. Nichols, and S.~K. Lele.
\newblock Unstructured large eddy simulations of supersonic jets.
\newblock \emph{AIAA Journal}, 55\penalty0 (4):\penalty0 1164--1184, 2017.

\bibitem[Br\`es et~al.(2018)Br\`es, Jordan, Jaunet, Rallic, Cavalieri, Towne,
  Lele, Colonius, and Schmidt]{BresJFM2018}
G.~A. Br\`es, P.~Jordan, V.~Jaunet, M.~Le Rallic, A.~V.~G. Cavalieri, A.~Towne,
  S.~K. Lele, T.~Colonius, and O.~T. Schmidt.
\newblock Importance of the nozzle-exit boundary-layer state in subsonic
  turbulent jets.
\newblock \emph{Journal of Fluid Mechanics}, 851:\penalty0 83--124, 2018.

\bibitem[Scarano(2001)]{Scarano2002}
F.~Scarano.
\newblock Iterative image deformation methods in piv.
\newblock \emph{Measurement Science Technology}, 13\penalty0 (1), 2001.

\bibitem[Freund(1997)]{Freund-97}
J.~B. Freund.
\newblock Proposed inflow/outflow boundary condition for direct computation of
  aerodynamic sound.
\newblock \emph{AIAA J.}, 35\penalty0 (4):\penalty0 740--742, 1997.

\bibitem[Mani(2012)]{mani12}
A.~Mani.
\newblock Analysis and optimization of numerical sponge layers as a
  nonreflective boundary treatment.
\newblock \emph{Journal of Computational Physics}, 231:\penalty0 704--716,
  2012.

\bibitem[Vreman(2004)]{Vreman:2004p754}
A.~Vreman.
\newblock An eddy-viscosity subgrid-scale model for turbulent shear flow:
  Algebraic theory and applications.
\newblock \emph{Physics of Fluids}, 16:\penalty0 3670--3681, Jan 2004.

\bibitem[Kawai and Larsson(2012)]{KawaiPoF2012}
S.~Kawai and J.~Larsson.
\newblock Wall-modeling in large eddy simulation: Length scales, grid
  resolution, and accuracy.
\newblock \emph{Physics of Fluids}, 24:\penalty0 015105, 2012.

\bibitem[Bodart and Larsson(2011)]{bodartwm2011}
J.~Bodart and J.~Larsson.
\newblock Wall-modeled large eddy simulation in complex geometries with
  application to high-lift devices.
\newblock In \emph{Annual Research Briefs}. Center for Turbulence Research,
  Stanford University, 2011.

\bibitem[Kaplan et~al.(2021)Kaplan, Jordan, Cavalieri, and
  Br\`es]{KaplanJFM2021}
O.~Kaplan, P.~Jordan, A.~V.~G. Cavalieri, and G.~Br\`es.
\newblock Nozzle dynamics and wavepackets in turbulent jets.
\newblock \emph{Journal of Fluid Mechanics}, 923:\penalty0 A22, 2021.

\bibitem[Michalke and Michel(1979)]{MICHALKE1979341}
A.~Michalke and U.~Michel.
\newblock Prediction of jet noise in flight from static tests.
\newblock \emph{Journal of Sound and Vibration}, 67\penalty0 (3):\penalty0
  341--367, 1979.
\newblock ISSN 0022-460X.

\bibitem[Citriniti and George(2000)]{CitrinitiGeorge2000}
J.~H. Citriniti and W.~K. George.
\newblock Reconstruction of the global velocity field in the axisymmetric
  mixing layer utilizing the proper orthogonal decomposition.
\newblock \emph{Journal of Fluid Mechanics}, 418:\penalty0 137--166, 2000.

\bibitem[Jung et~al.(2004)Jung, Gamard, and George]{JungJFM2004}
D.~Jung, S.~Gamard, and W.~K. George.
\newblock Downstream evolution of the most energetic modes in a turbulent
  axisymmetric jet at high reynolds number. part 1. the near-field region.
\newblock \emph{Journal of Fluid Mechanics}, 514:\penalty0 173--204, 2004.

\bibitem[McKeon and Sharma(2010)]{MckeonSharma}
B.~J. McKeon and A.~S. Sharma.
\newblock A critical-layer framework for turbulent pipe flow.
\newblock \emph{Journal of Fluid Mechanics}, 658:\penalty0 336--382, 2010.

\bibitem[Rodr\'iguez et~al.(2015)Rodr\'iguez, Cavalieri, Colonius, and
  Jordan]{RodriguezEuropean}
D.~Rodr\'iguez, A.~V.~G. Cavalieri, T.~Colonius, and P.~Jordan.
\newblock A study of wavepacket models for subsonic turbulent jets using local
  eingemode decompostion of piv data.
\newblock \emph{European Journal of Mechanics B/Fluids}, 49:\penalty0 308--321,
  2015.

\bibitem[Lesshafft and Huerre(2007)]{LesshaftHuerre2007}
L.~Lesshafft and P.~Huerre.
\newblock Linear impulse response in hot round jets.
\newblock \emph{Physics of Fluids}, 19\penalty0 (2), 2007.

\bibitem[Dergham et~al.(2013)Dergham, Sipp, and Robinet]{Derghametal2013}
G.~Dergham, D.~Sipp, and J-Ch. Robinet.
\newblock Stochastic dynamics and model reduction of amplifier flows: the
  backward facing step.
\newblock \emph{Journal of Fluid Mechanics}, 719:\penalty0 406--430, 2013.

\bibitem[Nogueira et~al.(2020)Nogueira, Cavalieri, Hanifi, and
  Henningson]{NogueiraetalTCFD2020}
P.~Nogueira, A.~V.~G. Cavalieri, A.~Hanifi, and D.~S. Henningson.
\newblock Resolvent analysis in unbounded flows: the role of free-stream modes.
\newblock \emph{Theoretical and Computational Fluid Dynamics}, 34:\penalty0
  163--176, 2020.

\bibitem[Crouch et~al.(2007)Crouch, Garbaruk, and Magidov]{Crouch_etal_JcP2007}
J.~D. Crouch, A.~Garbaruk, and D.~Magidov.
\newblock Predicting the onset of flow unsteadiness based on global
  instability.
\newblock \emph{Journal of Computational Physics}, 224\penalty0 (2):\penalty0
  924--940, 2007.

\bibitem[Oberleithner et~al.(2014)Oberleithner, Paschereit, and
  Wygnanski]{Oberleithner_etal_JFM2014}
K.~Oberleithner, C.~O. Paschereit, and I.~Wygnanski.
\newblock On the impact of swirl on the growth of coherent structures.
\newblock \emph{Journal of Fluid Mechanics}, 741:\penalty0 156--199, 2014.

\bibitem[Rukes et~al.(2016)Rukes, Paschereit, and
  Oberleithner]{Rukes_etal_2016}
L.~Rukes, O.~Paschereit, and K.~Oberleithner.
\newblock An assessment of turbulence models for linear hydrodynamic stability
  analysis of strongly swirling jets.
\newblock \emph{European Journal of Fluid Mechanics}, 59:\penalty0 205--218,
  2016.

\bibitem[Tammisola and Juniper(2016)]{Tammisola_Juniper_2016}
O.~Tammisola and M.~P. Juniper.
\newblock Coherent structures in a swirl injector at $re = 4800$ by nonlinear
  simulations and linear global modes.
\newblock \emph{Journal of Fluid Mechanics}, 3\penalty0 (5):\penalty0 620--657,
  2016.

\bibitem[Hwang and Cossu(2010)]{HwangCossu2010}
G.~Y. Hwang and C.~Cossu.
\newblock Amplification of coherent streaks in the turbulent couette flow: an
  input–output analysis at low reynolds number.
\newblock \emph{Journal of Fluid Mechanics}, 633:\penalty0 333--348, 2010.

\bibitem[Morra et~al.(2019)Morra, Semeraro, Henningson, and
  Cossu]{MorraetalArxiv2019}
P.~Morra, O.~Semeraro, D.~S. Henningson, and C.~Cossu.
\newblock On the relevance of reynolds stresses in resolvent analyses of
  turbulent wall-bounded flows.
\newblock \emph{Journal of Fluid Mechanics}, 867:\penalty0 969--984, 2019.

\bibitem[Pickering et~al.(2021)Pickering, Rigas, Schmidt, Sipp, and
  Colonius]{Pickeringetal_eddy_2021}
E.~Pickering, G.~Rigas, O.~T. Schmidt, D.~Sipp, and T.~Colonius.
\newblock Optimal eddy-viscosity models of coherent structures in turbulent
  jet.
\newblock \emph{Journal of Fluid Mechanics}, 917:\penalty0 A29, 2021.

\bibitem[Kuhn et~al.(2021)Kuhn, Soria, and Oberleithner]{Kuhn_etal_JFM2021}
P.~Kuhn, J.~Soria, and K.~Oberleithner.
\newblock Linear modelling of self-similar jet turbulence.
\newblock \emph{Journal of Fluid Mechanics}, 919:\penalty0 A7, 2021.

\bibitem[Chu(1965)]{Chu1965}
B.-T. Chu.
\newblock On the energy transfer to small scale disturbances in fluid flow
  (part 1).
\newblock \emph{Acta Mech.}, 1\penalty0 (3):\penalty0 215--234, 1965.

\bibitem[\r{A}kervik et~al.(2008)\r{A}kervik, Ehrenstein, Gallaire, and
  Henningson]{AKERVIK2008501}
Espen \r{A}kervik, Uwe Ehrenstein, Fran\c{c}ois Gallaire, and Dan~S.
  Henningson.
\newblock Global two-dimensional stability measures of the flat plate
  boundary-layer flow.
\newblock \emph{European Journal of Mechanics - B/Fluids}, 27\penalty0
  (5):\penalty0 501--513, 2008.
\newblock ISSN 0997-7546.

\bibitem[Arratia et~al.(2013)Arratia, Caulfield, and
  Chomaz]{arratia_caulfield_chomaz_2013}
C.~Arratia, C.~P. Caulfield, and J.-M. Chomaz.
\newblock Transient perturbation growth in time-dependent mixing layers.
\newblock \emph{Journal of Fluid Mechanics}, 717:\penalty0 90–133, 2013.
\newblock \doi{10.1017/jfm.2012.562}.

\bibitem[Hack and Moin(2017)]{hack_moin_2017}
M.~J.~P. Hack and P.~Moin.
\newblock Algebraic disturbance growth by interaction of orr and lift-up
  mechanisms.
\newblock \emph{Journal of Fluid Mechanics}, 829:\penalty0 112–126, 2017.
\newblock \doi{10.1017/jfm.2017.557}.

\bibitem[Karban et~al.(2022)Karban, Bugeat, Towne, Lesshafft, Agarwal, and
  Jordan]{Karban_etal2022}
U.~Karban, B.~Bugeat, A.~Towne, L.~Lesshafft, A.~Agarwal, and P.~Jordan.
\newblock An empirical model of noise sources in subsonic jets.
\newblock \emph{arXiv paper arXiv:2210.01866}, 2022.

\bibitem[Khorrami et~al.(1989)Khorrami, Malik, and Ash]{Khorrami_etal_JCP1989}
M.~R. Khorrami, M.~R. Malik, and R.~L. Ash.
\newblock Applications of spectral collocation techniques to the stability of
  swirling flows.
\newblock \emph{Journal of Computational Physics}, 81\penalty0 (206), 1989.

\end{thebibliography}

\end{document}